\documentclass[preprint]{aastex62}
\usepackage{subfigure}
\usepackage{amsmath}
\usepackage{xcolor}
\usepackage{amssymb}
\usepackage{rotating}
\usepackage[]{natbib}

\newcommand{\feh}{\hbox{$ [\mathrm{Fe}/\mathrm{H}]$}}
\newcommand{\alh}{\hbox{$ [\mathrm{Al}/\mathrm{H}]$}} 
\newcommand{\nih}{\hbox{$ [\mathrm{Ni}/\mathrm{H}]$}} 
 
\newcommand{\nife}{\hbox{$ [\mathrm{Ni}/\mathrm{Fe}]$}} 
\newcommand{\sife}{\hbox{$ [\mathrm{Si}/\mathrm{Fe}]$}} 
\newcommand{\mgfe}{\hbox{$ [\mathrm{Mg}/\mathrm{Fe}]$}} 
\newcommand{\bchisq}{\hbox{$\bar{\chi}^2$}}

\newcommand{\mgh}{\hbox{$ [\mathrm{Mg}/\mathrm{H}]$}} 
\newcommand{\alphafe}{\hbox{$ [\mathrm{\alpha}/\mathrm{Fe}]$}} 
 
\newcommand{\qlow}{Q$_{\rm low}$}
\newcommand{\qhigh}{Q$_{\rm high}$}

\newcommand{\mh}{\hbox{$ [{\rm m}/{\rm H}]$}} 
\newcommand{\mhDR}{\hbox{$ [{\rm m}/{\rm H}]_{\rm DR6}$}} 
\newcommand{\Mh}{\hbox{$ [{\rm M}/{\rm H}]$}} 
\newcommand{\Xh}{\hbox{$ [{\rm X}/{\rm H}]$}} 
\newcommand{\numax}{\hbox{$\nu_\mathrm{max}$}} 
\newcommand{\dnu}{\hbox{$\Delta\nu$}} 
\newcommand{\kms}{\hbox{$\mathrm{km\,s}^{-1}$}}
\newcommand{\SNR}{\hbox{SNR}}
\newcommand{\dex}{\hbox{dex}}
\newcommand{\loggf}{\hbox{$\log gf$}}
\newcommand{\logg}{\hbox{$\log g$}}
\newcommand{\teff}{\hbox{$T_\mathrm{eff}$}}
\newcommand{\rave}{\textsc{Rave}}
\newcommand{\ie}{i.e.}
\newcommand{\eg}{e.g.}

\shorttitle{\rave\ DR6 - II.: {stellar atmospheric parameters} and abundances}
\shortauthors{Steinmetz et al.}

\begin{document}
\title{The Sixth Data Release of the Radial Velocity Experiment (\rave) -- II: Stellar Atmospheric Parameters, Chemical Abundances and Distances}

\author[0000-0001-6516-7459]{Matthias Steinmetz}
\affiliation{Leibniz-Institut f{\"u}r Astrophysik Potsdam (AIP), An der Sternwarte 16, 
14482 Potsdam, Germany}

\author[0000-0002-1317-2798]{Guillaume Guiglion}
\affiliation{Leibniz-Institut f{\"u}r Astrophysik Potsdam (AIP), An der Sternwarte 16, 
14482 Potsdam, Germany}

\author[0000-0002-8861-2620]{Paul J. McMillan}
\affiliation{Lund Observatory, Department of Astronomy and Theoretical Physics, Lund 
University, Box 43, 22100 Lund, Sweden}

\author[0000-0002-6070-2288]{Gal Matijevi\v c}
\affiliation{Leibniz-Institut f{\"u}r Astrophysik Potsdam (AIP), An der Sternwarte 16, 14482 Potsdam, Germany}

\author[0000-0002-2366-8316]{Harry Enke}
\affiliation{Leibniz-Institut f{\"u}r Astrophysik Potsdam (AIP), An der Sternwarte 16, 14482 Potsdam, Germany}

\author[0000-0002-9035-3920]{Georges Kordopatis}
\affiliation{Universit{\'e} C{\^o}te d'Azur, Observatoire de la C{\^o}te d'Azur, CNRS, 
Laboratoire Lagrange, France}

\author[0000-0002-2325-8763]{Toma\v{z} Zwitter}
\affiliation{University of Ljubljana, Faculty of Mathematics and Physics, Jadranska 19, 
SI-1000 Ljubljana, Slovenia}

\author[0000-0003-0974-4148]{Marica Valentini}
\affiliation{Leibniz-Institut f{\"u}r Astrophysik Potsdam (AIP), An der Sternwarte 16, 
14482 Potsdam, Germany}

\author[0000-0003-1269-7282]{Cristina Chiappini}
\affiliation{Leibniz-Institut f{\"u}r Astrophysik Potsdam (AIP), An der Sternwarte 16, 
14482 Potsdam, Germany}

\author[0000-0003-2688-7511]{Luca Casagrande}
\affiliation{Research School of Astronomy \& Astrophysics, The Australian National University, Canberra, Australia}

\author{Jennifer Wojno}
\affiliation{The Johns Hopkins University, Department of Physics and Astronomy, 3400 N. 
Charles Street, Baltimore, MD 21218, USA}

\author[0000-0001-5261-4336]{Borja Anguiano} 
\affiliation{Department of Astronomy, University of Virginia, Charlottesville, VA, 22904, USA}

\author[0000-0002-4605-865X]{Olivier Bienaym\'e}  \affiliation{Observatoire astronomique de Strasbourg, Universit\'e de Strasbourg, CNRS, 
11 rue de l'Universit\'e, F-67000 Strasbourg, France }

\author{Albert Bijaoui}
\affiliation{Universit{\'e} C{\^o}te d'Azur, Observatoire de la C{\^o}te d'Azur, CNRS, 
Laboratoire Lagrange, France}

\author{James Binney}
\affiliation{Rudolf Peierls Centre for Theoretical Physics, Clarendon Laboratory, Parks Road, Oxford, OX1 3PU, UK}

\author[0000-0001-6516-7459]{Donna Burton}
\affiliation{Australian Astronomical Observatory, Siding Spring, Coonabarabran NSW 2357, Australia}
\affiliation{University of Southern Queensland (USQ), West Street Toowoomba Qld 4350 Australia}

\author{Paul Cass}
\affiliation{Australian Astronomical Observatory, Siding Spring, Coonabarabran NSW 2357, Australia}

\author{Patrick de Laverny}
\affiliation{Universit{\'e} C{\^o}te d'Azur, Observatoire de la C{\^o}te d'Azur, CNRS, 
Laboratoire Lagrange, France}

\author{Kristin Fiegert}
\affiliation{Australian Astronomical Observatory, Siding Spring, Coonabarabran NSW 2357, Australia}

\author[0000-0001-6280-1207]{Kenneth Freeman}
\affiliation{Research School of Astronomy \& Astrophysics, The Australian National University, Canberra, Australia}

\author{Jon P. Fulbright}
\affiliation{The Johns Hopkins University, Department of Physics and Astronomy, 3400 N. Charles Street, Baltimore, 
MD 21218, USA}

\author[0000-0003-4446-3130]{Brad K. Gibson}
\affiliation{E.A. Milne Centre for Astrophysics, University of Hull, Hull, HU6 7RX, 
United Kingdom}

\author[0000-0003-4632-0213]{Gerard Gilmore}  
\affiliation{Institute of Astronomy, Cambridge, UK}

\author[0000-0002-1891-3794]{Eva K.\ Grebel}
\affiliation{Astronomisches Rechen-Institut, Zentrum f\"ur Astronomie
der Universit\"at Heidelberg, M\"onchhofstr.\ 12--14, 69120 Heidelberg,
Germany}

\author[0000-0003-3937-7641]{Amina Helmi}  
\affiliation{Kapteyn, Astronomical Institute, University of Groningen, P.O. Box 800, 9700 AV Groningen, The Netherlands}

\author[0000-0002-2808-1370]{Andrea Kunder}  
\affiliation{Saint Martin's University, 5000 Abbey Way SE, Lacey, WA, 98503, USA}

\author[0000-0001-6805-9664]{Ulisse Munari}
\affiliation{INAF Astronomical Observatory of Padova, 36012 Asiago (VI), Italy}

\author{Julio F. Navarro}
\affiliation{Department of Physics and Astronomy, University of Victoria, Victoria, BC, 
Canada V8P5C2.}

\author[0000-0001-6516-7459]{Quentin Parker}
\affiliation{CYM Physics Building, The University of Hong Kong, Pokfulam, Hong Kong SAR, PRC}
\affiliation{The Laboratory for Space Research, Hong Kong University, 
Cyberport 4, Hong Kong SAR, PRC}

\author{Gregory R. Ruchti}
\altaffiliation{deceased}
\affiliation{The Johns Hopkins University, Department of Physics and Astronomy, 3400 N. Charles Street, Baltimore, MD 21218, USA}

\author{Alejandra Recio-Blanco}
\affiliation{Universit{\'e} C{\^o}te d'Azur, Observatoire de la C{\^o}te d'Azur, CNRS, 
Laboratoire Lagrange, France}

\author{Warren Reid}
\affiliation{Department of Physics and Astronomy, Macquarie University, Sydney, NSW 
2109, Australia}
\affiliation{Western Sydney University, Locked bag 1797, Penrith South, NSW 2751, 
Australia}

\author[0000-0003-4072-9536]{G. M. Seabroke}
\affiliation{Mullard Space Science Laboratory, University College London, Holmbury St Mary, Dorking, RH5 6NT, UK}

\author{Alessandro Siviero}
\affiliation{Dipartimento di Fisica e Astronomia G. Galilei, Universita' di Padova, 
Vicolo dell'Osservatorio 3, I-35122, Padova, Italy}

\author{Arnaud Siebert}
\affiliation{Observatoire astronomique de Strasbourg, Universit\'e de Strasbourg, CNRS, 
11 rue de l'Universit\'e, F-67000 Strasbourg, France }

\author{Milorad Stupar}  
\affiliation{Australian Astronomical Observatory, Siding Spring, Coonabarabran NSW 2357, Australia}
\affiliation{Western Sydney University, Locked Bag 1797, Penrith South, NSW 2751, Australia}  

\author[0000-0002-3590-3547]{Fred Watson}  
\affiliation{Department of Industry, Innovation and Science, 105 Delhi Rd, North Ryde, NSW 2113, Australia}

\author{Mary E.K.\ Williams}
\affiliation{Leibniz-Institut f{\"u}r Astrophysik Potsdam (AIP), An der Sternwarte 16, 
14482 Potsdam, Germany}

\author[0000-0002-4013-1799]{Rosemary F.G.\ Wyse}
\affiliation{The Johns Hopkins University, Department of Physics and Astronomy, 3400 N. Charles Street, Baltimore, 
MD 21218, USA}
\affiliation{Kavli Institute for Theoretical Physics, University of California, Santa Barbara, CA 93106, USA}

\author[0000-0003-4524-9363]{Friedrich Anders}
\affiliation{Institut de Ci\`encies del Cosmos, Universitat de Barcelona (IEEC-UB), 
Mart\'i i Franqu\`es 1, 08028 Barcelona, Spain}
\affiliation{Leibniz-Institut f{\"u}r Astrophysik Potsdam (AIP), An der Sternwarte 16, 
14482 Potsdam, Germany}

\author[0000-0003-2595-5148]{Teresa Antoja} 
\affiliation{Institut de Ci\`encies del Cosmos, Universitat de Barcelona (IEEC-UB), 
Mart\'i i Franqu\`es 1, 08028 Barcelona, Spain}

\author[0000-0001-7516-4016]{Joss Bland-Hawthorn}
\affiliation{Sydney Institute for Astronomy, School of Physics,
The University of Sydney, NSW 2006, Australia}

\author[0000-0002-9480-8400]{Diego Bossini}
\affiliation{Instituto de Astrof\'isica e Ci$\hat{e}$ncias do Espa\c{c}o, Universidade do Porto, CAUP, Rua das Estrelas, 4150-762 Porto,Portugal} 

\author[0000-0002-8854-3776]{Rafael A. Garc\'\i a} 
\affiliation{IRFU, CEA, Universit\'e Paris-Saclay, F-91191 Gif-sur-Yvette, France} 
\affiliation{AIM, CEA, CNRS, Universit\'e Paris-Saclay, Universit\'e Paris Diderot, Sorbonne Paris Cit\'e, F-91191 Gif-sur-Yvette, France} 

\author[0000-0002-0759-0766]{Ismael Carrillo} 
\affiliation{Leibniz-Institut f{\"u}r Astrophysik Potsdam (AIP), An der Sternwarte 16, 
14482 Potsdam, Germany}

\author[0000-0002-5714-8618]{William J. Chaplin} 
\affiliation{School of Physics and Astronomy, University of Birmingham, Edgbaston, Birmingham B15 2TT, UK}
\affiliation{Stellar Astrophysics Centre (SAC), Department of Physics and Astronomy, Aarhus University,  DK-8000 Aarhus C, Denmark} 

\author{Yvonne Elsworth}
\affiliation{School of Physics and Astronomy, University of Birmingham, Edgbaston, Birmingham, B15 2TT, UK}
\affiliation{Stellar Astrophysics Centre (SAC), Department of Physics and Astronomy, Aarhus University,  DK-8000 Aarhus C, Denmark}

\author[0000-0003-3180-9825]{Benoit Famaey}
\affiliation{Observatoire astronomique de Strasbourg, Universit\'e de Strasbourg, CNRS, 11 rue de l'Universit\'e, F-67000 Strasbourg, France }

\author[0000-0003-3333-0033]{Ortwin Gerhard}
\affiliation{Max-Planck-Institut f{\"u}r extraterrestrische Physik, Postfach 1312,
  Giessenbachstr., 85741 Garching, Germany}

\author{Paula Jofre}
\affiliation{N\'ucleo de Astronom\'ia, Facultad de Ingenier\'ia y Ciencias, Universidad Diego Portales, Ej\'ercito 441, Santiago de Chile}

\author[0000-0002-5144-9233]{Andreas Just}
\affiliation{Astronomisches Rechen-Institut, Zentrum f\"ur Astronomie
der Universit\"at Heidelberg, M\"onchhofstr.\ 12--14, 69120 Heidelberg,
Germany}

\author[0000-0002-0129-0316]{Savita Mathur}
\affiliation{Instituto de Astrof\'{\i}sica de Canarias, La Laguna, Tenerife, Spain}
\affiliation{Dpto. de Astrof\'{\i}sica, Universidad de La Laguna, La Laguna, Tenerife, Spain}

\author{Andrea Miglio}
\affiliation{School of Physics and Astronomy, University of Birmingham, Edgbaston, Birmingham, B15 2TT, UK}
\affiliation{Stellar Astrophysics Centre (SAC), Department of Physics and Astronomy, Aarhus University,  DK-8000 Aarhus C, Denmark}

\author[0000-0002-5627-0355]{Ivan Minchev}
\affiliation{Leibniz-Institut f{\"u}r Astrophysik Potsdam (AIP), An der Sternwarte 16, 
14482 Potsdam, Germany}

\author{Giacomo Monari}
\affiliation{Leibniz-Institut f{\"u}r Astrophysik Potsdam (AIP), An der Sternwarte 16, 14482 Potsdam, Germany}
\affiliation{Observatoire astronomique de Strasbourg, Universit\'e de Strasbourg, CNRS, 
11 rue de l'Universit\'e, F-67000 Strasbourg, France }

\author[0000-0002-7547-1208]{Benoit Mosser} 
\affiliation{LESIA, Observatoire de Paris, PSL Research University, CNRS, Sorbonne Universit\'e, Universit\'e Paris Diderot,  92195 Meudon, France}

\author{Andreas Ritter}
\affiliation{The Laboratory for Space Research, Hong Kong University, 
Cyberport 4, Hong Kong SAR, PRC}

\author[0000-0002-9414-339X]{Thaise S. Rodrigues} 
\affiliation{INAF Astronomical Observatory of Padova, 36012 Asiago (VI), Italy} 

\author[0000-0002-0894-9187]{Ralf-Dieter Scholz}
\affiliation{Leibniz-Institut f{\"u}r Astrophysik Potsdam (AIP), An der Sternwarte 16, 14482 Potsdam, Germany}

\author[0000-0002-0920-809X]{Sanjib Sharma}
\affiliation{Sydney Institute for Astronomy, School of Physics, The University of 
Sydney, NSW 2006, Australia}

\author{Kseniia Sysoliatina} 
\affiliation{Astronomisches Rechen-Institut, Zentrum f\"ur Astronomie
der Universit\"at Heidelberg, M\"onchhofstr.\ 12--14, 69120 Heidelberg,
Germany}

\correspondingauthor{Matthias Steinmetz}
\email{msteinmetz@aip.de}
\collaboration{(The \rave\  collaboration)}

\begin{abstract}
We present part 2 of the 6th and final Data Release (DR6 or FDR) of the Radial Velocity 
Experiment (\rave), a magnitude-limited ($9< I < 12$) spectroscopic 
survey of Galactic stars randomly selected in the southern hemisphere. The \rave\  
medium-resolution spectra ($R\sim7500$) cover the Ca-triplet region ($8410-8795$\,\AA) 
and span the complete time frame from the start of \rave\  observations on 12 April 2003 
to their completion on 4 April 2013. In the second of two publications, we present the 
data products derived from 518\,387 observations of 451\,783 unique stars  using a 
suite of advanced reduction pipelines focussing on stellar atmospheric parameters, in 
particular purely spectroscopically derived stellar atmospheric parameters (\teff, \logg, 
and the overall metallicity), enhanced {stellar atmospheric parameters} inferred via a Bayesian pipeline using 
Gaia DR2 astrometric priors, and asteroseismically calibrated {stellar atmospheric parameters} for giant 
stars based on asteroseismic observations for 699 K2 stars. In addition, we provide 
abundances of the elements Fe, Al, and Ni, as well as an 
overall $\alphafe$ ratio obtained using a new 
pipeline based on the \texttt{GAUGUIN} optimization method that is able to deal with 
variable signal-to-noise ratios. The \rave\ DR6 catalogs are cross matched with relevant 
astrometric and photometric catalogs, and are complemented by orbital parameters and effective 
temperatures based on the infrared flux method. 
The data can be accessed via the \rave\  Web site\footnote{{\tt http://rave-survey.org}} 
or the Vizier database. 
\end{abstract}



\section{Introduction}

Wide-field spectroscopic surveys of the stellar content of the Galaxy provide crucial information 
on the combined chemical and dynamical history of the Milky Way, and for the understanding of the 
formation and evolution of galaxies in a broader context. Spectroscopy  enables us to measure the 
radial velocities of stars, which, when combined with positions, distances and proper motions from 
astrometry, allows us to study Galactic dynamics in detail. Spectroscopy also enables us to measure 
atmospheric properties ({surface gravity} \logg\ and 
{effective temperature} \teff) of stars and the abundance of chemical elements in the 
stellar atmosphere, thus providing important clues on the chemical evolution of the Galaxy and of 
its stellar populations \citep[see, e.g.,][who also coined the term \emph{Galactic Archaeology} for 
this type of research]{freeman2002}. The combination of large wide-field spectroscopic surveys with 
massive and precise astrometric information as delivered by the Gaia mission \citep{GaiaMission} is 
particular powerful, as demonstrated by a large number of publications in the past two years.

The scientific potential of combining wide-field spectroscopy and astrometry has been motivation 
for a number of spectroscopic Galactic Archaeology surveys, starting with the Geneva-Copenhagen 
survey \citep[CGS,][]{nordstrom2004} and the Radial Velocity Experiment \citep[\rave,][]{steinmetz2003},  
followed by a meanwhile considerable number of surveys of similar or even larger size at lower (e.g., 
SEGUE, \citealt{yanny2009}; and LAMOST, \citealt{LAMOST})  and higher spectral 
resolution (e.g., APOGEE, \citealt{APOGEE}; GALAH, \citealt{GALAH}; and 
Gaia-ESO, \citealt{gilmore2012}). For a recent review on abundances derived from large spectroscopic surveys we refer to \cite{jofre2019}.

This publication addresses the determination of  {stellar atmospheric parameters}, chemical abundances, and distances 
in the context of the \rave\ survey, which over its 10 year observing campaign amassed this information 
based on more than half a million spectra. Together with the accompanying paper 
\citep[][henceforth DR6-1]{steinmetz2020a}, which is focusing on \rave\  spectra, error spectra, spectral 
classification, and radial velocity determinations, it 
constitutes the sixth and final data release (DR6) of \rave. In particular, DR6 provides a new set of 
{stellar atmospheric parameters} employing parallax information from Gaia DR2 \citep{GaiaDR2}, and a robust determination 
of the $\alpha$-enhancement $\alphafe$. 

The paper is structured as follows: 
in Section 2 we give a brief overview of the \rave\ survey and its collected data.  Section 3 
presents an update on the {stellar atmospheric parameter} determination 
and introduces a new catalog of stellar {stellar atmospheric parameters} inferred using a Bayesian 
pipeline with Gaia DR2 parallax priors following the procedure outlined in 
\cite{mcmillan2018}. In Section 4, a new optimization pipeline \texttt{GAUGUIN} is 
presented \citep{GAUGUIN10,GAUGUIN12,
guiglion2016} in order to extract $\alphafe$ ratios as well as individual abundances of Fe, Al, 
and Ni. Section 5 describes how orbital parameters of stars are derived from \rave\  
combined with Gaia DR2 astrometric information. \rave\ data validation including a 
comparison of \rave\  stellar {stellar atmospheric parameters} and abundances with external observational data 
sets is done in Section 6. Section 7 presents the \rave\  DR6 catalog, followed by a 
reanalysis of some previously published \rave\  results in order to demonstrate the 
capabilities of \rave\  DR6 (Section 8). Finally, Section 9 gives 
a summary and draws some conclusions.

\section{Survey Data and their reduction}

The motivation, history, specifications and performance of the \rave\ survey are presented in detail in 
the data release papers DR1, DR2, DR3, DR4, and DR5 \citep{steinmetz2006,zwitter2008,siebert2011,kordopatis2013,kunder2017} 
and a comprehensive summary is given in DR6-1 \citep{steinmetz2020a}. Here, we only summarize the main properties of the \rave\ survey. 

\rave\ was initiated in 2002 as a kinematically unbiased wide-area survey of the southern hemisphere with the primary goal 
to determine radial velocities of Milky Way stars \citep{steinmetz2003}. Thanks to the 6dF multi-object spectrograph on the 
1.23m UK Schmidt telescope at Siding Spring in Australia, up to 150 spectra could be simultaneously acquired over a field 
of view of 5.7\degr. Spectra were taken at an average resolution of $R=7 500$ over the IR Ca triplet region at $8410-8795$ \AA, 
which is similar in coverage and somewhat lower in resolution when compared to the spectral range probed by the Gaia RVS instrument \citep[$R_\text{RVS}=11 500$, ][]{cropper2018}.

The targets of RAVE are mainly drawn from a magnitude range $9<I<12$, where $I$ is Cousins $I$. At an exposure time of 
typically 1 hour, a signal-to-noise (\SNR) of $\SNR \approx 40$ can be achieved for targets between $I\approx 10-11$ 
(see DR6-1 for details). Since a 6dF fibre corresponds to $6.7\arcsec$ on the sky, \rave\ observing avoided the bulge 
region and disk regions at low Galactic latitude in order to minimize contamination by unresolved multiple sources within a single fiber. 

The input catalog of \rave\ was initially produced by a combination of the  Tycho-2 catalog \citep{Hog2000} and the Supercosmos 
Sky Survey \citep[SSS, ][]{Hambly2001b}. Later on, upon availability, the input catalogue was converted to the DENIS 
\citep{DENIS} and 2MASS \citep{2MASS} system.

Since \rave\ was designed as a survey with its main focus on studies of Galactic dynamics and Galactic evolution, the 
observing focus was to approach an unbiased target selection with a wide coverage of the accessible sky. Consequently 
most targets were only observed once. In order to account, at least statistically, for the effects of binarity, about 
4000 stars were selected for a series of repeat observations roughly following a logarithmic series with a cadence of 
separations of 1, 4, 10, 40, 100, and 1000 days (see DR6-1, Section 2.7 for details). 

During the overall observing campaign of \rave, which lasted from 12 April 2003 to 4 April 2013, 518,387 spectra for  
451,783 stars where successfully taken and reduced.  

The data reduction of \rave\  follows the sequence of the following pipeline:
\begin{enumerate}
    \item quality control of the acquired data on site with the RAVEdr software 
    package (paper DR6-1, Section 3.1).
    \item reduction of the spectra (DR6-1, Section 3.1).
    \item spectral classification (DR6-1, Section 4).
    \item determination of (heliocentric) radial velocities with \texttt{SPARV} 
    (\emph{`Spectral Parameter 
And Radial Velocity'}, DR6-1, Section 5).
    \item determination of {stellar atmospheric parameters} {\teff\, \logg, and \mh}\footnote{For proper definition and differences between \mh, \Mh, and \feh\ see Section \ref{subsec:MADERA}} with \texttt{MADERA} (\emph{`MAtisse and DEgas used in RAve'}, (Section \ref{subsec:MADERA}).
    \item determination of the effective temperature using additional photometric 
    information (\emph{InfraRed Flux Method} (\texttt{IRFM}), Section \ref{subsec:IRFM}).
    \item modification of the \rave\  {stellar atmospheric parameters} {\teff\, \logg, and \Mh}  derived spectroscopically  with additional photometry and Gaia DR2 parallax priors 
    using \texttt{BDASP} (\emph{Bayesian Distances Ages and Stellar Parameters}, DR6-2 Section \ref{subsec:BDASP}).
    \item determination of the abundance of the elements Fe, Al, and Ni, and an overall $\alphafe$ ratio with the pipeline \texttt{GAUGUIN} 
    (Section \ref{sec:GAUGUIN}).
    \item recalibration of the  {stellar atmospheric parameters} {\teff\, \logg, and \Mh,   for giant stars based on K2 asteroseismic information (Section \ref{subsec:seismo})  followed by the determination of the chemical abundances \feh\ and \mgh\ using the GAUFRE pipeline \citep{valentini2013}}.
\end{enumerate}

The output of these pipelines is accumulated in a PostgreSQL data base and accessible via the \rave\  website 
{\tt http://www.rave-survey.org} (Section \ref{sec:FDR} and DR6-1, Section 7).

\section{Stellar Atmospheric Parameters}
\subsection{Stellar Atmospheric Parameters from Spectroscopy}\label{subsec:MADERA}

\rave\ DR6 employs the exact same procedure to derive stellar atmospheric parameters 
from spectroscopy as DR5 \citep{kunder2017}. In short, the pipeline \texttt{MADERA} uses a combination of (i)  a 
decision tree \citep[DEGAS,][]{bijaoui2012}, which normalizes the spectrum iteratively as 
well as parameterizing the low \SNR\ spectra, and (ii) a projection algorithm 
\citep[MATISSE,][]{recio-blanco2006} which is used to obtain the {stellar atmospheric parameters} for 
the high \SNR\ spectra ($>30$). 

Both of the methods are used with the grid of 3580 synthetic spectra first 
calculated in the framework of \citet{kordopatis2011a} and adjusted for DR4 \citep{kordopatis2013} assuming the 
Solar abundances of \cite{grevesse2008} and \cite{asplund2005}.
This grid 
has been computed using the MARCS model atmospheres \citep{gustafsson2008} and 
Turbospectrum \citep{plez2012} under the assumption of local thermodynamic equilibrium 
(LTE). {The atomic data was taken from the 
VALD} \footnote{http://vald.astro.uu.se/} 
{database \citep{kupka2000}, with updated 
oscillator strenghts from \citet{gustafsson2008}}. 
The line-list has been calibrated primarily on the Solar spectrum of 
\citet{hinkle2003} and with adjustments to fit also the Arcturus 
spectrum to an acceptable level \citep[see][for further information]{kordopatis2011a}. Furthermore, the
grid excludes the cores of the Calcium triplet lines, as they can suffer, depending on spectral type,  
from non-LTE effects or emission lines owing to stellar activity. 
The grid has three free parameters: effective temperature, \teff, 
logarithm of the surface gravity, \logg, and metallicity\footnote{In the synthetic 
grid, all of the elements except the $\alpha$ are solar-scaled.}, \mh. These free 
parameters are hence the parameters that \texttt{MADERA} determines. 

We note that the $\alpha$ enhancement \alphafe\ varies across the grid, but is not a 
free parameter. Indeed, only one \alphafe\ value is adopted per \mh\ grid-point:
\begin{equation}
\label{eqn:MADERA_alpha}
\left[\frac{\alpha}{\mathrm{Fe}}\right] =\left\{
\begin{array}{rl}
+0.4\quad : & \left[\frac{\mathrm{m}}{\mathrm{H}}\right]\leq -1\\&\\
-0.4 \times \left[\frac{\mathrm{m}}{\mathrm{H}}\right] \quad :& 
-1<\left[\frac{\mathrm{m}}{\mathrm{H}}\right]<0\\&\\
0 \quad :& \left[\frac{\mathrm{m}}{\mathrm{H}}\right]\geq 0
\end{array}
\right.
\end{equation}

This implies that the \mh\ value of the grid can be thought of as the 
content of all the  metals in the star, except the $\alpha$ elements. 
The derived value of \mh\  from an observed spectrum, denoted by {$\mh_u$}, should hence 
be considered as an overall metallicity estimator assuming an $\alpha$-enhancement. {This should be discriminated from methods and grids based on the total metallicity \emph{including} an $\alpha$ enhancement, like those used in Section \ref{subsec:BDASP} and \ref{subsec:seismo}, we refer to those metallicity estimators as \Mh.} 

{Finally, with \feh\ we refer to direct measurements of the iron content by fitting iron line,} 

{e.g. with the GAUGUIN method (Section \ref{sec:GAUGUIN}) or when using high resolution data for validation}
{(see Section \ref{sec:Validation})}.

\subsubsection{\texttt{MADERA}'s quality flags}\label{subsubsec:MADERA_flags}

In addition to the {stellar atmospheric parameters} (\teff, \logg, \mh) and their associated 
uncertainties, the pipeline provides each spectrum with one of the five quality flags 
{(\texttt{algo\_conv\_madera})} 
given below\footnote{These flags are unchanged from those in DR4 and DR5.} to allow the 
user to filter, quite robustly, the results according to adopted criteria that are sound 
and objective (\eg, convergence of the algorithm):  
\begin{itemize}
\item
`0': The analysis was carried out as desired. The normalization process converged, 
as did MATISSE (for high SNR spectra) or DEGAS (for low \SNR\ spectra). There are 322,367 
spectra that fulfill this criterion. 

\item `1': Although the spectrum has a sufficiently high \SNR\ to use the projection 
algorithm, the MATISSE algorithm did not converge.  {Stellar atmospheric parameters} for stars with 
this flag are not reliable.  There are 17,639 spectra affected by this.

\item `2':  The spectrum has a sufficiently high \SNR\ to use the projection algorithm, 
but MATISSE oscillates between two solutions. The reported parameters are the mean of 
these two solutions. In general, the oscillation happens for a set of parameters that 
are nearby in parameter space and computing the mean is a sensible thing to do.  
However, this is not always the case, for example, if the spectrum contains artifacts. 
The mean may then not provide accurate  {stellar atmospheric parameters}. The 58,992 spectra with a 
flag of `2'  could be used for analyses, but with caution (a visual inspection of the 
observed spectrum and its solution may be required).  

\item `3': MATISSE gives a solution that is extrapolated {to values outside of} the parameter
range {defining} the learning grid ($T_{\rm eff}$ outside the range [3500,8000]\,K, $\log g$ {outside} 
of the range [0,5.5], metallicity outside the range [-5,+1]\,dex), and the solution is forced to be the one from
DEGAS.  For spectra having artifacts but high \SNR\ overall, this is a sensible
thing to do, as DEGAS is less sensitive to such discrepancies. This applies to 87,335
spectra. However, for the few hot stars that have been observed by \rave, adopting this 
approach is not correct.  A flag of `3' and a $T_{\rm eff} >$ 7750\,K is very likely to
indicate that this is a hot star with $T_{\rm eff} >$ 8000\,K and hence that
the parameters associated with that spectrum are not reliable. 
 
\item `4': This flag will only appear for low \SNR\ stars and metal-poor giants. 
Indeed, for metal-poor giants, the spectral lines available are neither strong enough 
nor numerous enough to have DEGAS successfully parameterize the star. Tests on 
synthetic spectra have shown that to derive reliable parameters the settings used to 
explore the branches of the decision tree need to be changed compared to the `standard' 
parameters adopted for the rest of the parameter space.  A flag `4' therefore marks 
this change in the setting for book-keeping purposes, and the 31,488 spectra associated 
with this flag \emph{should be safe} for any analysis.  
  
\end{itemize}

\subsubsection{Calibration of the {stellar atmospheric parameters}}\label{subsubsec:calibration}

Several tests performed for DR4 as well as the subsequent science
papers, have indicated that the stellar parameter pipeline is globally robust
and reliable. However, being based on synthetic spectra that may not match
the real stellar spectra over the entire parameter range, the direct outputs
of the pipeline need to be calibrated on reference stars in order to minimize
possible systematic offsets. 

To calibrate the DR6 outputs of the pipeline, the same calibration data-set and 
polynomial fit compared to literature values has been used as for DR5. For completeness 
reasons, we review the relations in the following subsections, but refer the reader to 
the DR4 and DR5 papers for further details. We  performed  tests with additional 
subsets (coming from e.g.~asteroseismic surface gravities) or/and more complex 
polynomials to calibrate the pipeline's output and obtained  results that 
did not show any significant improvement over the approach that was adopted in DR4 and DR5. 

\paragraph{Metallicity calibration} 
The calibration relation for DR6 is: 
\begin{equation}\label{eq:mh_cal}
\begin{split}
\mh_{\rm DR6}= \mh_{\rm u} + (0.276 - 0.044~{\logg_{\rm u}}\\
 +0.002~{ \logg_{\rm u}^2} -0.248~\mh_{\rm u}\\
 +0.007~{ \mh_{\rm u}}~{ \logg_{\rm u}}- 0.078~{
 \mh_{\rm u}^2)},\bigskip
\end{split}
\end{equation}
where $\mh_{\rm DR6}$ is the calibrated metallicity, and $\mh_{\rm u}$ and $\logg_{\rm u}$ are, 
respectively, the un-calibrated  metallicity and surface 
gravity (both the raw output from the pipeline).
The adopted calibration corrects for a rather constant underestimation of 0.2 \dex\ at the lowest 
metallicities, while also correcting trends in the more metal-rich regimes, where the giant stars exhibit higher offets than the dwarfs.
As already described in the earlier  DR papers,  
this relation has been 
calibrated against \feh\ values from the literature. This implies that $\mh_{\rm DR6}$ is a proxy 
for \feh\ \emph{only} if \emph{all} of the elements in the targeted star are 
solar-scaled and if the $\alpha$ abundances are following the same relation as adopted 
for the synthetic grid at the \feh\ value of the star of interest.  $\mh_{\rm DR6}$ should therefore be rather thought of as a metallicity 
indicator, \ie\ to depend on a combination of elements. $\mh_{\rm DR6}$ is, however, not equal to the overall metallicity of the star, as 
discussed above - see also equation~\ref{eq:Salaris} in section~\ref{subsec:BDASP}). 

\paragraph{Surface gravity calibration}

The following quadratic expression defines our
surface gravity calibration:
\begin{equation}
\logg_{\rm DR6}=\logg_{\rm u} + (0.515-0.026~{\logg_{\rm u}} -0.023~{\logg_{\rm u}^2}).
\end{equation}
This relation increases gravities of supergiants by $\sim 0.5$~dex, and of dwarfs by $\sim 0.75$~dex.

\paragraph{Effective temperature calibration} 

The adopted calibration for effective
temperature is
\begin{equation}
\rm T_{\rm eff, DR6}=T_{\rm eff,u} + (285\,\text{K}-0.073{T_{\rm eff,u}}+40\,\text{K}\times{\logg_{\rm u}}).
\end{equation}

Corrections reach up to 200~K for cool dwarfs, but are generally much smaller. 

\subsection{Infrared Flux Method Temperatures}\label{subsec:IRFM}

Effective temperatures from the infrared flux method 
\citep[\texttt{IRFM},][]{casagrande2006,casagrande2010} are derived in a manner similar to that carried out 
in  \rave\  DR5, where a detailed description can  be found. Briefly, our
implementation of the \texttt{IRFM} uses APASS and 2MASS photometry to recover 
stellar bolometric and infrared fluxes. The ratio of these two fluxes for a given star 
is compared to that predicted
from theoretical models,  for a given set of  {stellar atmospheric parameters}, and an
iterative approach is used to converge on the final value of \teff. The advantage of
comparing observed versus model fluxes in the infrared is that this region
is largely dominated by the continuum and thus very  sensitive to 
\teff, while the dependence on surface gravity and metallicity is minimal. 
Here, we adopt for each star the calibrated $\logg_{\rm DR6}$ and $\mh_{\rm DR6}$ from the 
\texttt{MADERA} pipeline, but if  we were instead to use the parameter values from the \texttt{SPARV} pipeline 
the derived value of \teff\ would differ by a few tens of Kelvin at most. Since the \texttt{IRFM} 
simultaneously determines bolometric fluxes and effective temperatures, stellar angular
diameters can be derived, and are also provided in DR6. Extensive comparison with
interferometric angular diameters to validate this method is discussed in
\citet{casagrande2014}. We are able to provide 
\teff\ from the \texttt{IRFM} for over 90\% of our sample, while for about 6\% of them we have to resort to
color-temperature relations derived from the \texttt{IRFM}. For less than 3\% of our
targets effective temperatures could not be determined due to the lack of
reliable photometry.

In a photometric method such as the \texttt{IRFM}, reddening can have a 
non-negligible impact. We rescale the reddening from \cite{schlegel1998} as described in 
\rave\ DR5, but for stars with $\log(g)<3.5$ and $E(B-V)>0.3$ we now use the relation
$E(B-V)=0.918(H-W2-0.08)/0.366$ from the Rayleigh–Jeans color excess
method \citep[RJCE,][]{majewski2011}.

\subsection{Distances, Ages, Stellar Atmospheric Parameters with Gaia priors}\label{subsec:BDASP}
 
\rave\ DR6 includes for the first time  {stellar atmospheric parameters} derived using the Bayesian 
framework demonstrated in \cite{mcmillan2018}, along with derived distances, ages, and 
masses, which have also been derived for previous data releases. We refer to the method 
as the \texttt{BDASP} code. This follows 
the pioneering work deriving (primarily) distances by \cite{burnett2010} and 
\cite{binney2014}. 

This method takes as its input the  {stellar atmospheric parameters} \teff\ (taken from the \texttt{IRFM}, 
Section~\ref{subsec:IRFM}), \logg\ taken from the \texttt{MADERA} pipeline, an estimate of the 
overall metallicity taken from the \texttt{MADERA}  (see below), $J$, $H$, 
and $K_s$ magnitudes from 2MASS, and, for the first time, parallaxes from Gaia DR2. 
For a detailed description of the method, the interested reader should refer to 
\cite{mcmillan2018}, where parallaxes from Gaia DR1 \citep{gaiadr1} were used for the 
219,566 \rave\  sources which entered the TGAS part of the Gaia catalog 
\citep{michalik2015}. Here we simply give a brief overview and note differences in the 
methodology used here and by \cite{mcmillan2018}.

\texttt{BDASP} applies the simple Bayesian statement
\[
P(\hbox{model}|\hbox{data})=\frac{P(\hbox{data}|\hbox{model})P(\hbox{model})}{P(\hbox{data})},
\]
where in our case ``data'' refers to the inputs described above for a single stars, and 
``model'' comprises a star of specified initial mass
${\cal M}$, age $\tau$, metallicity \Mh, and location, observed through a specified 
line-of-sight extinction. $P(\hbox{data}|\hbox{model})$ is determined assuming 
uncorrelated Gaussian uncertainties on all inputs, and using PARSEC isochrones 
\citep{bressan12} to find the values of the  {stellar atmospheric parameters} and absolute 
magnitudes of the model star. The uncertainties of the  {stellar atmospheric parameters} are assumed 
to be the quadratic sum of the quoted internal uncertainties and the external 
uncertainties, as found for \rave\  DR5 \cite[Table 4]{kunder2017}.  $P(\hbox{model})$ 
is our prior, and $P(\hbox{data})$ is a normalization that we can safely ignore. We 
adopt the `Density' prior from \cite{mcmillan2018}, which is the least informative 
prior considered in that study. Even with the, significantly less precise, Gaia DR1 
parallax estimates, the choice of prior had a limited impact on the results, and this is 
reduced still further because of the very high precision of the Gaia DR2 parallax 
estimates. 

As discussed above, the \texttt{MADERA} pipeline provides \mh, the metal content except the alpha 
elements, which is calibrated against \feh. To provide an estimate of 
the overall metallicity \Mh\ we assume that we can scale all abundances with \mh\ 
except those of $\alpha$-elements, which we assume all scale in the same way assumed by 
\texttt{MADERA} (i.e.~following equation~\ref{eqn:MADERA_alpha}). A proxy for the overall metallicity  
including $\alpha$ elements, denoted by \Mh, can then be inferred by applying a 
modified version of the \cite{salaris1993} formula,  derived using the same technique as in 
\citet{valentini2018}:
\begin{equation}
\label{eq:Salaris}
    \left[\frac{\mathrm{M}}{\mathrm{H}}\right]=\left[\frac{\mathrm{m}}{\mathrm{H}}\right]_{\rm DR6} + \log_{10}(C\times10^{\alphafe}+(1-C)),
\end{equation}
with $C=0.661$.  Within this approximation [M/H] corresponds to the composition assumed by the PARSEC isochrones. 

As was made clear at the time of Gaia DR2, astrometric measurements from Gaia have 
small but significant systematic errors (including an offset of the parallax 
zero-point), which vary across the sky on a range of scales, and are dependent on 
magnitude and color \citep{lindegren2018,arenou2018}. This has been demonstrated for a 
variety of comparison samples since Gaia DR2
\citep{Sahlholdt2018,Graczyk2019,Stassun2018,Zinn2018,Khan2019}.
The parallaxes used in \texttt{BDASP} are 
therefore corrected for a parallax zero-point of $-54\, \mu{\rm as}$, following the 
analysis of \cite*{schoenrich2019}. This study determined this zero-point offset for 
stars with Gaia DR2 radial velocities, which cover a similar magnitude range to \rave, 
and have a larger zero-point offset than the fainter quasars considered by 
\cite{lindegren2018}. We also add a systematic uncertainty of $43\, \mu{\rm as}$ in 
quadrature with the quoted parallax uncertainties to reflect a best estimate of the 
small-scale spatially varying parallax offsets as found by \cite{lindegren2018}.

In DR5 we provided an improved description of the distances to stars with a 
multi-Gaussian fit to the probability density function (pdf) in distance modulus. This 
was particularly important for stars with multi-modal pdfs, for example stars where 
there was an ambiguity over whether they were subgiants or dwarfs. With the addition of 
Gaia parallaxes, these ambiguities have become rare. Fewer than one percent of the sources 
required multi-Gaussian pdfs under the selection criteria used in \cite{mcmillan2018}. {These are generally pdfs with a narrow peak associated with the red 
clump and an overlapping broader one associated stars ascending the red giant branch, 
rather than truly multi-modal pdfs \citep[see][for discussion and examples]{binney2014}}.
In the interests of simplicity, we therefore do not provide these multi-Gaussian pdfs 
with DR6. Extinction is taken into account in the same way as in \cite{binney2014} and 
\cite{mcmillan2018},  and is relatively weak for the majority of RAVE stars. The median 
extinction we find corresponds to $A_V\sim0.2\,{\rm mag}$, which is
$\sim0.06\,{\rm mag}$ in the $J$-band (the band that suffers the most extinction of 
all those we consider).

We can use the Gaia DR2 parallaxes to validate our Bayesian distance finding method in 
the same way as \cite{mcmillan2018} did with the TGAS parallaxes: comparing the 
parallax estimates using \texttt{BDASP} without including the Gaia parallax, to the Gaia DR2 
parallax. Since these are independent estimates, we would expect that if we take the 
difference between these values divided by their combined uncertainty (the two 
uncertainties summed in quadrature), it will be distributed as a Gaussian, with average 
zero and standard deviation unity. 

\begin{figure*}
\begin{center}
\includegraphics[height=5cm]{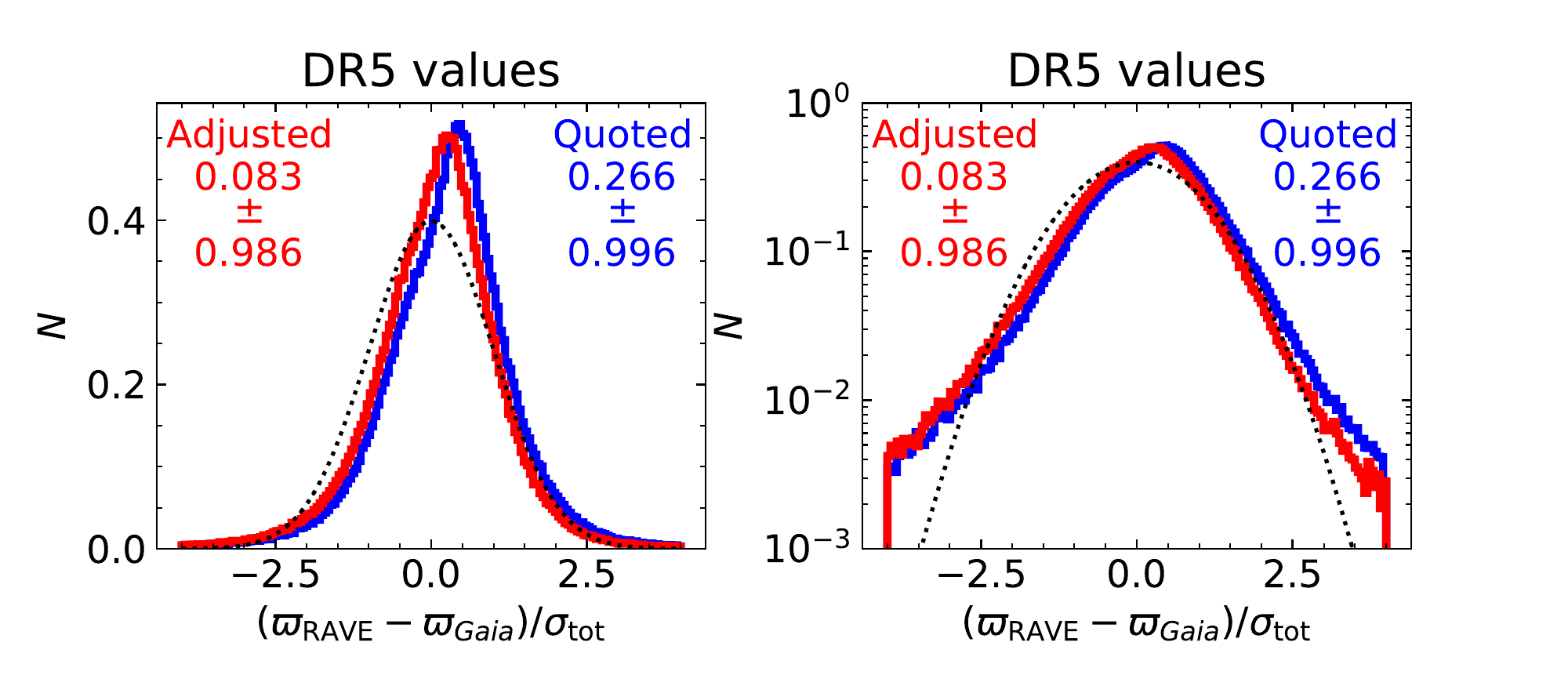}
\includegraphics[height=5cm]{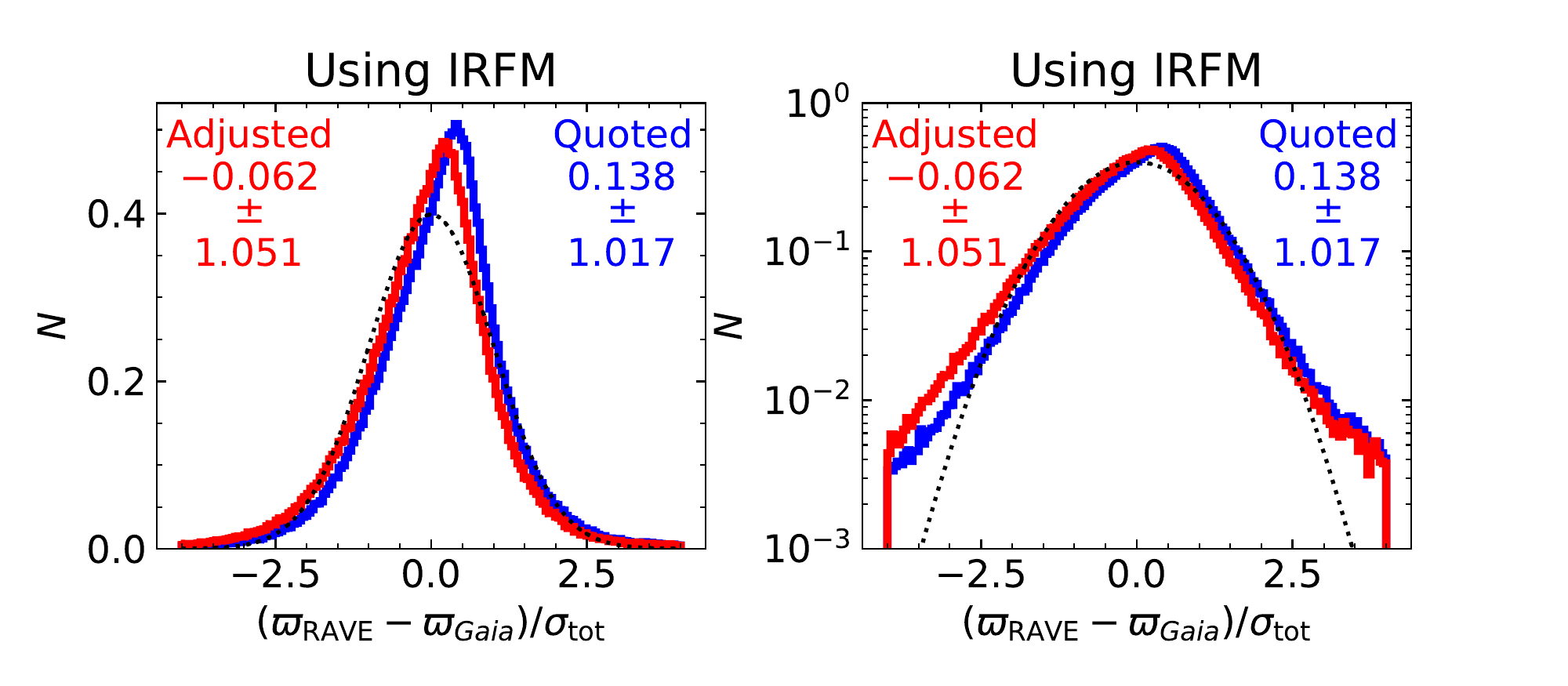}
\caption{\label{fig:BDASPvsGaia}
Comparison of Gaia DR2 parallaxes and the purely spectro-photometric 
parallaxes derived by \texttt{BDASP} for \rave\  DR5 using \texttt{MADERA} \teff\ (top) and using \texttt{IRFM} 
\teff\ (bottom). The colors indicate whether we are using the parallaxes as quoted by 
Gaia DR2 (blue), or adjusted for a parallax zero-point of $-54\, \mu{\rm as}$, and 
systematic uncertainty of $43\, \mu{\rm as}$ (red). The black dotted line is a normal 
distribution of mean zero and standard deviation unity. The right-hand panels show the 
same data as the left-hand ones, but with a logarithmic y-axis to emphasize the tails 
of the distribution. The numbers given in the top corners of each panel are the mean 
and standard deviation of the values (considering only values between $-4$ and $4$).}
\end{center}
\end{figure*}

In Figure~\ref{fig:BDASPvsGaia} we plot histograms showing this comparison for the 
parallax estimates from \rave\  DR5 (using the same techniques described here, and using 
\texttt{MADERA} \teff), or using \texttt{BDASP} but taking the \texttt{IRFM} \teff\ as input \citep[as this was 
shown to be a better approach by][]{mcmillan2018}. In both cases we show the 
comparison to the quoted values from Gaia DR2, and a comparison to the `adjusted' 
values, where we have corrected the Gaia parallaxes for their assumed zero-point offset 
and systematic uncertainties. This figure demonstrates that the parallax zero-point 
offset of Gaia is significant, even for these relatively nearby stars. Including the 
zero-point offset brings the Gaia parallaxes more in line with those from \rave. We 
also see that the use of \texttt{IRFM} temperatures significantly improves the \texttt{BDASP} parallaxes. 
In all cases we see more outlying values than we would expect for a Gaussian 
distribution. One could, therefore, reasonably argue that we should be using the median, 
rather than a sigma-clipped mean (as in Figure~\ref{fig:BDASPvsGaia})  to quantify the 
bias of the values. The median values, compared to the zero-point adjusted Gaia 
parallaxes, are 0.136 and 0.011 using MADERA \teff\ or IRFM \teff, respectively.

\begin{figure*}
\begin{center}
\plotone{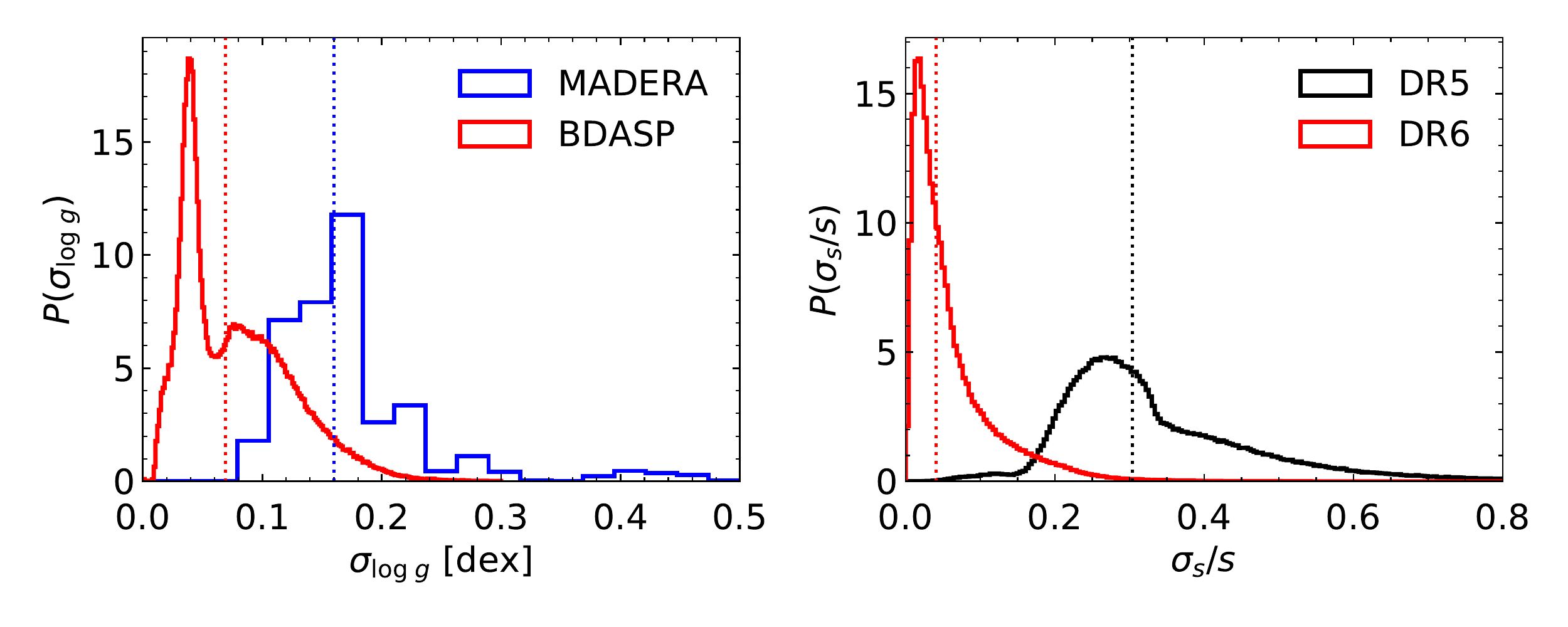}
\caption{\label{fig:BDASP_precision}
Histograms of quoted uncertainties, demonstrating the improved precision when 
including information from Gaia DR2 parallax values in the \texttt{BDASP} framework. Left: 
Comparison of the quoted $\log g$ uncertainty from  \texttt{MADERA} alone (blue) as compared to 
those from \texttt{BDASP} (which has \texttt{MADERA} $\log g$ as one of its inputs: red). Right: 
Comparison of the relative uncertainties in distance found by \texttt{BDASP} in \rave\ DR5 
(which did not have Gaia parallaxes as input: black) and DR6 (which does: red).}
\end{center}
\end{figure*}

We can compare the precision we achieve for  {stellar atmospheric parameters} with \texttt{BDASP} in \rave\ DR6 
to the precision achieved with MADERA. The most interesting of these is the precision 
in \logg, where the Gaia parallax provides the greatest value. This comparison is shown 
in Figure~\ref{fig:BDASP_precision}, and we see that we improve by more than a factor
of two (median precision $0.16\,$\dex\ for \texttt{MADERA}, $0.07$\,\dex\ for \texttt{BDASP}). 
The dwarf stars, which are nearby and therefore have precise parallax estimates from Gaia,
dominate the narrow peak at the smaller \texttt{BDASP} $\log g$ uncertainties of $\sim0.04\,{\rm dex}$ 
while giants make up most of the broader peak of larger $\log g$ uncertainties.
We can also compare the distance uncertainty \texttt{BDASP} found without Gaia DR2 parallax input 
in \rave\ DR5, as compared to the distance uncertainty now. Here we find a dramatic 
improvement, from a typical distance uncertainty of 30 percent in DR5 to one of 4 
percent in DR6. Furthermore, Gaia DR2 parallaxes are available for 99.8\% of \rave\ spectra,
as opposed to only 49\% in \rave\ DR5/Gaia DR1. At this point the only significant gain in 
precision of using spectro-photometric information to derive distances for RAVE stars is 
for the red clump and 
high on the giant branch \citep[the latter being known to be problematic for RAVE:][]{mcmillan2018}.
Otherwise the distance estimates are, to a fairly close approximation, derived directly from 
Gaia parallaxes, so it makes little difference whether these distances, or the ones derived 
directly from Gaia parallaxes alone are used.
This reflects the extraordinary precision of Gaia DR2, and emphasizes 
the value of combining the Gaia data with \rave.

\subsection{Asteroseismically calibrated red giant catalog}\label{subsec:seismo}

The surface gravity provided by asteroseismology ($\logg_{\rm S}$) is now widely used for 
testing the accuracy of the \logg\ measured from spectroscopy. The seismic {$\logg_{\rm S}$}\ can 
be easily computed starting from the scaling relations, two relations that directly 
connect stellar mass and radius to the effective temperature (\teff) and two seismic 
observables \dnu\ ({average frequency separation}) and 
\numax ({frequency of maximum oscillation power}). 
The seismic {$\logg_{\rm S}$}\ depends only on \teff\ and \numax, 
and it is defined as:

\begin{equation}
\label{eq:seismologg}
\logg_{\rm S}= \logg_{\odot} + \log\left(\frac{\numax}{\numax_\odot}\right)
+\frac{1}{2} \log\left(\frac{\teff}{\teff_\odot}\right)
\end{equation}

where the solar values are $\logg_\odot$=4.44\,\dex, $\numax_\odot$= 3090$\,\mu$Hz, and 
$\teff_\odot=5777$\,K \citep{huber2011}.

Large spectroscopic surveys as APOGEE, Gaia-ESO, LAMOST, and GALAH observed seismic 
targets with the purpose of testing and calibrating, if necessary, the \logg\ measured 
by their spectroscopic pipelines. Thanks to the recommissioned {\it Kepler} satellite, 
the K2 mission \citep{K2}, \rave\  had the opportunity to incorporate seismic data starting in DR5 
\citep{kunder2017}, where a set of 87 red giant stars, observed by K2 in Campaign 1, 
were used as calibrators and for an {\it ad hoc\/} calibration for red giant stars only 
\citep{valentini2017}. In DR6, 699 red giants observed during the first six K2 
campaigns are used (see Table \ref{tab:seismo}, showing the number of targets per 
campaign). This allowed for an improved coverage of the parameter space (in particular, effective  
temperature and metallicity).  We use the procedure as outlined in DR5, but use $\teff_\mathrm{,IRFM}$ as the 
prior for the effective temperature. We also allow a larger flexibility 
interval ($\pm 450\,$ K instead of 350 K as in \cite{valentini2017}. The calibration 
adopted in this case turned out to be very similar to the one in DR5, confirming the 
robustness of the method, given the larger seismic data set. For the catalog presented in 
the later part of this work (Section \ref{subsec:DR6_seismo}):

\begin{eqnarray}
\label{Eq:corrlogg}
{\logg_\mathrm{S}}=\logg_u-0.76^{0.80}_{0.74}\times\logg_u+1.98^{2.06}_{1.90}\qquad,
\end{eqnarray}
where $\logg_u$ is the \emph{un-calibrated} \logg\ delivered by the \texttt{MADERA} pipeline.
Further details on the seismic and spectroscopic data analysis are presented in Valentini 
et al. (2020, in prep).

Asteroseismology can be also used for providing estimates of the mass of red giants, and hence their 
age (since the age of a red giant corresponds to the time it spent on the main 
sequence, and therefore its mass). In  Valentini et al.~(2020, in preparation) we 
derive mass, radius, and distance of the K2-\rave\  stars using PARAM 
\citep{rodrigues2017}, a Bayesian tool that infers stellar properties using both atmospheric 
and seismic parameters as input. 

\begin{deluxetable}{l|c}
\tablecaption{List of the number of K2-\rave\  calibrating stars in the first six K2 fields.\label{tab:seismo}}
\tablehead{\colhead{K2 Field} & \colhead{N. of \rave\  Targets}} 
\startdata
      C1 & 87 \\
        C2 & 116 \\
        C3 & 288 \\
        C4 & -- \\
        C5 & -- \\
        C6 & 208 \\ \hline
\enddata
\end{deluxetable}


\section{Chemical abundances with \texttt{GAUGUIN}}\label{sec:GAUGUIN}

The spectral region studied by RAVE contains,  along with the Ca triplet, a considerable number 
of spectral lines that can be exploited for abundance determination of individual elements. In 
\cite{boeche2011}, 604 absorption lines for \ion{N}{1}, \ion{O}{1}, \ion{Mg}{1}, \ion{Si}{1}, 
\ion{S}{1}, \ion{Ca}{1}, \ion{Ti}{1}, \ion{Ti}{2}, \ion{Cr}{1}, \ion{Fe}{1}, \ion{Fe}{2}, 
\ion{Co}{1}, \ion{Ni}{1}, \ion{Zr}{1}, and the CN molecule could be identified in the spectra 
of the Sun and Arcturus, respectively. By means of a curve of growth analysis, \cite{boeche2011} 
could devise an automated pipeline to measure individual abundances for seven species (Mg, Al, 
Si, Ca, Ti, Fe and Ni) based on an input for \teff, \logg\ and an overall metallicity \mh\ with 
an accuracy of about $\approx 0.2$ \dex\ for abundance levels comparable to that in solar type stars. 
Subsequently, this code was developed further and is now publicly available under the name SP\_Ace 
\citep{Sp_Ace}. The shortcoming of the method was its loss of sensitivity for abundances considerably 
below the solar level, and also that individual error estimates were difficult to obtain. For \rave\ 
DR6 we changed this strategy considering the following considerations:

\begin{enumerate}
     \item Since \rave\ was primarily designed to be a Galactic archeology survey, and considering the 
limitations imposed by resolution, \SNR, wavelength range and accuracy of the deduced 
{stellar atmospheric parameters} \teff\ and \logg, our main focus is not to obtain precise measurements of individual stars 
but rather to obtain reliable trends for populations of stars.
     \item As analyzed in detail in \cite{kordopatis2011a} and         \cite{kordopatis2013}, the 
\ion{Ca}{2} wavelength range at $R \la 10000$ suffers from considerable spectral degeneracies which, if 
not properly accounted for, can result in considerable biases of automated parameterization pipelines. 
    Our approach, therefore, relies on the \texttt{MADERA} derived values for \teff, \logg, and \mh\ as 
input values. Alternatively also the {stellar atmospheric parameters} derived from the \texttt{BDASP} pipeline could be employed 
as input parameter and for the convenience of the reader we provide them also in Section \ref{sec:FDR}. 
Our preference lies, however, in the \texttt{MADERA} input values as they are purely spectroscopically 
derived and thus maximize the internal consistency between the derived atmospheric parameters and the 
inferred abundances.
    \item 
   To derive individual abundances of non-$\alpha$ elements (here: Fe, Al and Ni) we fit the absorption lines for individual species by varying the metallicity around the value for \mhDR.
    \item For $\alpha$-elements, however, a different approach is needed. Here, we vary the overall $\alphafe$ 
overabundance for a given $\mh$ so to optimize the match between the \rave\ spectrum and that in the template library. A fit of the overall spectrum allows us to take advantage of the maximum amount of information ($\alpha-$element lines, including the \ion{Ca}{2} triplet).
    
\end{enumerate}

As we will illustrate in Sections \ref{subsec:quality} and \ref{subsubsec:GAUGUIN_val}, this approach is 
capable of  providing crucial chemical information for lower metallicity stars, for which the \ion{Ca}{2} triplet is still prominent. 

The practical implementation of the aforementioned strategy employs the optimization pipeline \texttt{GAUGUIN} 
 \citep{bijaoui2012, guiglion2018a} to match a \rave\ spectrum to  a pre-computed synthetic spectra grid via a Gauss-Newton algorithm.

\subsection{The \texttt{GAUGUIN} method}

\texttt{GAUGUIN} was originally developed in the framework of the Gaia/RVS analysis 
developed within the Gaia/DPAC for the estimation of the stellar atmospheric 
parameters \citep[for the mathematical basis, see][]{GAUGUIN10}. For first 
applications, see \cite{GAUGUIN12, GSPspec16}. A natural extension of 
\texttt{GAUGUIN}'s applicability to the derivation of stellar chemical 
abundances was then initiated within the context of the Gaia/RVS 
\citep[DPAC/Apsis pipeline,][]{Apsis13}, the AMBRE Project 
\citep{AMBRE13,phdthesisguiglion,guiglion2016,guiglion2018a}, 
and the Gaia-ESO Survey \citep{gilmore2012}. 
Currently, \texttt{GAUGUIN} is integrated into the Apsis pipeline at the Centre 
National d'\'Etudes Spatiales (CNES), for Gaia-RVS spectral analysis 
\citep{bailerjones2013, recio-blanco2016, andrae2018}. 

\texttt{GAUGUIN} determines chemical abundance ratios for a given star by comparing 
the observed spectrum to a set of synthetic spectra. In order to both have a fast method 
and be able to deal with large amounts of data, it is best to avoid synthesizing model 
spectra on-the-fly. \texttt{GAUGUIN} is therefore based on a pre-computed grid of synthetic 
spectra, that we interpolate to the  {stellar atmospheric parameters} of the star, in order to 
create a set of synthetic models for direct comparison with the observation.

The triplet of calibrated $\{\teff,\logg,\mh\}$ from \texttt{MADERA} is used as  
 input {stellar atmospheric parameters} for \texttt{GAUGUIN}. 
 
 \subsubsection{Preparation of the observed \rave\ spectra for abundance analysis}
We perform 
an automatic adjustment of the whole radial-velocity-corrected \rave\  spectral 
continuum provided by the \texttt{SPARV} pipeline (see DR6-1). For a given star defined by 
$\teff$, $\logg$, and $\mh$, 
we linearly interpolated a synthetic 
spectrum using the respective 
spectral grid (see next sections). Removing the line features by sigma-clipping, 
we performed a polynomial 
fit on the ratio between the observed and the interpolated spectra continua. 
We use a simple gradient for this polynomial fit (\ie, first order), as the best choice for the problem.
Tests showed that at  $R\sim 7500$ resolution, 
using a second- or third-order polynomial fit leads to systematic shifts 
of the continuum by 1.5-2.0\% for typical giant-branch stars 
($\teff<5\,000\,$K, $\logg<3$), and by 0.5-1.0\% for hot stars ($\teff>5\,500\,$K, $\logg<3.5$), owing to the presence of the strong 
Calcium triplet.

We made further tests to explore the impact 
of a bad continuum placement, based on a first order fit. 
Such test were performed on synthetic spectra of an Arcturus-like 
(giant) star and a Sun-like (dwarf) star. We shifted the continuum 
by 2\% for Arcturus and by 1\% for the Sun. The measured abundances 
($\alphafe$, Ni, Al, and Fe) are then biased by approximately 
0.033 \dex\ for the dwarfs and 0.055 \dex\ for the giants.

\subsection{Abundance determination of [Fe/H], [Al/H], and [Ni/H]}\label{madera_grid}

In order to maintain consistency between the \texttt{MADERA} 
{stellar atmospheric parameters}  and chemical abundances of non-$\alpha$ elements, 
we employed the synthetic spectral grid as used 
by \texttt{MADERA} \citep[Section \ref{subsec:MADERA}, see also][]{kordopatis2013, kunder2017}. 
For the elemental abundance analysis we restricted the range of effective temperatures 
to be within  $4\,000\le\teff\le7\,000\,$K (in steps of 250\,K), thus avoiding stars 
that are too cool 
(owing to considerable mismatches between spectral templates and the \rave\ spectrum) 
or too hot (with spectral features that are too weak). 
We kept the same ranges in \logg\ and \mh\ as the \texttt{MADERA} grid \ie, 
$0.5\le\logg\le5.0$ (in steps of 0.5\,\dex) and 
$-5.0\le\mh\le1.0$\,\dex\ in steps of 0.25\,\dex. 
The spectral resolution of the synthetic spectra matches that of the observational data
($R\sim7\,500$), with  binning of $0.35\,$\AA. We refer the reader to 
Section 3.4 of \cite{kordopatis2013} for more details concerning this grid of synthetic spectra.

The intermediate-resolution and  wavelength domain of the \rave\  
spectra provides a unique scenario for determining chemical abundances, which is in synergy 
with the processing of the Gaia mission. In this framework, we were able to obtain chemical 
abundances of 3 elements: Fe, Al, and Ni. {In order to get the abundance \Xh\ for each of these 3 elements, we vary for a given \rave\ spectrum the metallicity around the metallicity $\mh_{\rm DR6}$ at fixed $\teff_{\rm , DR6}$ and $\logg_{\rm DR6}$ until the best match to an absorption line of element X is achieved. In the following, we refer to the varied metallicity parameter as $\mu$. In practice, we create a 1D grid of synthentic spectra $S1_i(\lambda)=S(\mu_i,\lambda)$ for 7 grid points $\mu_i$: the central point is obtained by a trilinear interpolation from the eight neigboring grid points in \teff,\logg, and \mh\ from the \texttt{MADERA} 3D grid of synthetic spectra. The other six grid points of the 1D grid are then obtained by applying the same interpolation procedure to the atmospheric parameters sets with $\mu=\mh\pm 0.2$, $\mh\pm 0.4$, and $\mh\pm 0.6$, respectively, but keeping \teff\ and \logg\ unchanged. Then the best matching spectrum is found by minimizing the quadratic distance between the observed spectrum $S(\lambda)$ and and a synthetic one $S1(\mu,\lambda)$. The latter one is obtained by interpolation in $\mu$ on the 1D grid $S1_i(\lambda)$. The procedure is applied over a narrow wavelength range (typically from 4 to 9 pixels) around each spectral line \citep[see][and below]{guiglion2016}.} For each line of element X, 
we computed a {$\chi^2$} between the observed spectrum and its best 
synthetic spectrum. We averaged those {$\chi^2$} values, weighted by the number 
of pixels used for the fit, providing then a mean {$\chi^2$}. To search 
for the best lines, we made a careful examination of spectral 
features in the \rave\  wavelength region. A more detailed discussion of this 
procedure can be found in \citet{guiglion2018b}. The resulting selection of   
lines for the chemical abundance analysis with \texttt{GAUGUIN} are given in 
Table \ref{tab:lines_abund_gauguin}, and have been astrophysically calibrated 
by \citet{kordopatis2011b}. For a given element with several 
spectral lines, we averaged the individual line abundance 
measurements thanks to a sigma-clipped mean.

\startlongtable
\begin{deluxetable}{lrrr}
\tablecaption{\label{tab:lines_abund_gauguin}Ion, wavelength {(line)}, excitation potential {($\chi_e$)} 
and $\loggf$ values for the spectral lines used in the chemical 
abundance pipeline \texttt{GAUGUIN} (data are from \citealt{kordopatis2011b}, 
{see also Section \ref{subsec:MADERA} of the present paper}.)}.
\tablehead{\colhead{Ion} & \colhead{line [\AA]} & \colhead{$\chi_e$} & \colhead{$\loggf$}} \startdata
\ion{Al}{1} & $8\,772.865$ & 4.022 & -0.39 \\
\ion{Al}{1} & $8\,773.897$ & 4.022 & -0.20 \\
\ion{Fe}{1} & $8\,514.794$ & 5.621 & -2.13 \\
\ion{Fe}{1} & $8\,515.108$ & 3.018 & -2.13 \\
\ion{Fe}{1} & $8\,526.669$ & 4.913 & -0.71 \\
\ion{Fe}{1} & $8\,582.257$ & 2.990 & -2.36 \\
\ion{Fe}{1} & $8\,592.951$ & 4.956 & -0.91 \\
\ion{Fe}{1} & $8\,611.803$ & 2.845 & -2.06 \\
\ion{Fe}{1} & $8\,621.601$ & 2.949 & -2.47 \\
\ion{Fe}{1} & $8\,688.624$ & 2.176 & -1.33 \\
\ion{Fe}{1} & $8\,698.706$ & 2.990 & -3.32 \\
\ion{Fe}{1} & $8\,699.454$ & 4.955 & -0.54 \\
\ion{Fe}{1} & $8\,713.187$ & 2.949 & -3.08 \\
\ion{Fe}{1} & $8\,713.208$ & 4.988 & -1.04 \\
\ion{Fe}{1} & $8\,757.187$ & 2.845 & -2.09 \\
\ion{Fe}{1} & $8\,763.966$ & 4.652 & -0.33 \\
\ion{Ni}{1} & $8\,579.978$ & 5.280 & -0.94 \\
\ion{Ni}{1} & $8\,636.995$ & 3.847 & -1.94 \\
\ion{Ni}{1} & $8\,702.489$ & 2.740 & -3.19 \\
\ion{Ni}{1} & $8\,770.672$ & 2.740 & -2.79 \\
\enddata
\end{deluxetable}

\subsection{Determination of $\alphafe$ ratios}\label{ges_grid}

Because $\alphafe$ is not a free parameter for a given metallicity in the synthetic spectral grid used by \texttt{MADERA} (Section \ref{subsec:MADERA}),  
we adopted the 2014 version of the 4D Gaia-ESO Survey (GES) synthetic spectra 
grid (de Laverny et al, in preparation), which provides high resolution synthetic 
spectra as a function of 4 input variables: \teff, \logg, \mh, and \alphafe.
The synthetic spectra grid adopted for the derivation of the $\alpha$ 
abundances is the one specifically computed for the Gaia-ESO Survey \citep[see  descriptions in][and Heiter et al. 2019, submitted]{smiljanic2014}.
In summary, the grid consists of $11\,610$ 1D LTE high-resolution synthetic spectra 
(sampled at 0.0004 nm) for non-rotating FGKM spectral type stars, covering the 
\ion{Ca}{2} triplet region. The GES 
atomic and molecular linelists (Heiter et al. 2019, submitted) were adopted 
for the computation of the synthetic spectra. The global metallicity ranges 
from $\mh=-5.0$ to $+1.0$ dex and five different $\alphafe$ enrichments 
 {are} considered for each metallicity value. 
 The effective temperature covers 
the domain $3\,600\le\teff\le8\,000\,$K (in steps of 200\,K from $3\,600$ to 
$4\,000$K, and 250\,K beyond), while the surface gravity covers the range 
$0.0\le\logg\le5.5$ (in steps of 0.5\,\dex). The grid computation adopts
almost the same methodology as the one used for the AMBRE Project \citep{AMBRE13} 
described in \citet{delaverny2012}. MARCS model atmospheres \citep{gustafsson2008} 
and the Turbospectrum code for radiative transfer \citep{alvarez1998, plez2012} 
are used, together with the Solar chemical abundances of \cite{grevesse2007}. {The grid also employs} consistent $\alphafe$ enrichments in 
the model atmosphere and the synthetic spectrum calculation together with an 
empirical law for the microturbulence parameter \citep[][and 
Bergemann et al., in preparation]{smiljanic2014}. Plane-parallel and spherical assumptions have 
been used in the atmospheric structure and flux computations for 
dwarfs ($\logg>3.5$) and giants ($\logg \le 3.5$), respectively.

\begin{figure}
\begin{center}
\plotone{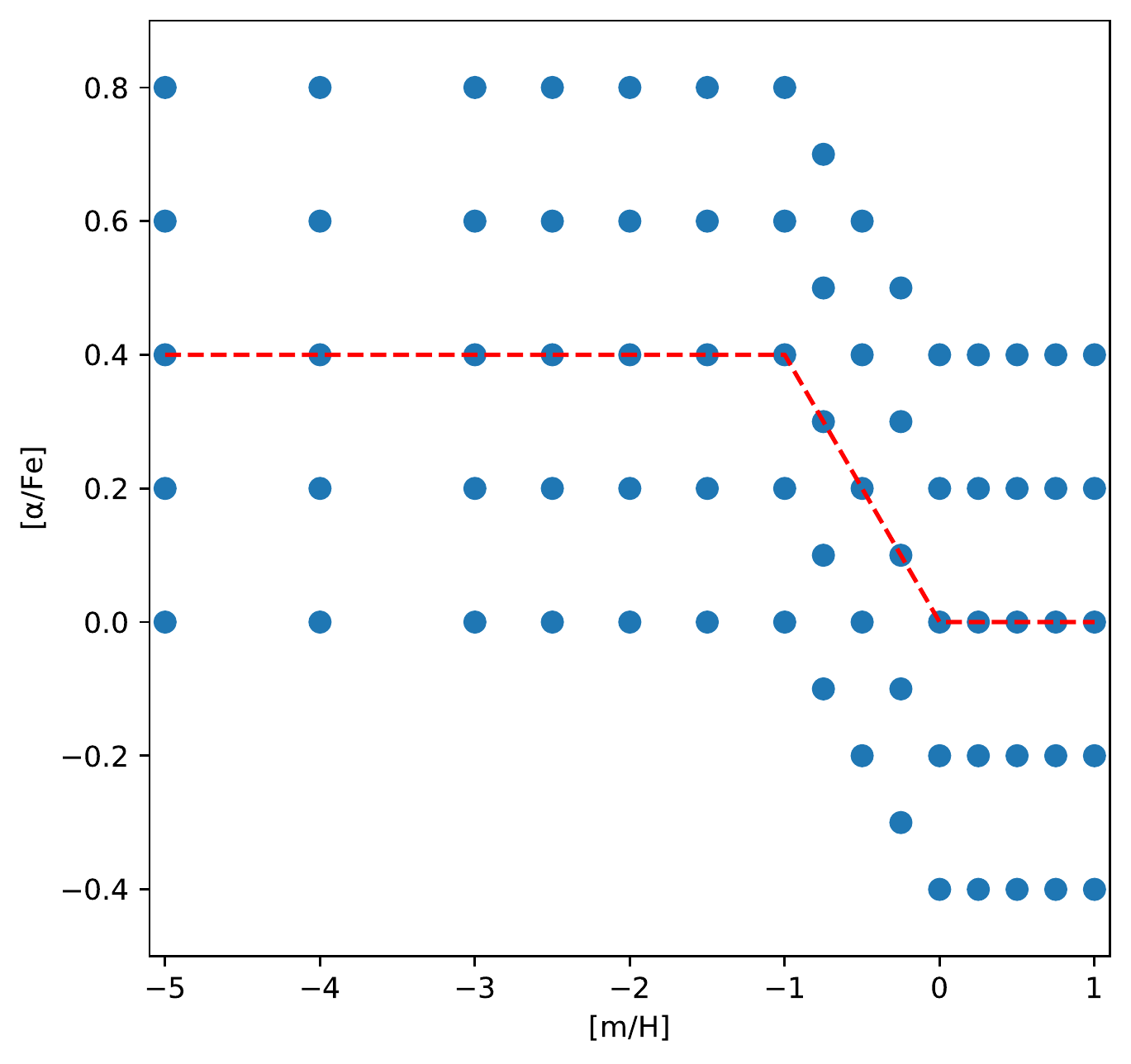}
\caption{\label{fig:ges_4D_grid_alphafe_vs_mh_coverage}$\alphafe$ vs. metallicity 
coverage of the GES synthetic spectra grid. The red dashed line shows the \texttt{MADERA} grid 
coverage used to determine {stellar atmospheric parameters} and Fe, Al and Ni abundances.
}
\end{center}
\end{figure}

As mentioned previously, the overall metallicity and $\alphafe$ ratios follow the same relation as Equation \ref{eqn:MADERA_alpha}, 
but the grid includes extra $\alphafe$ enrichments at each metallicity, as illustrated in 
Figure \ref{fig:ges_4D_grid_alphafe_vs_mh_coverage}. 
The spectral resolution of the GES synthetic spectra have 
been degraded in order to match that of the observational data ($R\sim7\,500$) 
with a binning of $0.35\,$\AA.

\begin{figure}
\begin{center}
\plotone{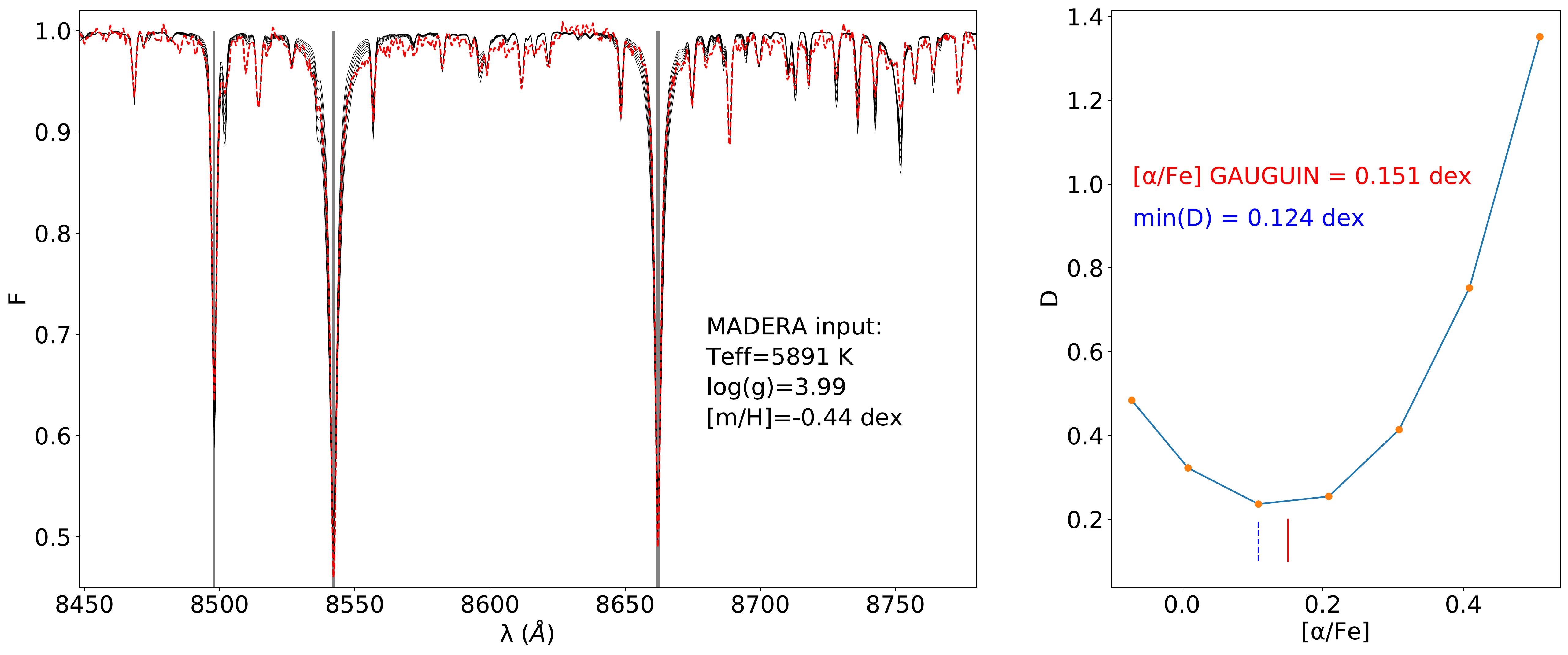}
\caption{\label{fig:detail_gauguin}Example of $\alphafe$ ratio 
measurement via \texttt{GAUGUIN} for the target 20070423\_1650m57\_114, 
with \texttt{MADERA} input. Grey zones are avoided in the fit (\ion{Ca}{2} cores). 
The left hand panel shows, in black, the set of 
synthetic spectra interpolated to the  {stellar atmospheric parameters} of the star, 
with 7 values of $\alphafe$. The observed spectrum is shown in red. 
The right-hand panel displays the distance between the models and 
the observed spectrum, as a function of the $\alphafe$ of each 
synthetic spectrum. The rough minimum of the quadratic distance (blue dashed line) 
is refined by the Gauss-Newton algorithm, leading to a different minimum 
(red line) and hence abundance. 
We notice that in this case only 7 models are use for the distance computation 
because of the grid edge effect in this metallicity regime.}
\end{center}
\end{figure}

{In order to get the  \alphafe\ abundance ratios, we follow the analogous procedure as in Section \ref{madera_grid}:
we create a 1D grid $S2_i(\lambda)=S(\alphafe_i,\lambda)$ by trilinear interpolation from the eight neighboring grid points of the GES 4D grid of synthetic spectra to the calibrated \texttt{MADERA} stellar atmospheric parameters $\teff_{\rm , DR6}$, $\logg_{\rm DR6}$, and $\mh_{\rm DR6}$ of the underlying \rave\ star. The initial \alphafe\ of the input spectrum is assumed to follow Equation \ref{eqn:MADERA_alpha}. A 1D grid with 9 elements is then created by applying the analogous interpolation to atmospheric parameters sets with $\alphafe=\alphafe_{\rm initial}\pm 0.1$, $\alphafe_{\rm initial}\pm 0.2$, $\alphafe_{\rm initial}\pm 0.3$, and $\alphafe_{\rm initial}\pm 0.4$, respectively, but keeping \teff, \logg, and \mh\ unchanged. We then compute the quadratic distance 
between the observed spectrum $S(\lambda)$ and each of the interpolated synthetic spectra 
of $S2_i(\lambda)$ over the whole spectral range. We exclude the cores of the \ion{Ca}{2} triplet lines as they can 
suffer from deviations owing to NLTE effects or chromospheric emission lines depending on the spectral type.
An example of such a 1D grid is shown in Figure \ref{fig:detail_gauguin}, 
for a \rave\ spectrum.}

For both steps, derivation of the elemental abundances of Al, Ni and Fe, as well as for the $\alpha$ overabundance, a rough minimum of the quadratic distance is given by the closest 
point to the true minimum (see 
Figure \ref{fig:detail_gauguin}, right panel, dashed line). It is then refined 
using a Gauss-Newton algorithm \citep{bijaoui2012}, as illustrated by the 
red dashed line in Figure \ref{fig:detail_gauguin}. We provide a 
$\chi^2$ fit between the observed spectrum and a synthetic one, 
computed for the \texttt{GAUGUIN} abundance solution.

\texttt{GAUGUIN} was implemented combining C++ and IDL\footnote{Interactive Data 
Language}, allowing it to derive $60$ $\alphafe$ ratios per second, 
and $1\,200$ individual abundances per second. {For 
the analysis of the whole data set with \texttt{GAUGUIN}} 
(normalization, abundances + errors) the overall 
computation time was 29 hours, on a single CPU-core.

\subsection{Calibration of \texttt{GAUGUIN} [Fe/H], [Al/H], [Ni/H] ratios}

The synthetic spectra adopted to derive [Fe/H], [Al/H], [Ni/H] ratios 
are calibrated with respect to the Sun and verified with respect to 
Arcturus and Procyon \citep{kordopatis2011a, kordopatis2011b}. From line-to-line, 
small mismatches can occur between the observed Solar and Arcturus 
spectra and their respective synthetic spectrum. We therefore chose 
to apply a zero-point correction to the \texttt{GAUGUIN} abundances. 
To do so, we determined with \texttt{GAUGUIN} the chemical abundances for the 
Sun and Arcturus, using the high resolution library of \citet{hinkle2003}, 
degraded to match the \rave\  spectral resolution. The input  {stellar atmospheric parameters} of both stars were chosen to be consistent with those obtained by 
\texttt{MADERA}. The \texttt{MADERA} zero-point correction was derived by feeding \texttt{GAUGUIN} with
the \emph{un-calibrated} {stellar atmospheric parameters} derived by \texttt{MADERA} from the Solar and 
Arcturus spectra: 
Sun \{$\teff_{,u}=5\,578\,$K, 
$\logg_{u}=4.09$, $\mh_{u}=-0.24\,$\dex\}; 
Arcturus \{$\teff_{,u}=4\,318$K, 
$\logg_{u}=2.04$, $\mh_{u}=-0.35\,$\dex\}\footnote{corresponding 
to calibrated values of \{$\teff=5\,619\,$K, $\logg=4.11$, $\mh=-0.06\,$\dex\} for the 
Sun and \{$\teff=4\,370$K, $\logg=2.41$, $\mh=-0.08\,$\dex\} for Arcturus, 
respectively.}. The averaged zero-point corrections that we apply to the \texttt{GAUGUIN}-derived 
[Fe/H], [Al/H] and [Ni/H]   abundances are presented in Table \ref{tab:zero_point_abund_gauguin}. 
The \feh\ corrections are minor as \texttt{GAUGUIN} tends to track  
the input metallicity very well. Arcturus zero-point abundances have been applied 
to giants ($\logg\le3.5$), while the Solar zero-point abundances have been 
applied to dwarfs ($\logg>3.5$), line by line. We note that such zero-point 
corrections will shift the global patterns in the [X/Fe]$\,vs.\,$\feh\ plane, 
but their slope will remain mainly unchanged owing to very small $\feh$ corrections.
\begin{deluxetable*}{lrrrr}
\tablecaption{\label{tab:zero_point_abund_gauguin}Zero-point corrections that 
we added to \texttt{GAUGUIN} \Xh\ abundances, for the giants (Arcturus) and dwarfs 
(Sun).}
\tablehead{\colhead{Elem.} & \colhead{$\text{Corr}_{\text{Arc}}$} & \colhead{$\text{Corr}_{\text{Sun}}$} }
\startdata
Fe & +0.01 & +0.02 \\
Al & -0.13 & -0.17 \\
Ni & -0.16 & -0.19 \\
\enddata
\end{deluxetable*}

\subsection{Individual errors on [$\alpha$/Fe], [Fe/H], [Al/H] and [Ni/H]}\label{subsec:gauguin_errors}

We provide individual error estimates on the 
\texttt{GAUGUIN} abundance ratios, while for the previous releases only a global 
error was provided \citep{kunder2017}. To do so, we considered two main 
sources of uncertainty: propagation of the errors of the 
{stellar atmospheric parameters} $\sigma_{p}(p\pm e_p)$, and the internal error of 
\texttt{GAUGUIN} due to noise $\sigma_{\text{int}}(\SNR)$ (internal precision, 
see the adopted procedure below). 
We combined them in a quadratic sum and we obtained the total uncertainty of 
the \texttt{GAUGUIN} chemical abundances:

\begin{equation}\label{equ:eq_gaug_sig}
\sigma = \sqrt{\sigma_{p}^2(p \pm e_p) + \sigma_{\text{int}}^2(\SNR)}.
\end{equation}

We detail the way we computed $\sigma_{\text{int}}(\SNR)$ in the following section.

\subsubsection{The precision of \texttt{GAUGUIN} [$\alpha$/Fe], [Fe/H], [Al/H] 
and [Ni/H] abundances}\label{subsec:gauguin_precision}

\begin{figure}
\begin{center}

\includegraphics[width=1.0\textwidth]{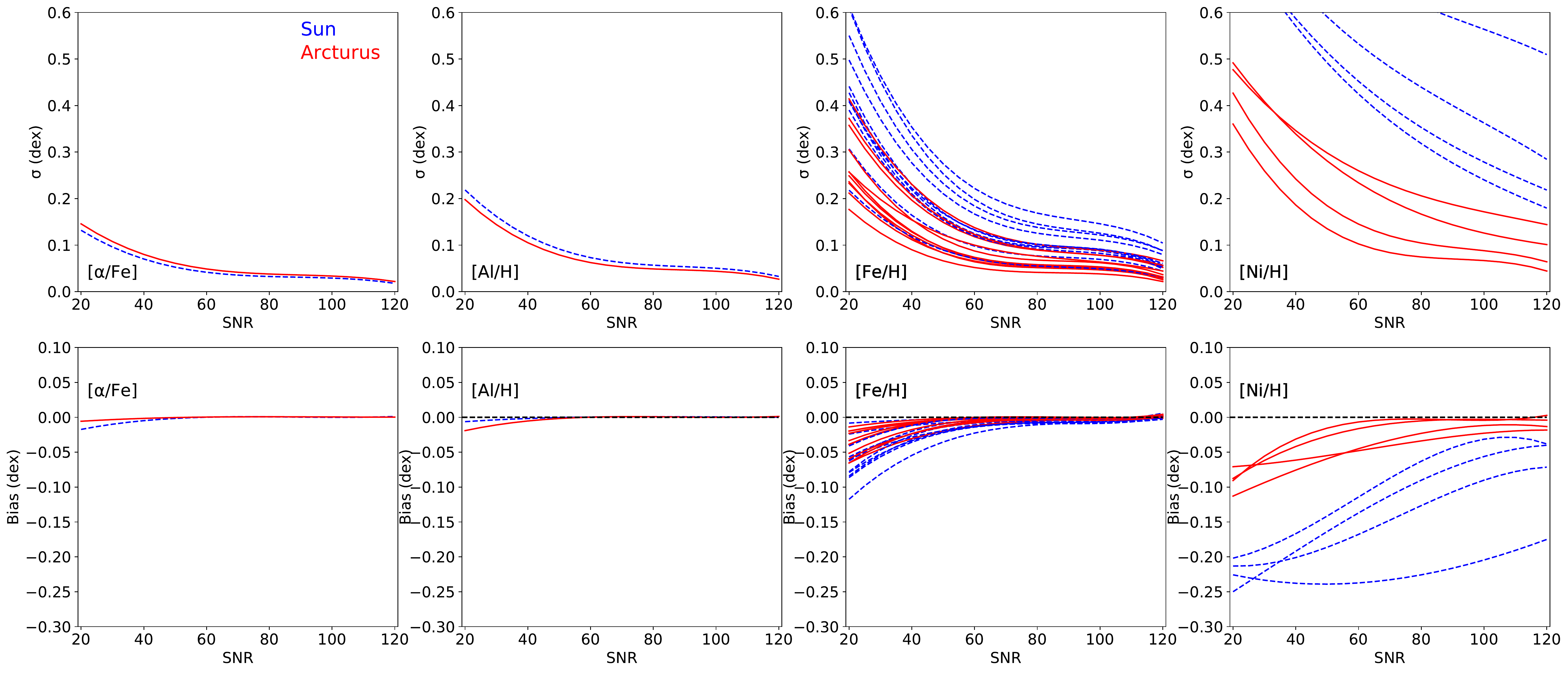}
\caption{\label{fig:gauguin_precision_accuracy}The internal precision (top row) 
of \texttt{GAUGUIN} and the internal accuracy (bottom row) as a function of Gaussian $\SNR$, 
for the overall $\alphafe$ and the individual lines of 
[Fe/H], [Ni/H], and [Al/H] (one curve per {spectral} line). 
The results are given for tests performed on both 
Solar-like stars (blue, dashed lines) and Arcturus-like stars (red, full lines).}
\end{center}
\end{figure}

The top row of Figure \ref{fig:gauguin_precision_accuracy} presents  the internal precision 
$\sigma_{\text{int}}(\SNR)$ as a function of $\SNR$ 
for $\alphafe$, [Al/H], [Fe/H], and [Ni/H], derived by \texttt{GAUGUIN}. 
The internal precision was characterized by taking 500 measurements of 
the abundance from noisy synthetic spectra of Sun-like ($\teff=5\,750\,$K, $\logg=4.5$, $\mh=+0.0$, 
$\alphafe=+0.0$) and Arcturus-like stars ($\teff=4\,250\,$K, $\logg=1.5$, $\mh=-0.5$, 
$\alphafe=+0.3$), adopting $\SNR=5$ to $120$, with  steps of $\Delta\SNR=5$. 
We computed a simple standard deviation of the 500 abundance measurements, 
{at a given $\SNR$ and for a given spectral line}. 
Figure~\ref{fig:gauguin_precision_accuracy} clearly shows that the internal error is 
larger for dwarf stars 
than it is for giants. 
The overall $\alphafe$ based on the overall fit of the spectrum appears to be pretty robust, 
with low $\sigma$ (high precision). For [Ni/H] in dwarfs, the internal error varies 
strongly from one {spectral} line to another. 

\subsubsection{The accuracy of \texttt{GAUGUIN} [$\alpha$/Fe], [Fe/H], [Al/H] 
and [Ni/H] abundances}\label{subsec:gauguin_accuracy}

We investigate the ability of \texttt{GAUGUIN} to determine accurate abundances in the presence of 
noise. We adopt the same strategy as in Section \ref{subsec:gauguin_precision}, 
measuring abundances in synthetic spectra of Arcturus- and Sun-like stars. 
The bottom panel of Figure \ref{fig:gauguin_precision_accuracy} shows the bias as the difference between 
the average over 500 measurements of \Xh\ by \texttt{GAUGUIN} and the expected abundance, 
for individual lines. For a typical giant like Arcturus, we see that the bias 
tends naturally to be zero for $\SNR>50$, except for some Ni lines which tend to 
settle around a bias of $0.03\,$\dex\ at high $\SNR$. For a Sun-like star, the bias behaves 
very well for Fe for $\SNR>40$. For both stars, \texttt{GAUGUIN} creates no 
systematics for [Al/H], even at very low $\SNR$, the single spectral line being 
unblended and strong even in the Sun. On the other hand, in a Sun-like 
star, Ni exhibits large systematics with respect to Fe and Al 
because of its weak spectral lines. We conclude that our \texttt{GAUGUIN}-derived values intrinsically do 
not suffer from large systematics. We point out that in dwarfs, [Ni/H] 
values should be treated with caution, and can suffer from large systematics, 
even at large $\SNR$. 

\subsubsection{Total uncertainty of \texttt{GAUGUIN} [$\alpha$/Fe], [Fe/H], [Al/H] 
and [Ni/H] abundances}\label{subsec:gauguin_tot_error}

\begin{figure*}
\begin{center}
\plotone{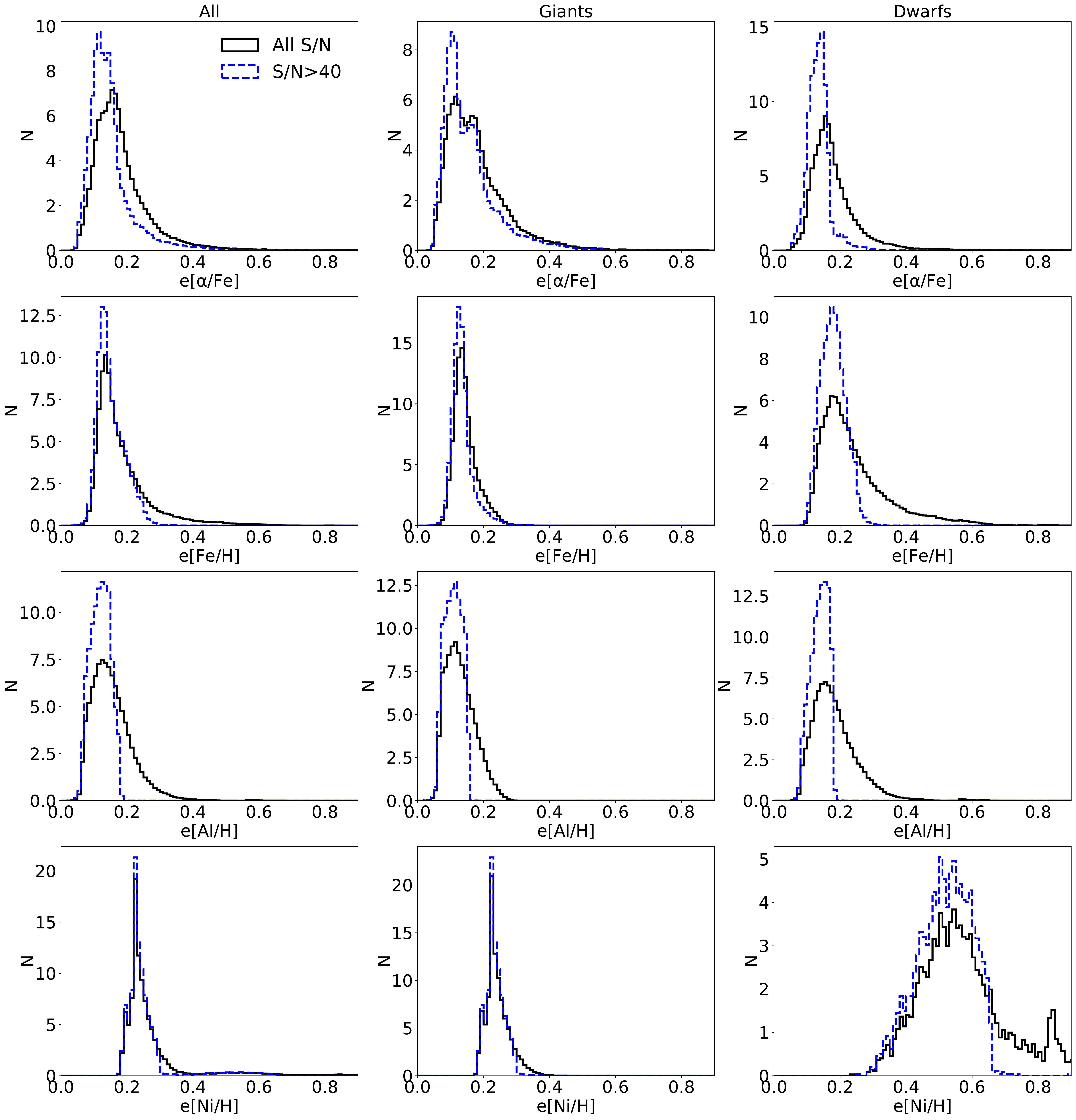}
\caption{\label{fig:gauguin_4d_total_error_alphafe_madera_all_giants_dwarfs}The 
distribution of total uncertainties for the $\alphafe$ ratios measured by \texttt{GAUGUIN}, 
with \texttt{MADERA} input. Left: whole sample. Middle: giants ($\log g < 3.5$). 
Right: dwarfs ($\log g \ge 3.5$).}
\end{center}
\end{figure*}

Figure \ref{fig:gauguin_4d_total_error_alphafe_madera_all_giants_dwarfs} 
shows the total uncertainty of the \texttt{GAUGUIN} $\alphafe$, [Fe/H], [Al/H], [Ni/H] 
ratios, derived using Equation~\ref{equ:eq_gaug_sig}, using \texttt{MADERA}  {stellar atmospheric parameters} as input. We show only stars with 
{a quality flag equal to "0" (as described in Section 
\ref{subsubsec:MADERA_flags})}. We observe that while the total uncertainties 
of $\alphafe$ abundances are very similar between dwarfs and giants 
($\sim0.16\,$dex), the total uncertainties of the other abundances are 
systematically larger for dwarfs.

When discarding stars with $\SNR<40$, we tend to remove the tail towards 
larger errors of the distributions. The typical errors for giants 
are of the order of $0.13\,$\dex\ for Fe and Al, and $>0.2\,$\dex\ for 
Ni. 
We give the median errors of each distribution in 
Table \ref{tab:gauguin_erros_median}. We note that even at high $\SNR$, 
Ni suffers from larger uncertainties for dwarf stars. We strongly 
recommend the reader to use the individual total errors in order to 
select the most reliable \texttt{GAUGUIN} abundances for their specific science 
application.

\begin{table}
\centering
\begin{tabular}[c]{c c c c c}
\hline
Elem & \SNR & All & Giants & Dwarfs \\
\hline 
\hline
$\alphafe$ &  all  & 0.16 & 0.16 & 0.17 \\
           & $>40$ & 0.13 & 0.13 & 0.13 \\
$[$Fe/H$]$ &  all  & 0.16 & 0.14 & 0.21 \\
           & $>40$ & 0.14 & 0.13 & 0.18 \\
$[$Al/H$]$ &  all  & 0.14 & 0.12 & 0.17 \\
           & $>40$ & 0.12 & 0.11 & 0.14 \\
$[$Ni/H$]$ &  all  & 0.24 & 0.23 & 0.56 \\
           & $>40$ & 0.23 & 0.23 & 0.52 \\
\hline
\end{tabular}
\caption{\label{tab:gauguin_erros_median}Total \texttt{GAUGUIN} error for all 
elements, in the whole sample, giants and dwarfs, respectively. The error 
is presented adopting \texttt{MADERA} inputs. We refer the reader to 
\figurename~\ref{fig:gauguin_4d_total_error_alphafe_madera_all_giants_dwarfs} 
for a view of the distribution.}
\end{table}

\subsubsection{Further sources of uncertainty}

We conclude this discussion by  testing the sensitivity of \texttt{GAUGUIN}-derived abundances to micro-turbulence, rotational velocity, 
and radial velocity.

\begin{itemize}
 \item Micro-turbulence ($\xi$) is included in the GES synthetic 
spectra grid used by \texttt{GAUGUIN}, following a calibrated relation based 
on $\teff$, $\logg$, and $\mh$. Tests based on 
synthetic spectra revealed that the error on the \texttt{GAUGUIN} $\alphafe$ due to 
an error of $1\,\kms$ in $\xi$ is of the order of $0.01$ \dex\ for both Arcturus-like 
and Solar-like stars. For individual [Fe/H], [Al/Fe], [Ni/H], this error reaches 0.02 
\dex, \ie\ much smaller than the accuracy limit given by the resolution and 
\SNR\ limit of the RAVE spectra (typically 0.15-0.20 \dex\ uncertainty 
on chemical abundances).
The effects of micro-turbulence on the chemical abundances published in DR6 are 
thus negligible.

 \item We investigate how stellar rotation affects the \texttt{GAUGUIN} $\alphafe$ ratios, 
as such physical effects are not included in \texttt{GAUGUIN} or \texttt{MADERA}. 
We measure $\alphafe$, Fe, Al, and Ni in the synthetic spectra of two 
Arcturus- and Solar-like stars, for which we convolved the spectra with 
increasing rotational velocities (from 1 to 10 \kms). 
Our tests  reveal that such neglect of rotation is reasonable, as the induced 
systematic errors on the $\alphafe$ are only of the order of 
$0.009-0.017\,$\dex~for a typical rotational velocity of $5\,\kms$. 
This error is only of the order of $0.01-0.02\,$\dex~for [Fe/H], [Al/H], 
and [Ni/H]. As before, most of the RAVE stars should fall well below this limit.

 \item We tested the sensitivity of \texttt{GAUGUIN} when the observed 
spectrum and the set of synthetic spectra are not in the same rest-frame. 
The typical accuracy of \rave's  RV is $<2\,\kms$, corresponding 
to $16\%$ of a pixel. We perform the test on synthetic spectra of the Sun and 
Arcturus, and estimate that the error on our four abundances due to such a shift 
in the ``observed'' spectrum leads to errors of the order of $0.015\,$\dex\ for 
both spectral types. We note that this error is negligible at high \SNR, but 
tends to increase by a factor of two for $\SNR<40$. In this regime we 
expect larger uncertainties in the RV determination, and for RV errors 
of 4-5\,\kms, the uncertainty on \texttt{GAUGUIN} abundances increases by a factor 
of two. We encourage the reader to filter stars with large RV uncertainties, 
as mentioned in Sect~\ref{sec:Validation}.
\end{itemize}

\subsection{Sample selection and quality of fit}
\label{subsec:quality}

The internal error analysis presented above intrinsically assumes 
that the morphology of the observed \rave\ spectrum matches that of the 
synthetic grid. However, this condition does not necessarily {need} to be fulfilled, for example 
owing to a peculiarity of the \rave\ spectrum, either of an astrophysical nature (e.g., 
significantly deviant abundance pattern of the underlying star or shortcomings of the 
synthetic grid, particularly in the less studied ranges of the parameter space), or for  technical reasons (improper continuum normalization \eg, owing to fringing).

\begin{figure*}
\begin{center}
\plotone{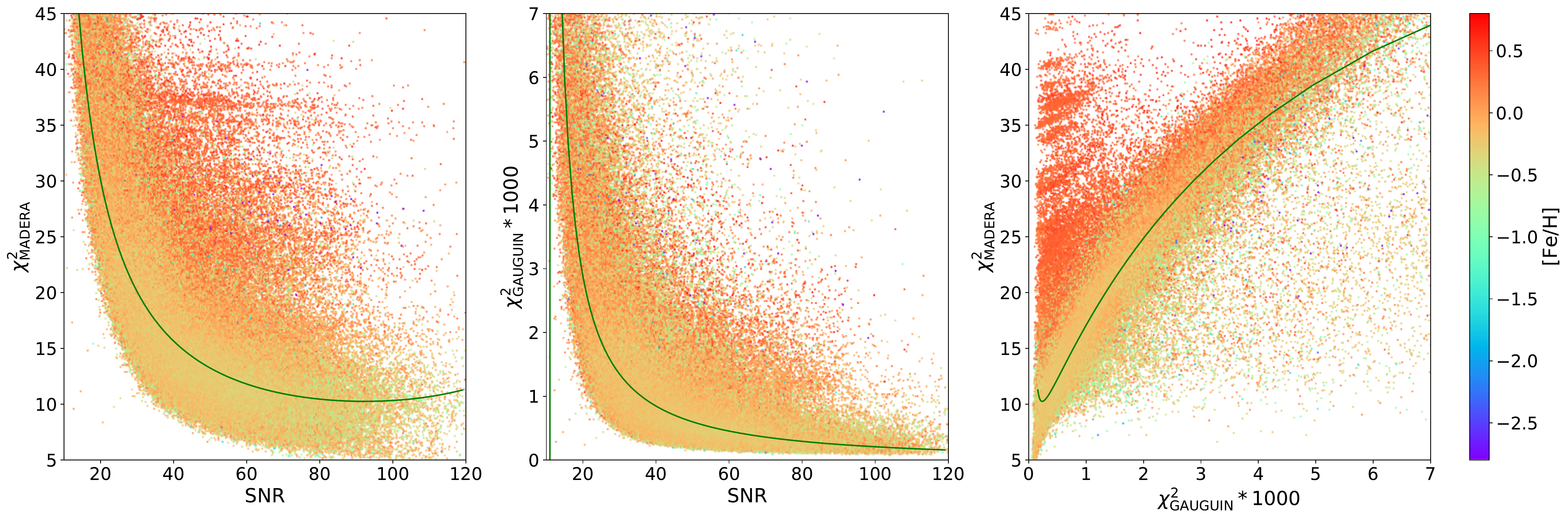}
\caption{\label{fig:goodfit} Left: $\chi^2$ values vs. \SNR\ for the \texttt{MADERA} pipeline. 
The green line corresponds to the median values as approximated by 
Equation \ref{eqn:chisq_madera}. The distribution is color coded by the metallicity $\mhDR$. 
Middle: $\chi^2$ values vs. \SNR\ for the \alphafe\ determination with the \texttt{GAUGUIN} pipeline. 
The green line corresponds to the median values as approximated by Equation \ref{eqn:chisq_gauguin}. 
The distribution is color coded by the metallicity $\mhDR$. Right: $\chi^2$ for the \texttt{MADERA} 
and \texttt{GAUGUIN} pipeline plotted against each other. }
\end{center}
\end{figure*}

The quality of the match between an actual \rave\ spectrum and the synthetic grids 
employed by the \texttt{MADERA} and \texttt{GAUGUIN} pipeline can be characterized by the two $\chi^2$ 
values provided by the \texttt{MADERA} and \texttt{GAUGUIN} pipelines (see also Figure \ref{fig:goodfit}): 
A poor match of the \texttt{MADERA} pipeline will result in large residuals between the \rave\ 
and the template spectrum, which in turn will result in a poor fit of \texttt{GAUGUIN} and/or in 
excessive (and likely unphysical) deviation in $\alphafe$. In addition, poor \SNR\ will 
naturally also lead to unreliable determinations using \texttt{MADERA} and/or \texttt{GAUGUIN}.

Figure \ref{fig:goodfit} illustrates this effect by showing, as a function of the \SNR\ 
and color coded by metallicity \mhDR, the $\chi^2_{\rm MADERA}$ (left) and 
$\chi^2_{\rm GAUGUIN}$ value for all objects for which \texttt{GAUGUIN} provides a converged 
solution for \alphafe. The {green} line corresponds to the median {$\chi^2$} as a function of \SNR, 
approximated by the following two relations:

\begin{equation}
\label{eqn:chisq_madera}
    \chi^2_{{median},{\rm MADERA}}(\SNR)=\frac{10^5}{-1.23\,\SNR^2 + 227\,\SNR\ + 725}
\end{equation}
and
\begin{equation}
\label{eqn:chisq_gauguin}
    \chi^2_{{median},{\rm GAUGUIN}}(\SNR)=\frac{1}{0.232\,\SNR^2 + 27.7\,\SNR\ -302}\qquad,
\end{equation}
for $10<\SNR<150$.
respectively. The right plot shows the $\chi^2$ values for \texttt{MADERA} and \texttt{GAUGUIN} against 
each other. The majority of the data points fall within a smooth distribution around the median value of 
either pipeline. 
Furthermore, as indicated by the right plot, usually both pipelines have a comparable quality of fit, \ie\ 
stars for which the \texttt{MADERA} pipeline provides results within the main distribution of the quality 
of fit also fall within the main distribution for \texttt{GAUGUIN}. However, both pipelines show a sizeable 
number of stars with considerably poorer fits than average even for very high \SNR\ values, usually associated 
with (\texttt{MADERA}-derived) very high super-solar metallicity. This is particularly prominent in the 
results for the \texttt{MADERA} pipeline. This finding is not very surprising as the aforementioned outliers 
predominantly correspond to very cool stars with a very dense forest of absorption lines, often also based on 
molecular lines, \ie\ where the proper modeling of synthetic spectra and the matching to medium resolution 
medium \SNR\ data is particularly challenging.

A simple and convenient way to characterize the simultaneous fit of \texttt{GAUGUIN} and \texttt{MADERA} can be defined via

\begin{equation}
\label{eqn:quality}
    \bchisq = \mu_{\rm MADERA}\,\frac{\chi^2_{\rm MADERA}}{\chi^2_{{median}, \rm MADERA}(\SNR)} + \mu_{\rm GAUGUIN}\frac{\chi^2_{\rm GAUGUIN}}{\chi^2_{{median}, \rm GAUGUIN}(\SNR)},
\end{equation}
\noindent 
with $\mu_{\rm MADERA}$ and $\mu_{\rm GAUGUIN}$ being being two arbitrary weighing factors. \bchisq\ is basically an 
effective $\chi^2$ for the combined fit, and its inverse, \ie, $Q=1/\bchisq$ can be seen as a quality parameter, \ie, 
a low value of \bchisq\ (a high value of $Q$) corresponds to a good fit. For the following we assume $\mu_{\rm MADERA}=1.25$ 
and $\mu_{\rm GAUGUIN}=1$, \ie\ we are a bit more restrictive with respect to the quality of the \texttt{MADERA} metallicity 
because of the poorer fits for some of the metal-rich stars.

\begin{figure*}
\begin{center}
\plotone{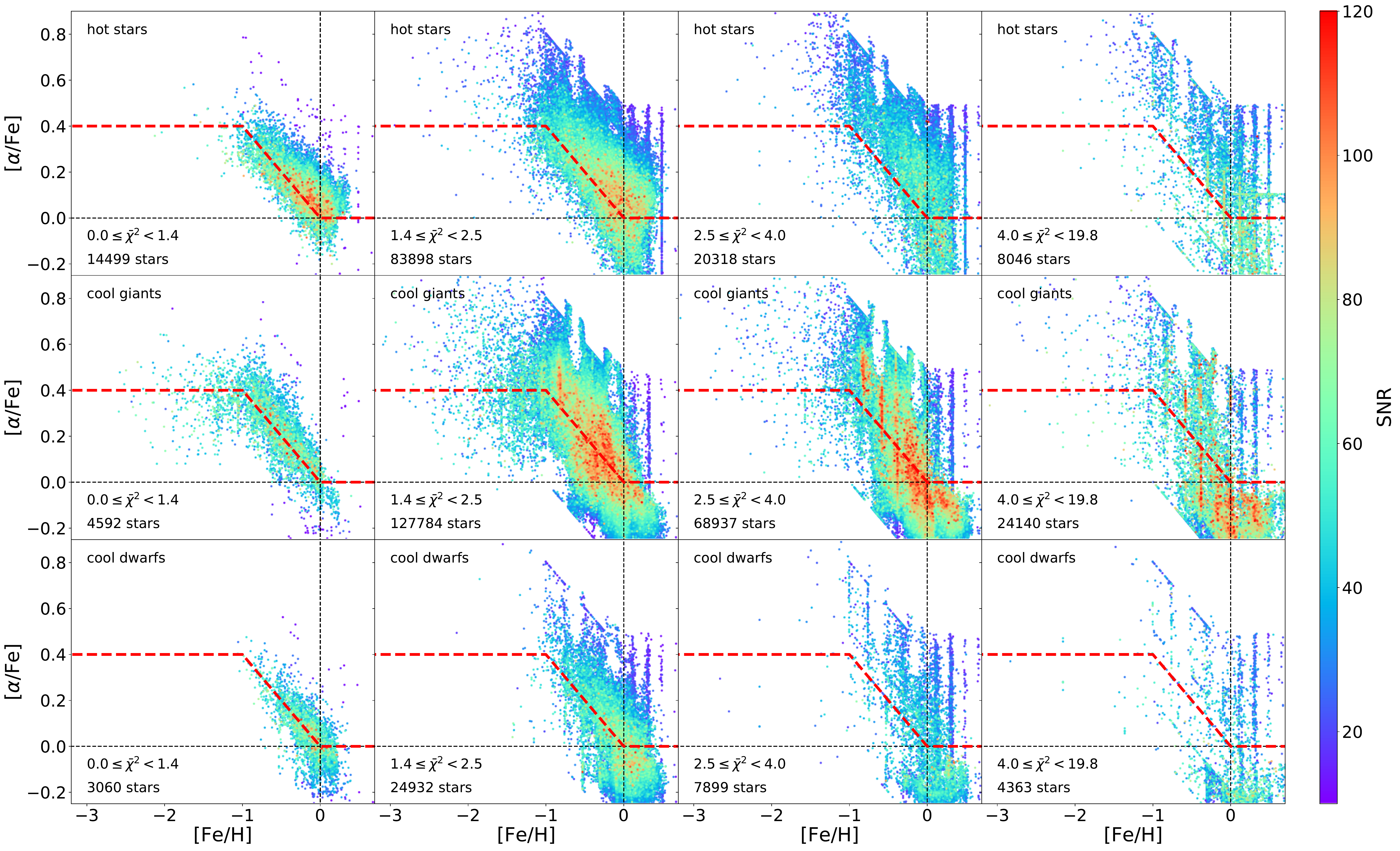}
\caption{\label{fig:al_fe_vary_SNR} \alphafe\ vs.~iron as a function of \SNR\ and the quality of the fit. The top row 
shows hot stars ($\teff > 5500$\,K), the middle row cool giants ($\teff \leq 5500$\,K, $\logg < 3.5$), the lower row 
cool dwarfs ($\teff \leq 5500$\,K, $\logg \geq 3.5$). The four columns from left to right show subsamples with a 
successively poorer quality parameter \bchisq\ of the \texttt{GAUGUIN} and \texttt{MADERA} fit (\ie, successively higher 
values). The color coding corresponds to the \SNR\ of the spectrum.}
\end{center}
\end{figure*}

In Figure \ref{fig:al_fe_vary_SNR} we consider the \alphafe\ vs.~iron relation for 4 different bins in the quality 
parameter \bchisq, namely $0\leq \bchisq < 1.4$, $1.4\leq \bchisq < 2.5$, $2.5 \leq \bchisq <4 $, and $\bchisq\geq 4$. 
The ({\feh},\alphafe) values are color-coded by \SNR. We separate the population of stars into three categories based on 
the \texttt{BDASP} stellar parameters, hot stars ($\teff\ > 5500$\,K, top row), cool giants ($\teff\ > 5500$\,K and 
$\logg < 3.5$, middle row), and cool dwarfs ($\teff\ > 5500$\,K and $\logg > 3.5$, bottom row). 

The high quality ($\bchisq<1.4$) sample shows the expected behavior: for metallicities above ${\feh}\approx-1$, all stars 
follow the \alphafe\ vs {\feh} relation of the Galactic disk. Owing to the scatter in the abundance determination of about 
0.15 dex even for the highest quality sample, a separation into two disk components cannot obviously be seen (see, 
however, section \ref{subsec:MW_Tomography}). Cool dwarfs, which owing to their low luminosity are mainly in the 
immediate neighborhood of the Sun, mainly have abundances comparable to the Sun, hot stars that can be identified to 
larger distances extend the \alphafe\ vs.~\mh\ relation towards considerably lower abundances. For cool giants, the 
data extends well into the metal poor regime $\mh<-1$, and the transition to a constant $\alpha$ overabundance is nicely 
traced. Relaxing the quality criteria keeps these characteristics at first for all three populations, but increases the 
scatter. For the high \SNR\ end of the distribution, fairly confined and well defined relations can still be traced. A 
further relaxation of the quality parameter ($\bchisq>2.5$) results in populating the area near the edges of the GES grid, 
in particular for the lower \SNR\ data.

\begin{figure*}
\begin{center}
\plotone{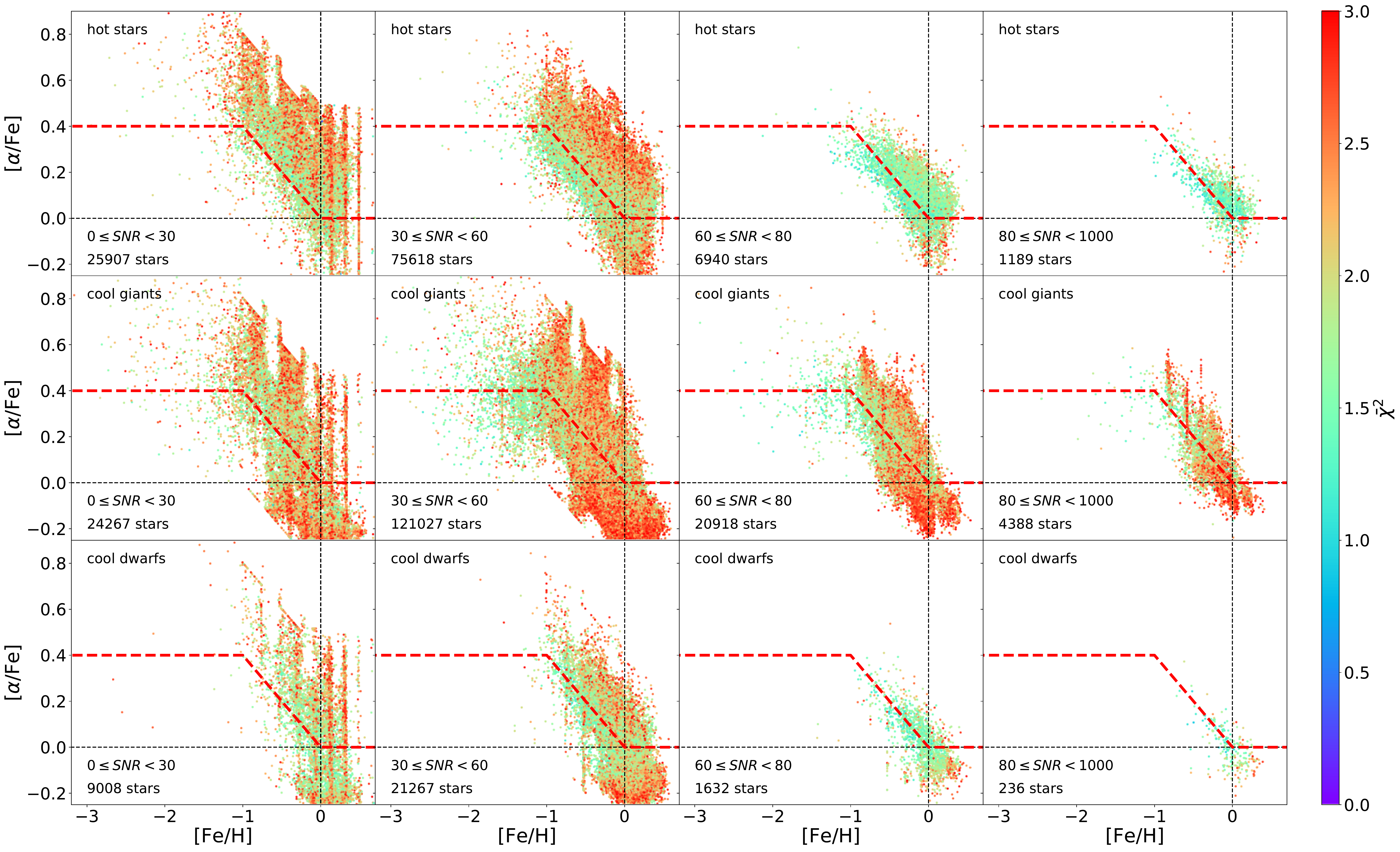}
\caption{\label{fig:al_fe_vary_Q} Same as Figure \ref{fig:al_fe_vary_SNR} but columns representing bins with increasing 
\SNR\ (from left to right). The color coding corresponds to the \bchisq\ value of the fit. }
\end{center}
\end{figure*}

The aforementioned behavior is also reflected in Figure \ref{fig:al_fe_vary_Q}, which illustrates it, now projected in bins 
of different \SNR\ (and color coding by $\bchisq$). In particular, it shows a systematic shift of the \alphafe\ vs.~\feh\ 
relation with decreasing \SNR, with the lower \SNR\ subset seeming to have systematically higher $\alphafe$. This effect can 
be understood, as the pipeline reacts to the increasing noise level by interpreting this as a higher $\alpha$ abundance. 
Again this effect is more pronounced in situations where the match between the \rave\ spectrum and template is poorer.

\begin{figure*}
\begin{center}
\plotone{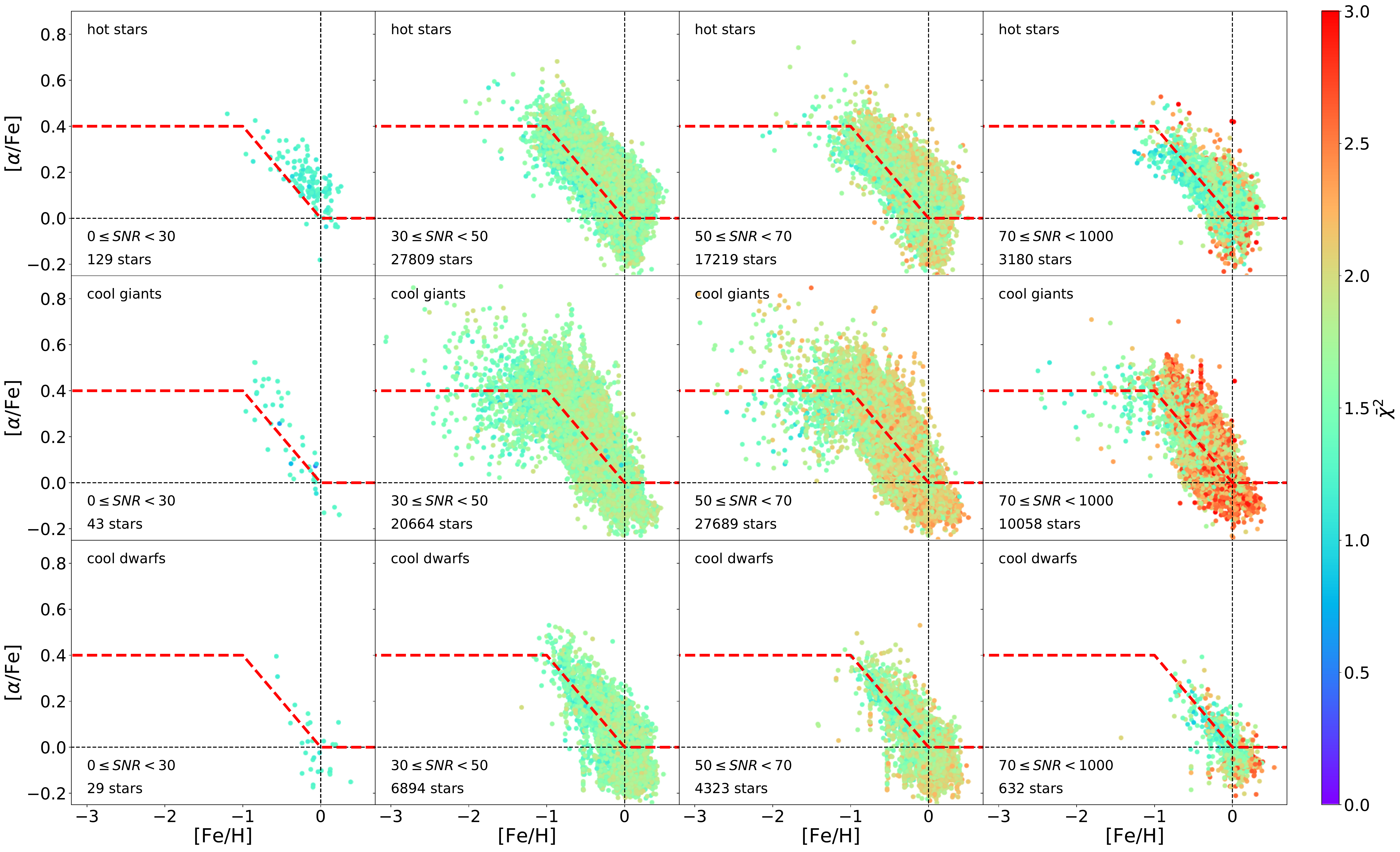}
\caption{\label{fig:al_fe_good_q} Same as Figure \ref{fig:al_fe_vary_Q} but only stars  that fulfill the quality 
criterion in Equation \ref{eqn:quality} are shown.}
\end{center}
\end{figure*}

Overall, $\bchisq<2.5$ provides satisfactory results for hot stars, while for cooler stars and low \SNR\ still some clustering 
at the grid boundaries, in particular at fairly negative \alphafe\ and high metallicity, can be observed. Such a quality cut 
also removes most targets from the first year of \rave\ observations that were still contaminated by light from 2nd order 
spectra (see DR6-1, Section 2.4). The residual presence of questionable abundance measurements at negative \alphafe\ and high 
metallicity can be suppressed by requiring a more stringent (lower) \bchisq\ value in particular for lower \SNR\ values. For 
example, the constraint $\bchisq<1.4$ for $\SNR < 40$ basically removes all stars with $ \alphafe <-0.1$ at  
$\feh > -0.4$. This is demonstrated in Figure \ref{fig:al_fe_good_q}, in which a critical threshold 
of $\bchisq < \bchisq_{\rm crit}=1.1\times\sqrt{\SNR/10-1.5}$ is applied.

\section{Orbits}\label{sec:Orbits}
For the convenience of users of \rave\  DR6 we provide the kinematic and orbital 
properties corresponding to each  observed star   
(see also Table~\ref{tab:DR6_Orbits} in Sec.~\ref{subsec:cat_orbits} for a list of the derived quantities). In each case we use as input the 
position on the sky and proper motion found by Gaia DR2, the radial velocity found by \texttt{SPARV}, 
and the \texttt{BDASP} derived distance. We take the quoted uncertainty on distance, along with the 
radial velocity error given as \texttt{hrv\_error\_sparv}. The proper motion 
uncertainties are found by summing in quadrature the quoted Gaia uncertainties and an 
estimate of their systematic uncertainties \citep[66 $\mu{\rm as}\,{\rm yr}^{-1}$, 
estimated from the small-scale spatially varying measured proper motions of quasars by][]{lindegren2018}.

Heliocentric positions and velocities are given in the standard coordinate system used in the solar neighborhood, \ie, the 
X direction is towards the Galactic center ($l=0,b=0$), the Y axis is in the direction of Galactic rotation ($l=90^{\circ},b=0$), 
and the Z axis points at the north Galactic pole ($b=90^{\circ}$). 

The orbital properties are found in the best-fitting Milky Way potential 
from \cite{mcmillan2017}. The Sun is assumed to lie at $R_0=8.21\,{\rm kpc}$, 
where the circular velocity is $233.1\,{\rm km}\,{\rm s}^{-1}$ and at a height 
above the plane $z_0=0.014\,{\rm kpc}$ \citep{binney1997}. The velocity of the 
Sun with respect to the local standard of rest is taken from \cite*{schoenrich2010}. We place the Sun at $\phi=0^{\circ}$ 
in our Galactocentric coordinate system, which (combined with the requirement that z increases towards the north Galactic 
pole) means that the Sun and stars of the Galactic disk have \emph{negative} $v_\phi$, and therefore negative angular momentum around the Galaxy model's symmetry axis.
We note that since this potential is axisymmetric, the orbital parameters we 
derive are increasingly untrustworthy as the influence of the Galactic 
bar becomes more significant 
\citep[e.g., within the bar's corotation radius, now thought to be of the order $5$ to $6\,{\rm kpc}$ from the Galactic center:][]{Sormani2015,Portail2017,Sanders2019}

The quoted values are derived as a Monte Carlo integral over the uncertainties. Many of 
the orbital properties have the unfortunate characteristic that a finite change in one of 
the measured quantities produces an infinite change in the orbital property (since a finite 
change in position and/or velocity can put the star on an unbound orbit which would have, for 
example, infinite radial action or apocentric radius), 
meaning that the expectation (mean) value is inevitably infinite. For this reason we describe the 
output of the Monte Carlo integral in terms of the median value and the difference between 
the median and the percentiles corresponding to $\pm1\sigma$ (i.e. $15.9$ and $84.1$ 
percent) such that one can quote, for example, the energy as ${\tt Energy}^{+{\tt 
EnergyPlus}}_{-{\tt EnergyMinus}}$. 

The values are found using the {\sc GalPot} software \citep{dehnen1998}\footnote{available 
at  \url{https://github.com/PaulMcMillan-Astro/GalPot}}, and the orbital actions $J_r, J_z$ are found 
using the St\"ackel Fudge \citep{binney2012} as packaged in 
{\sc agama} \citep{vasiliev2019} -- the third action $J_\phi$ is the same as the quoted value of the angular momentum.


\section{Validation of \rave\  DR6 parameters}\label{sec:Validation}

The data product of large surveys like \rave\ is always a compromise between the 
quality of the individual data entry and the area and depth of the survey. This applies 
to design decisions (like the applied exposure time/targeted \SNR) as well as to the 
decision which data to keep in the sample and which ones to exclude. Our policy for \rave\  
is to provide the maximum reasonable data volume possible, which allows the user to 
consider the tails of the distribution function. \textbf{The exact choice of the (sub)sample 
used for a particular case has to be made by the user based on the 
criteria needed for the respective science application!} Here, we only can give some 
first guidelines/recommendations regarding the {selection of proper sub samples}. For a description 
of the various parameters in the following paragraph we refer to the tables in Section 6 of DR6-1 and Section \ref{sec:FDR} in this publication.

\begin{enumerate}
\item \textbf{Radial Velocities:} Stars 
with 
$\texttt{correlationCoeff} > 10$ have a small scatter in the repeat measurements of 
their heliocentric radial velocity. The distribution peaks near 
$0.0\,$\kms, and the tail toward very large velocity differences is reduced by 90\% 
compared to the uncut sample, indicative of a high confidence measurement \citep[see, e.g.,][and Section 6 of DR6-1]{kordopatis2013}. We refer to 
the data set defined by these criteria as the core sample, or RV. 
\item \textbf{{Stellar atmospheric parameters}:} As a minimum requirement, the quality flags \texttt{algo\_conv\_madera}  of the \texttt{MADERA} pipeline (see Section \ref{subsubsec:MADERA_flags}) should be $\neq 1$, additionally to the aforementioned
criteria for the RV measurement. Higher confidence parameters (at the expense of a 
reduction in sample size) can be obtained by additionally requiring 
\texttt{algo\_conv\_madera}$ \ne 2$, or even \texttt{algo\_conv\_madera}$ =0$, by requiring that 
stellar spectra are classified as a certain type and/or by imposing constraints on the 
\SNR.
\item \textbf{Abundances:} basically the same considerations apply here as for the  {stellar atmospheric parameters}, but in addition a quality cut of $\bchisq\la 2.5$ should be applied, with a possibly even stronger constraint for targets with low \SNR.
\end{enumerate}

For the stellar parameter and abundance validation against external sources in this and the following section, we define five samples: 

\begin{itemize}
    \item Full: The full set of the \rave\ DR6 data base for which the pipelines deliver a result.
    \item RV: The subset of the full data base that fulfills the basic quality criterion for radial velocities (see above and DR6-1 Section 6).
    \item MD: The subset of the RV data base that fulfills the basic quality criterion for stellar parameter determination with the \texttt{MADERA} pipeline, i.e.,
    $\mathtt{algo\_conv\_madera} \neq 1$.
    \item BD: The subset of the MD data base that has Gaia DR2 distances and for which \texttt{BDASP} {stellar atmospheric parameters} could be derived.
    \item \qlow: The subset of the MD data base that fulfills the basic quality criterion $\bchisq<2.5$ for elemental abundance determination with the \texttt{GAUGUIN} pipeline.
    \item \qhigh: The subset of the MD data base that fulfills the basic quality criterion $\bchisq<\bchisq_{\rm crit}(\SNR)$ (see Section \ref{subsec:quality}) for elemental abundance determination with the \texttt{GAUGUIN} pipeline. 

\end{itemize}

\begin{figure*}
\begin{center}
\plotone{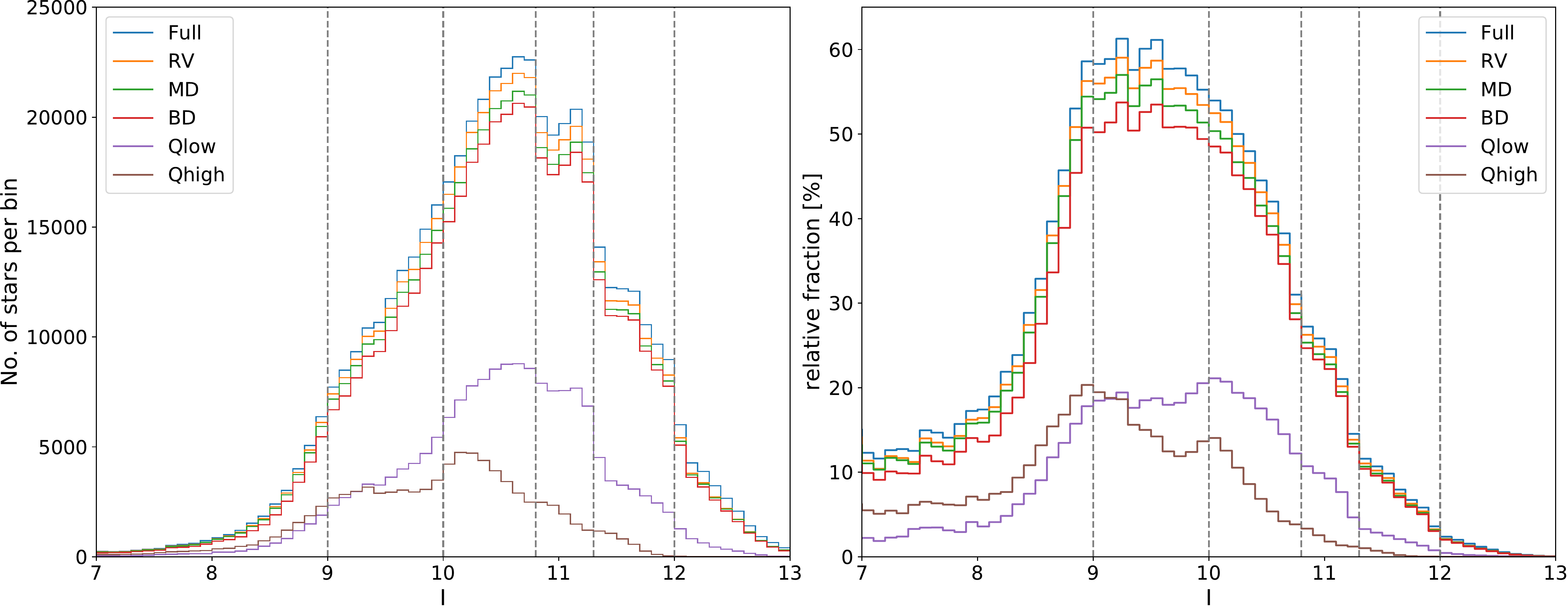}
\caption{\label{fig:completeness}Left: histogram of number of objects {per $I$ magnitude bins of $0.1$} for the various \rave\ DR6 samples as defined at the beginning 
of Section \ref{sec:Validation}.  The magnitude {range} used per field plate exposure are indicated with dashed lines (see DR6-1, Section 2.2). 
Right: completeness fraction of the respective sample relative to the number of 2MASS stars, as a function of $I$ magnitude. }
\end{center}
\end{figure*}

Figure \ref{fig:completeness} shows the number of objects {in $I$ magnitude bins of $0.1$} (left) and the fraction of 2MASS targets in the 
respective magnitude bin (right) that have an corresponding \rave\ measurement, for each of these samples. In the bright 
magnitude bin of \rave\ ($9<I<10)$, about 55\% of the 2MASS targets have a reliable RV measurement in \rave, about 50\% have 
reliable {stellar atmospheric parameters}, and about 20\% (15\%) have an \alphafe\ estimate in the \qlow\ (\qhigh) sample.

Where additional \SNR\ constraints are added (\eg, to show the Kiel diagram for different \SNR\ cuts in the next subsection), the 
lower limit of the \SNR\ is added to the sample name. For example, MD40 is the subset of the RV data base that fulfills the basic 
quality criterion for stellar parameter determination with the \texttt{MADERA} pipeline and for which the individual spectra have 
a \SNR\ {(\texttt{snr\_med\_sparv}, defined as the inverse of the median of the error spectrum -- see Section 3.2 of DR6-1)} of at least 40. 

The number of spectra and unique objects for the aforementioned samples are given in Table \ref{tab:ver_samples} and Table 
\ref{tab:ver_samples_unique}, respectively. Dwarfs and giants are divided based on their \texttt{BDASP} \logg\ values, \ie, 
$\logg_{\rm BDASP} \leq 3.5$ for giants and $\logg_{\rm BDASP} > 3.5$ for dwarfs.

\begin{deluxetable*}{l|rrrrrrr}
\tablecaption{\rave\ subsamples (spectra) used in this publication for validation and first science applications. \label{tab:ver_samples}}
\tablehead{\colhead{Sample} & \colhead{RV} & \colhead{MADERA} & \colhead{BDASP} & \colhead{\alphafe} & \colhead{Fe} & \colhead{Al} & \colhead{Ni}}
\startdata
        Full &  518,387  &  517,821  &  494,695  &  430,142  &  328,317  &  315,036  &  66,778\\
        RV & 497,828  &  497,708  &  477,827  &  425,948  &  324,856  &  312,645  &  65,651\\
        MD &  480,254  &  480,254  &  460,749  &  410,873  &  313,605  &  302,423  &  64,136\\
        BD &  460,749  &  460,749  &  460,749  &  401,927  &  307,301  &  296,096  &  63,389\\
        -- dwarfs & &  &  199,047  &  169,792  &  97,737  &  106,510  &  2,746\\
        -- giants & &  &  261,702  &  232,135  &  209,564  &  189,586  &  60,643\\
        \qlow & 166,867  &  166,867  &  162,646  &  166,867  &  122,663  &  127,864  &  15,291\\
        -- dwarfs & & &  &  92,446  &  57,530  &  65,700  &  1,286\\
        -- giants & & &  &  70,200  &  62,080  &  58,882  &  13,834\\
        \qhigh & 121,812  &  121,812  &  118,737  &  121,812  &  106,110  &  106,146  &  24,042\\
        -- dwarfs & & &  &  59,725  &  46,015  &  48,908  &  1,451\\
        -- giants & & & &  59,012  &  57,459  &  54,575  &  22,314\\
\enddata
\end{deluxetable*}

\begin{deluxetable*}{l|rrrrrrr}
\tablecaption{\rave\ subsamples (unique objects) used in this publication for validation and first science applications. \label{tab:ver_samples_unique}}
\tablehead{\colhead{Sample} & \colhead{RV} & \colhead{MADERA} & \colhead{BDASP} & \colhead{\alphafe} & \colhead{Fe} & \colhead{Al} & \colhead{Ni}}
\startdata
        Full & 451,783  &  451,358  &  431,060  &  380,319  &  292,196  &  281,379  &  61,824\\
        RV &  436,340  &  436,249  &  418,485  &  376,912  &  289,203  &  279,337  &  60,742\\
        MD &  423,021  &  423,021  &  405,524  &  365,117  &  280,205  &  271,112  &  59,371\\
        BD &  405,524  &  405,524  &  405,524  &  357,161  &  274,565  &  265,432  &  58,686\\
        -- dwarfs & &  & 173,514  &  150,487  &  87,205  &  95,476  &  2,572 \\
        -- giants & &  &  232,147  &  206,788  &  187,429  &  170,010  &  56,118\\
        \qlow &  153,634  &  153,634  &  149,781  &  153,634  &  113,188  &  118,223  &  14,596\\
        -- dwarfs & & &  &  84,502  &  52,563  &  60,266  &  1,196\\
        -- giants & & & &   65,311  &  57,857  &  54,973  &  13,239\\
        \qhigh &  110,768  &  110,768  &  107,995  &  110,768  &  96,558  &  96,805  &  22,286\\
        -- dwarfs & & & &  53,905  &  41,510  &  44,338  &  1,327\\
        -- giants & & & &  54,112  &  52,683  &  50,071  &  20,707\\
\enddata
\end{deluxetable*}

\subsection{Kiel diagrams of the \rave\ DR6 catalog}\label{subsec:val_kiel}
\begin{figure*}
\begin{center}
\plotone{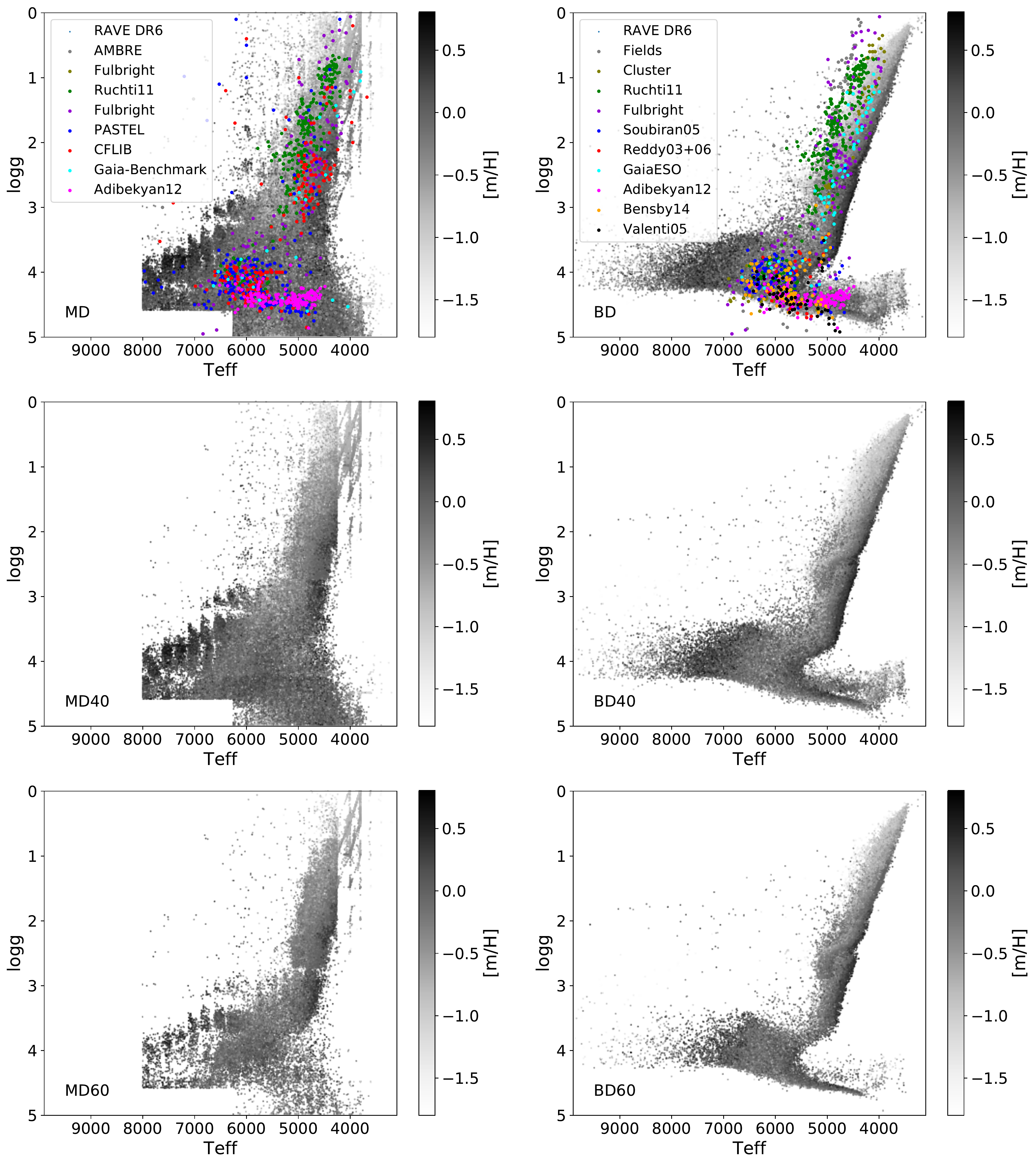}
\caption{\label{fig:kiel_compare}
Kiel diagrams for \texttt{MADERA} (left) and \texttt{BDASP} (right) for various SNR levels. In 
the top row, the calibration sample is overplotted, color coded by the origin as given by the key in the plot.}
\end{center}
\end{figure*}

Figure \ref{fig:kiel_compare} shows the \rave\ sample defined by the different quality cuts in 
the \teff\ vs \logg\ plane (the ``Kiel diagram''), where the blue scale  is coded by the metallicity \mh. 
The quality cuts applied are (from top to bottom) for the left column (i) the MD sample,  (ii) the MD40 sample, and 
(iii) MD60 sample, while for the right column, (i) the BD sample,  (ii) the BD40 sample, and 
(iii) BD60 sample is shown, respectively. In the top {left frame (MADERA)}, the {stellar atmospheric parameters} of the \emph{calibration} sample are also plotted, 
color coded by their origin (see Appendix \ref{sec:external}). {In the top right frame (BDASP) we also show the validation sample used below} {for verifying the output of the GAUGUIN pipeline}.As one can see, calibration {and validation} samples nicely 
cover the most relevant areas of the Kiel diagram, namely the main sequence, turn-off 
stars and subgiants, and the giant branch and red clump region. For the 
\cite{ruchti2011} sample, which was designed as a follow-up study of low metallicity 
candidates drawn from earlier \rave\ data releases, the shift towards higher 
temperature when compared to the higher-metallicity Gaia-ESO {DR5}\footnote{Available on http://casu.ast.cam.ac.uk/casuadc/} 
sample is clearly visible, a feature that is 
nicely reproduced for the \texttt{MADERA} and \texttt{BDASP} pipelines. Furthermore, 
the pixelization of 
the \texttt{MADERA} pipeline (for a discussion see DR4, Section 6.3) is clearly visible.

The results for the \texttt{BDASP} pipeline show a considerable sharpening of the distribution, 
with the main sequence and the position of the red clump being clearly defined. In 
particular, the region for $\logg > 3.5$ strongly benefits from the inclusion of Gaia 
parallaxes, as these dwarf stars are predominantly at lower distances ($d\lesssim 0.8$ 
kpc) and thus benefit from the accuracy of the parallax measurement.

\begin{figure}
\begin{center}
\plotone{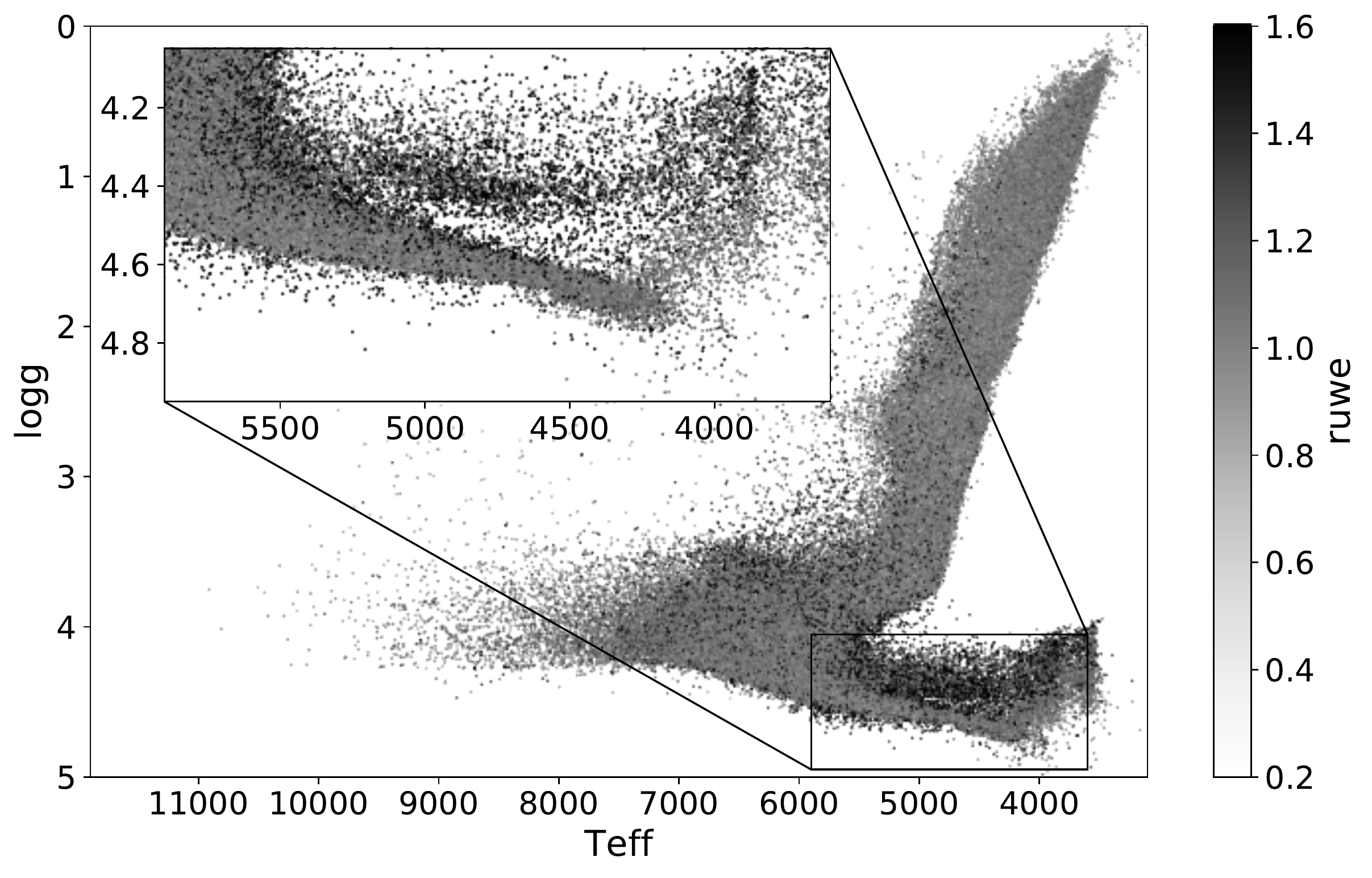}
\caption{\label{fig:kiel_goodness}
Kiel diagram using the \texttt{BDASP} {stellar atmospheric parameters} for the BD00 sample, color encoded by the re-normalized 
unit weight error (RUWE) of the Gaia DR2 astrometric solution.}
\end{center}
\end{figure}

For higher SNR cuts, a clearly defined sequence nearly parallel to the main sequence 
but shifted by about {+0.2}\,\dex\ in $\log g$ becomes visible. This parallel sequence is mainly a 
product of unresolved binaries. These binaries form a second track in the color 
magnitude diagram, about 0.7 mag brighter than main sequence stars of the same color. 
\texttt{BDASP} assumes every target to be a single star, and therefore finds a poor match between 
stars on this track and main sequence stars, instead finding close matches with 
pre-main-sequence stars which have lower $\log g$ at the same $\teff$. This explanation 
can be given more credence by the following observation: In Figure 
\ref{fig:kiel_goodness} we color code the Kiel diagram by Gaia's re-normalized unit 
weight error \citep[RUWE, ][]{RUWE},  which is an indication of the quality of  the 
Gaia DR2 astrometric fit. We see that the RUWE is noticeably higher in the parallel 
sequence, which is consistent with the astrometry being perturbed by the binary motion 
of the stars.

\begin{figure*}
\begin{center}
\plotone{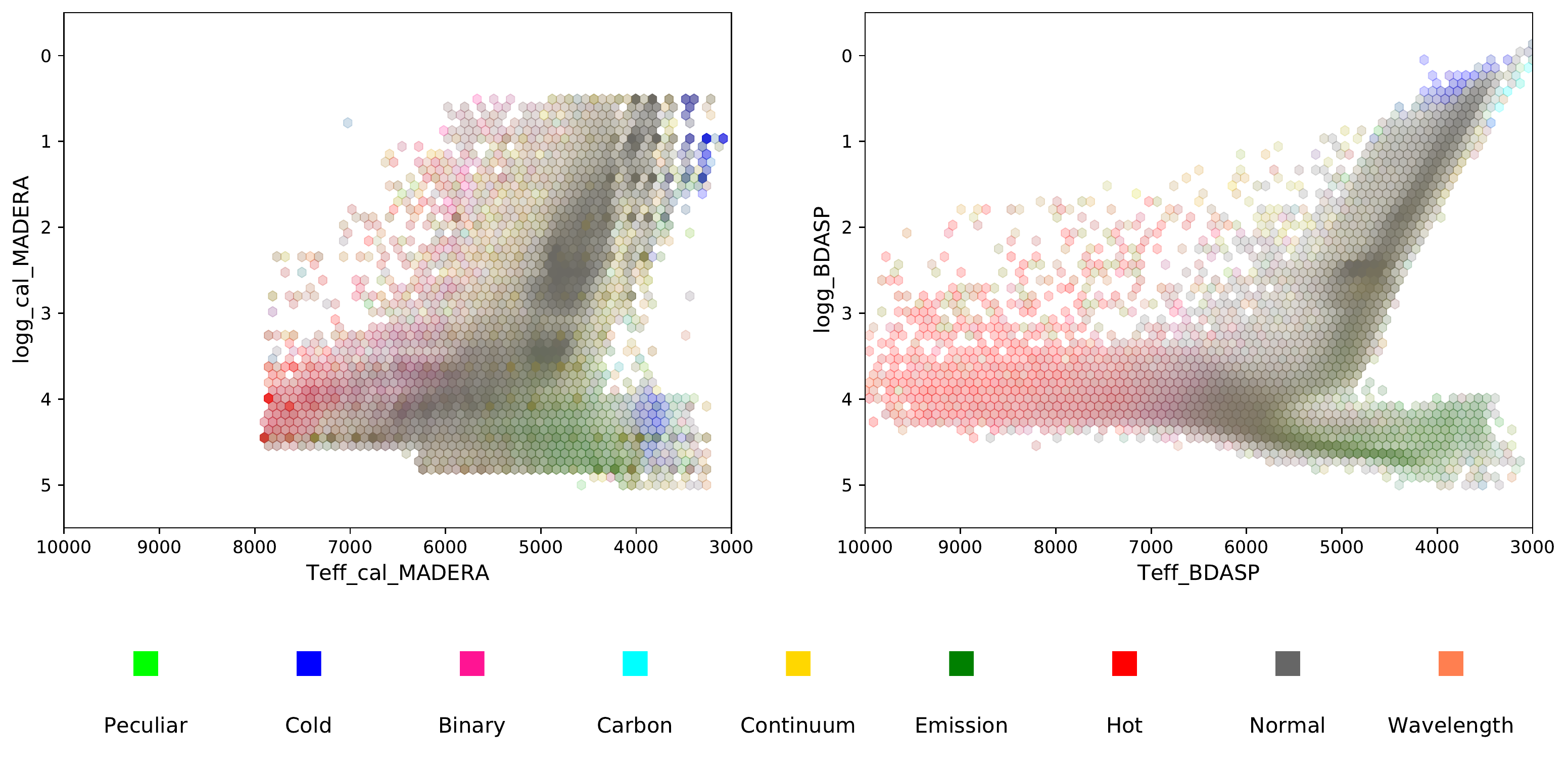}
\caption{\label{fig:kiel_class}
Kiel diagrams for MD sample (left) and the BD sample (right) color coded by the automated classification of the stars.}
\end{center}
\end{figure*}

We can both illustrate and verify the automated classification scheme (see Section 
4 in DR6-1) by showing where stars of different classifications lie in the 
\logg\ vs \teff\ plane (Figure \ref{fig:kiel_class}): The classification scheme nicely 
shows the transition to hot stars above a temperature of $\teff\approx 7000\,$K owing 
to the presence of strong Paschen line features, which dominate over the Ca triplet 
feature. On the main sequence, at effective temperatures below $5000\,$K, chromospheric emission lines become 
more prevalent in these cool and active stars \citep{zerjal2013}. 
At temperatures below $4000\,$K, molecular lines lead to a classification of the star 
as cool or as having carbon features, in particular near the tip of the giant 
branch. A slightly pinkish color in the sequence parallel to the main sequence for 
temperatures above 4500\,K also indicates a binary origin of stars in this part of 
the \logg\ vs \teff\ plane, for temperatures below 4500\,K, the emission line 
characteristics dominates the classification also in this part of the parallel 
sequence.

\subsection{Validation against external observations}\label{subsec:ext_val}

\subsubsection{Validation of stellar atmospheric parameters}\label{subsubsec:Stellar_val}

For an extensive validation of the pure spectroscopic \texttt{MADERA} {stellar atmospheric parameters} and their 
limitations we refer to the DR4 and DR5 publications.

\begin{figure}
\begin{center}
\plotone{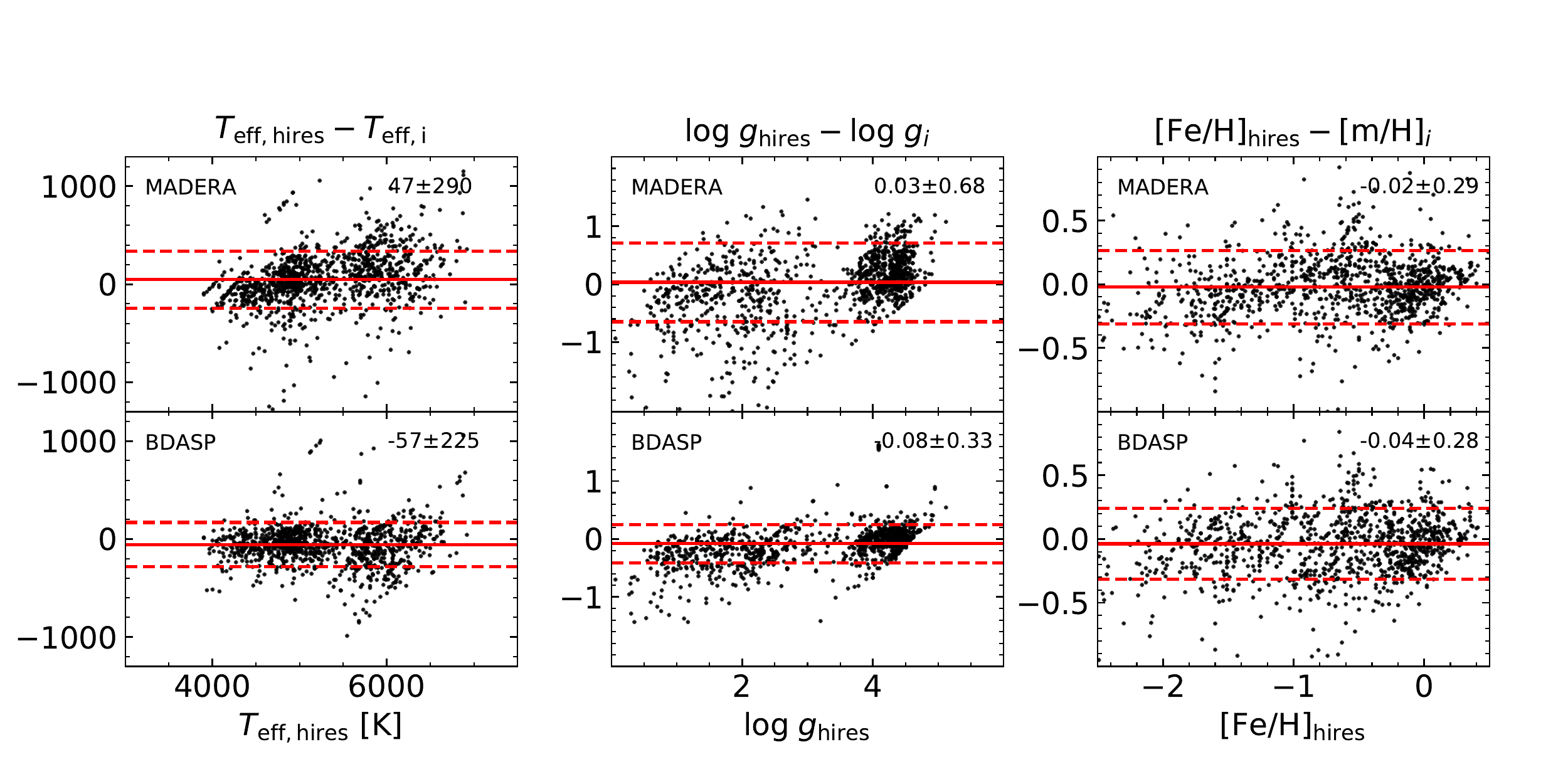}
\caption{\label{fig:StellarParmsExternal}Comparison of the residual (defined as high-resolution observation minus \rave\ DR6) of the 
stellar parameter pipelines \texttt{MADERA} (top row) and \texttt{BDASP} (bottom row) against the 
value of the given stellar parameter derived from the high-resolution data. Left column: \teff, middle column: \logg, right column: 
\mh\ {(for MADERA and BDASP) and \feh\ (for validation data), respectively}. The solid red line indicates the average residual, the dashed red lines show  the $\pm 
1\sigma$ deviation and the corresponding values are shown in the upper right corner of each 
frame.}
\end{center}
\end{figure}

\begin{figure}
\begin{center}
\plotone{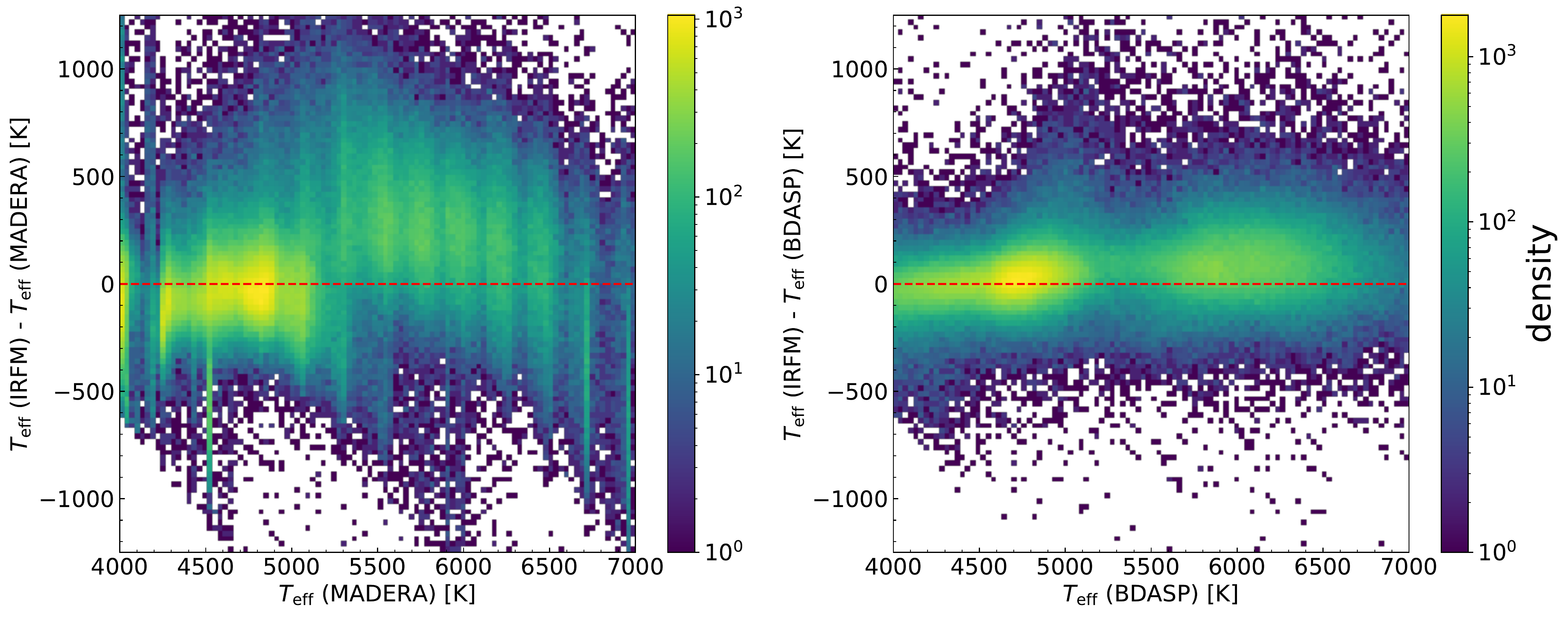}
\caption{\label{fig:irfm_compare}
Comparison between the effective temperatures derived from the \texttt{IRFM} against
those obtained with \texttt{MADERA} (left panel) and \texttt{BDASP} (right panel), for the MD20 
sample. For \texttt{MADERA}, giants with $\teff < 5200$\,K have temperatures that agree well 
with \texttt{IRFM} temperatures, but there is a systematic offset for  main-sequence/turnoff
stars. The pixelization, an artifact of the \rave\  stellar parameter pipeline \texttt{MADERA}, 
is apparent as vertical bands. The \texttt{BDASP} pipeline, which uses $T_\mathrm{eff,IRFM}$ as 
input, shows no significant systematic offsets.}
\end{center}
\end{figure}

Figure \ref{fig:StellarParmsExternal} compares \teff, \logg, and \mh\ derived from each of the \texttt{MADERA} and \texttt{BDASP} pipelines for the MD20 and BD20 sample 
with the values derived from 1094 external high-resolution observations 
 (see appendix \ref{sec:external}). {For the BDASP sample, the metallcity \Mh\ has} {been scaled to \mh\ using the inverse of equation \ref{eq:Salaris}}. Note that this comparison is not fully independent, 
as {some of} the external data set has been used to calibrate the outcome of the \texttt{MADERA} pipeline 
(see Section \ref{subsec:MADERA}, Appendix \ref{sec:external}). The effective temperatures of both methods give 
similar results in terms of uncertainties. However, the \texttt{MADERA} pipeline is more 
affected if low \SNR\ \rave\ targets are included in the comparison (for the 
MD00 sample, the standard deviation for \texttt{MADERA} increases to 320K, while the 
value for \texttt{BDASP} remains basically unchanged). A closer inspection of the \texttt{MADERA} plot 
also reveals a tendency to somewhat overestimate the temperatures between $5000$ and 
$6000$\,K by $\approx250$\,K. This becomes more visible when the \texttt{MADERA} \teff\ are compared with 
the temperatures derived via the infrared flux temperatures (see Figure 
\ref{fig:irfm_compare}). As expected, no such trend is visible in the effective 
temperature of the \texttt{BDASP} pipeline, which has used $T_\mathrm{eff, IRFM}$ as an input value. 

The surface gravities \logg\ demonstrate the full potential of the parallax constraint 
from Gaia DR2. The derivation of \logg\ has always be a major challenge for \rave\  
because of the short wavelength interval and well known degeneracies 
\citep{kordopatis2011a}. The \texttt{MADERA} 
pipeline results for \logg\ are on average unbiased, but exhibit a scatter of about {0.68} 
\dex, while the \texttt{BDASP} pipeline can considerably reduce the uncertainty to only 
{0.33}\,\dex\ and produce \logg\ values that are unbiased, compared to asteroseismic 
estimates (see Section \ref{subsec:seismo_val}). Indeed, a comparison of the structure of 
the \logg\ vs \teff\ diagram with external data from the GALAH 
and APOGEE surveys (Section \ref{subsec:Galah_val}), plus the analysis of the repeat 
observations in Section \ref{subsec:repeat_val} below, and the comparison with the 
asteroseismic information (Section \ref{subsec:seismo_val}) all lead to the conclusion 
that as far as \logg\ is concerned, much of the variation between the values obtained 
with \texttt{BDASP} and the external measurements may well need to be attributed to uncertainties 
in the external calibration sample.

In terms of the metallicity \mh, both pipelines perform equally well, and indeed we 
recommend using the \texttt{MADERA} \mh\ as the metallicity estimate, as it is the only one 
that is directly derived spectroscopically.

\subsubsection{Validation of \texttt{GAUGUIN} abundances}\label{subsubsec:GAUGUIN_val}

\paragraph{$\alphafe$ ratios}

\begin{figure*}
\begin{center}
\plotone{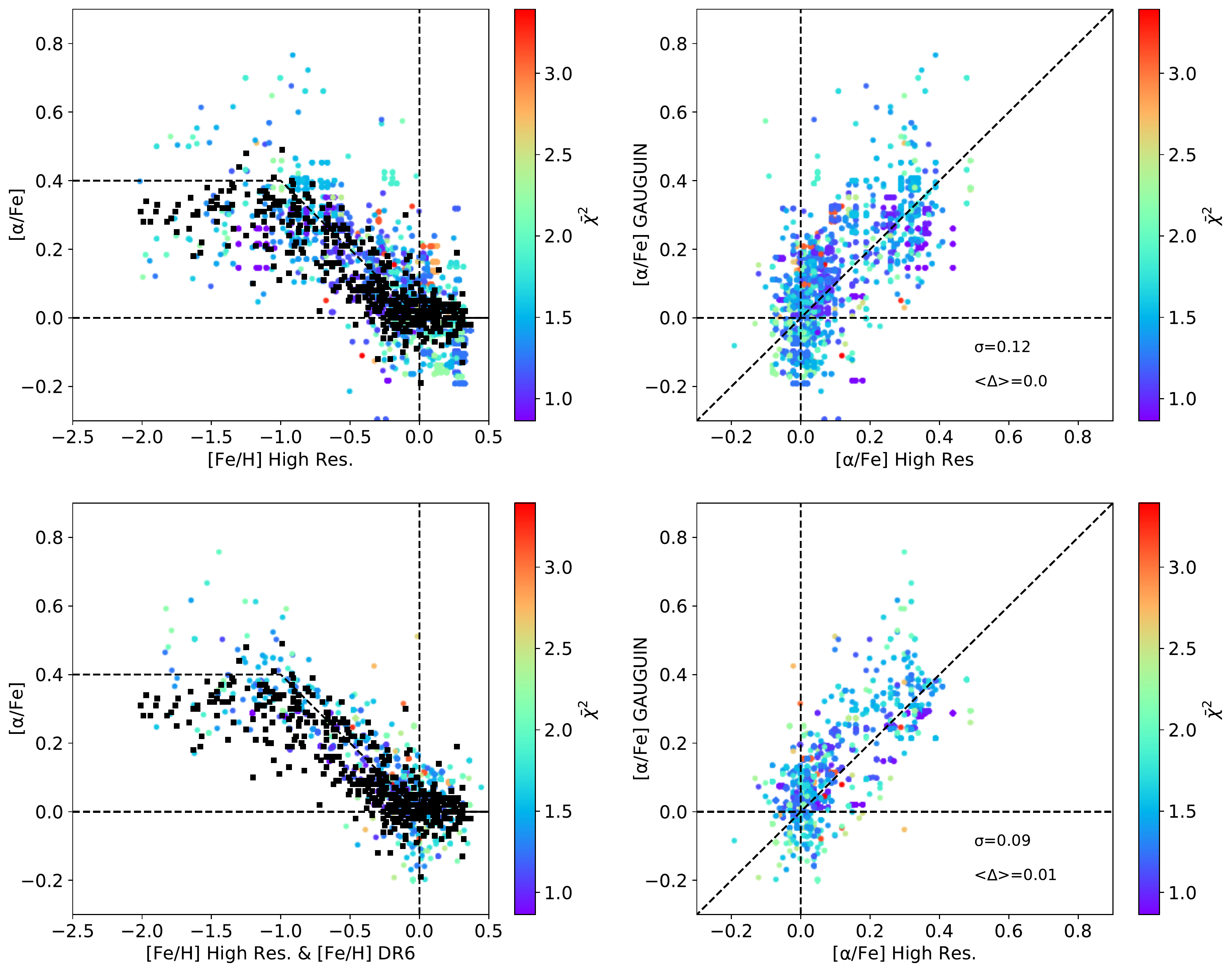}
\caption{\label{fig:gauguin_vs_reference_sample_2}Top row, left panel: 
\texttt{GAUGUIN} $\alphafe$ as a function of [Fe/H] (dots, color-codded with \bchisq) 
computed using the   {stellar atmospheric parameters of the calibration 
sample}. The black squares 
correspond to the calibration sample. The right panel plots the \alphafe\ ratios 
obtained with \texttt{GAUGUIN} against those of the calibration sample. Bottom row: same, 
but using \texttt{MADERA} inputs for computing \texttt{GAUGUIN} $\alphafe$ ratios. $\Delta$ 
denotes the difference between the $\alphafe$ ratios derived via the \texttt{GAUGUIN} 
pipeline and that derived from high-resolution observations of the reference star. 
The mean difference and its standard deviation are shown in the bottom right corner 
of the right column.}
\end{center}
\end{figure*}

Figure \ref{fig:gauguin_vs_reference_sample_2} (top row)  compares the  $\alphafe$ 
ratios obtained with \texttt{GAUGUIN} 
using the {stellar atmospheric parameters} \teff, \logg, and \feh\ 
of the calibration sample (see Section \ref{subsubsec:Stellar_val} and DR5 Section 7)
against the $\alphafe$ ratios of the calibration sample (defined as the average of 
[Si/Fe] and [Mg/Fe]). The abundances trace the pattern observed in the external 
reference stars very well, with a scatter of about 0.12 dex, and almost no bias. 
The bottom row of Figure \ref{fig:gauguin_vs_reference_sample_2} shows the analogous 
comparison when \texttt{MADERA} \teff, \logg, and \mh\ are used as input 
parameters. The abundances trace the pattern observed in the external reference 
stars still well, with a scatter somewhat increasing for lower-metallicity stars. 
Outliers can be directly mapped to large differences in \teff, \logg, and \mh\  
between the calibration sample and the corresponding \texttt{MADERA} values. 
This is consistent with the poor \bchisq\ values for those outliers.

\paragraph{The [Fe/H], [Al/H], and [Ni/H] ratios}

\begin{figure*}
\begin{center}
\plotone{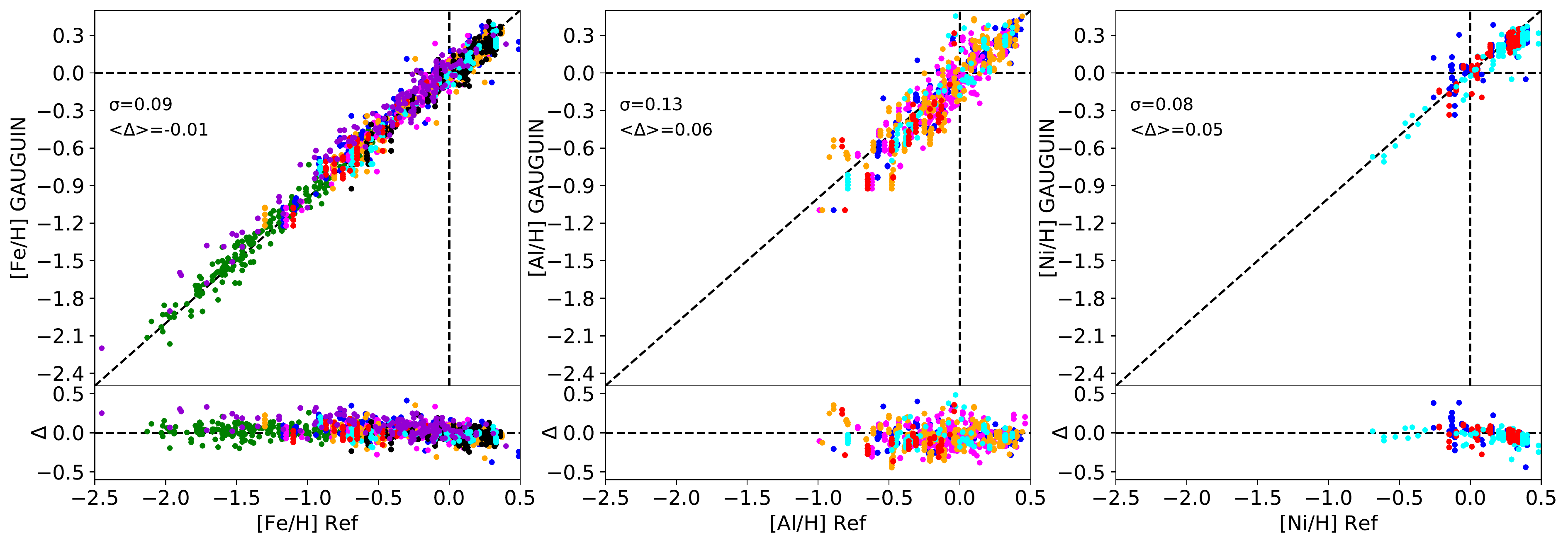}
\plotone{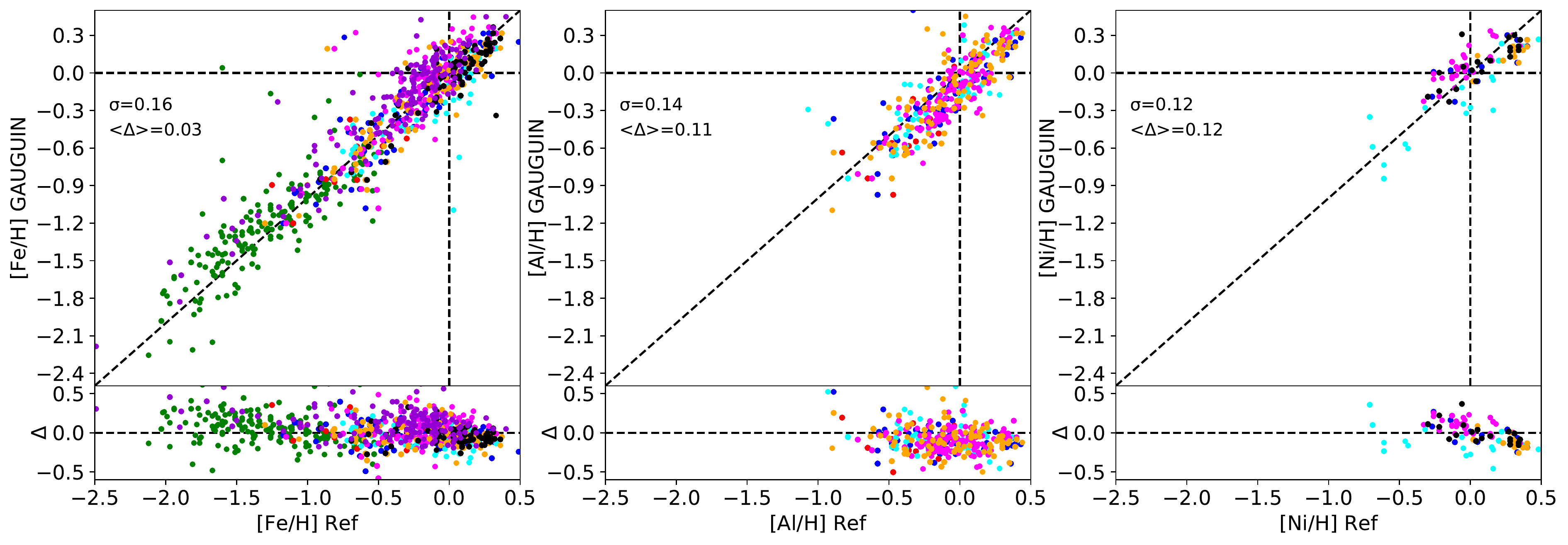}
\caption{\label{fig:gauguin_vs_reference_sample_1}Top: \texttt{GAUGUIN} chemical abundances 
of [Fe/H], [Ni/H], and [Al/H] computed using calibration sample 
{stellar atmospheric parameters}, as a function of the chemical abundances of the calibration sample. 
We have adopted the same color code for the external sample as was used in 
Figure \ref{fig:kiel_compare}. $\Delta$ denotes the difference between the  abundance 
of a given element derived via the \texttt{GAUGUIN} pipeline and that derived from 
high-resolution observations of the reference star. The mean difference and its 
standard deviation are shown in the upper left corner of each frame. 
Bottom: same, but using \texttt{MADERA} {stellar atmospheric parameters} as input for \texttt{GAUGUIN}.}
\end{center}
\end{figure*}

The top row of Figure \ref{fig:gauguin_vs_reference_sample_1} compares 
[Fe/H], [Al/H], and [Ni/H] obtained with the \texttt{GAUGUIN} pipeline against those of the  calibration sample. 
\texttt{GAUGUIN} was fed with the {stellar atmospheric parameters} 
of the calibration sample.  For [Fe/H], the bias seems to 
slightly increase towards the [Fe/H]-poor regime, with a dispersion of 0.09 dex. 
For [Al/H] and [Ni/H] ratios, we notice a weak scatter as well, 0.13 and 0.08, respectively. For [Ni/H] we 
unfortunately have only very few stars. 
The biases and scatters observed here can be due to several factors, such as the
different linelists and spectral resolution of RAVE and the reference studies.

In the bottom panel of Figure \ref{fig:gauguin_vs_reference_sample_1}, we compare 
the \texttt{GAUGUIN} [Fe/H], [Al/H], and [Ni/H] values derived with \texttt{MADERA} input
to those of the calibration sample (same stars as in the top row). 
For [Fe/H] and [Al/H], we notice an increase of the scatter for the Fe-poor regime.  Basically, the comparison gives 
fairly satisfactory results, with an increased dispersion, driven by different input {stellar atmospheric parameters}.

\paragraph{Abundance trends in the Kiel diagram}

\begin{figure*}
\begin{center}
\includegraphics[width=1.1\textwidth,angle=90]{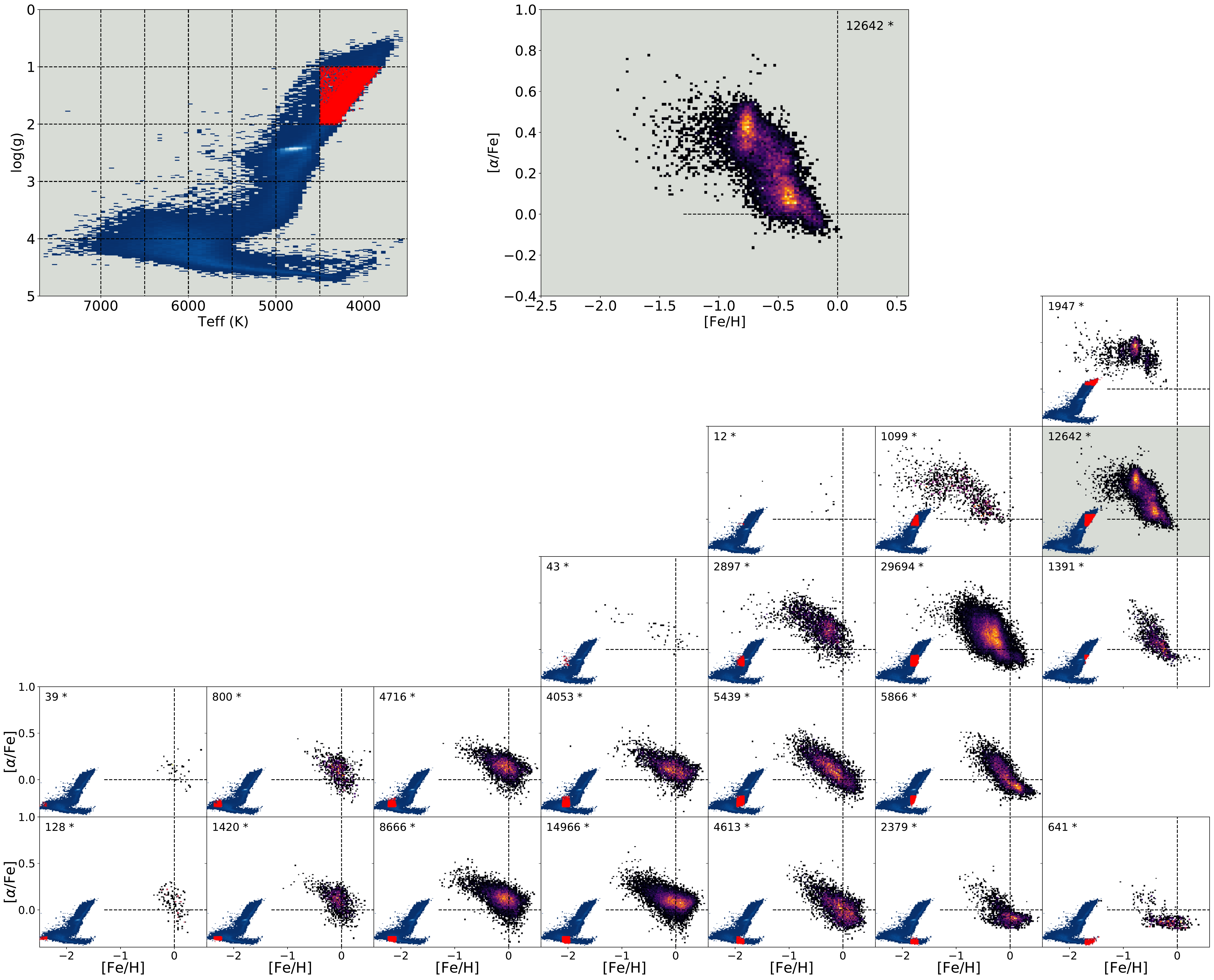}
\caption{\label{fig:Fig19_Gauguin_alpha_vs_FeH}$\alphafe$ as a function 
of [Fe/H] for the \qhigh\ sample, binned in $\teff$ ($\Delta\teff = 500\,$K) and $\logg$ ($\Delta\logg = 1\,$\dex). 
In each panel, we added a Kiel diagram 
in order to help the reader locating the subsample in the $\teff-\logg$ plane. For illustration purposes, the two 
inlays in the upper left corner magnify the subpanel for $\teff < 4500\,$K and $1<\logg<2$. 
In total, \alphafe\ and \feh\ abundances for $103\,474$ stars are shown.}
\end{center}
\end{figure*}

\begin{figure*}
\begin{center}
\includegraphics[width=1.1\textwidth,angle=90]{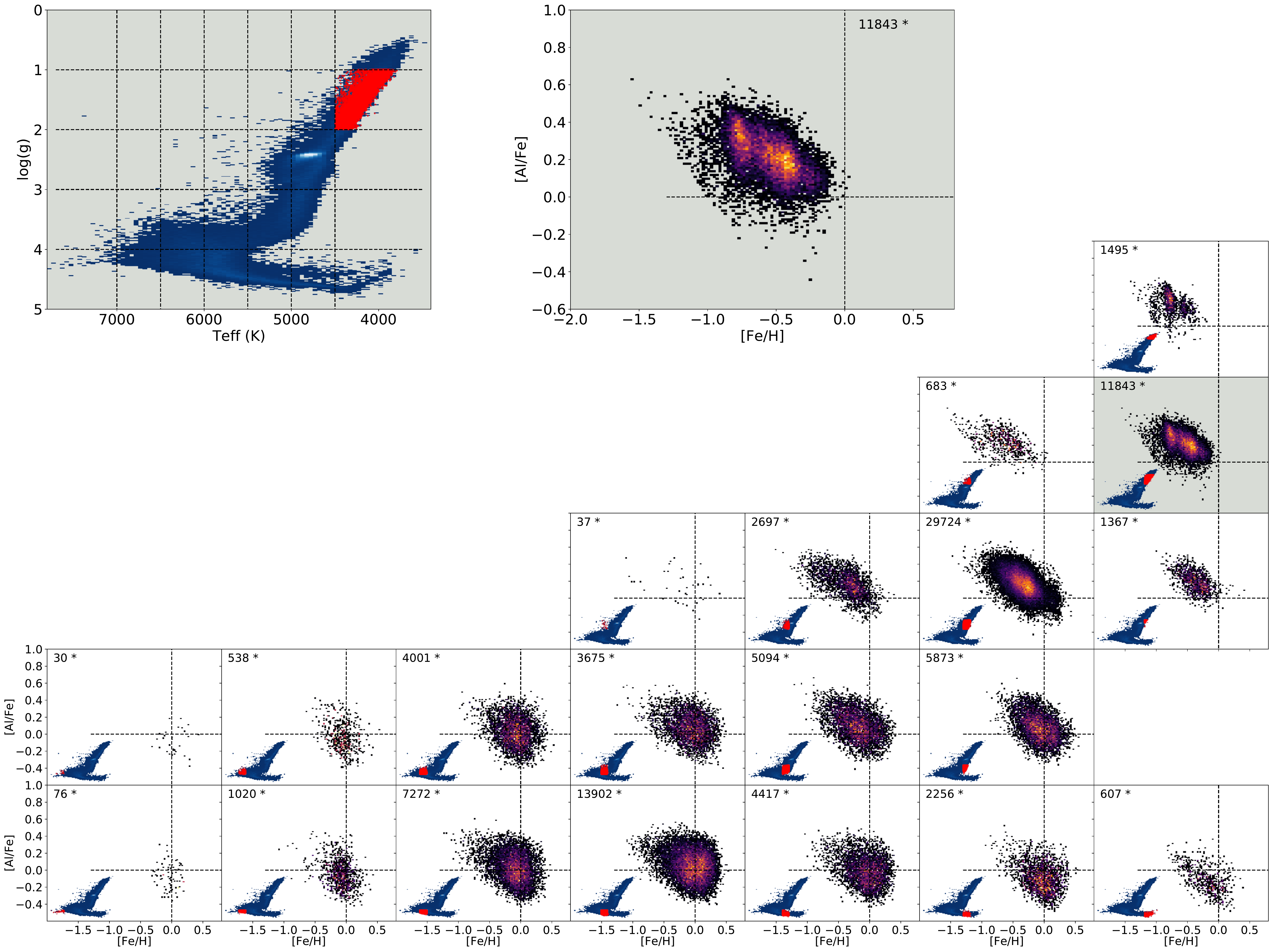}
\caption{\label{fig:Fig20_Gauguin_Al_vs_FeH} Same as Figure \ref{fig:Fig19_Gauguin_alpha_vs_FeH}, but showing [Al/Fe] as a function 
of [Fe/H] for the \qhigh sample.  In total, [Al/Fe] and \feh\ abundances for $96\,607$ stars 
are shown.}
\end{center}
\end{figure*}

\begin{figure*}
\begin{center}
\includegraphics[width=1.1\textwidth,angle=90]{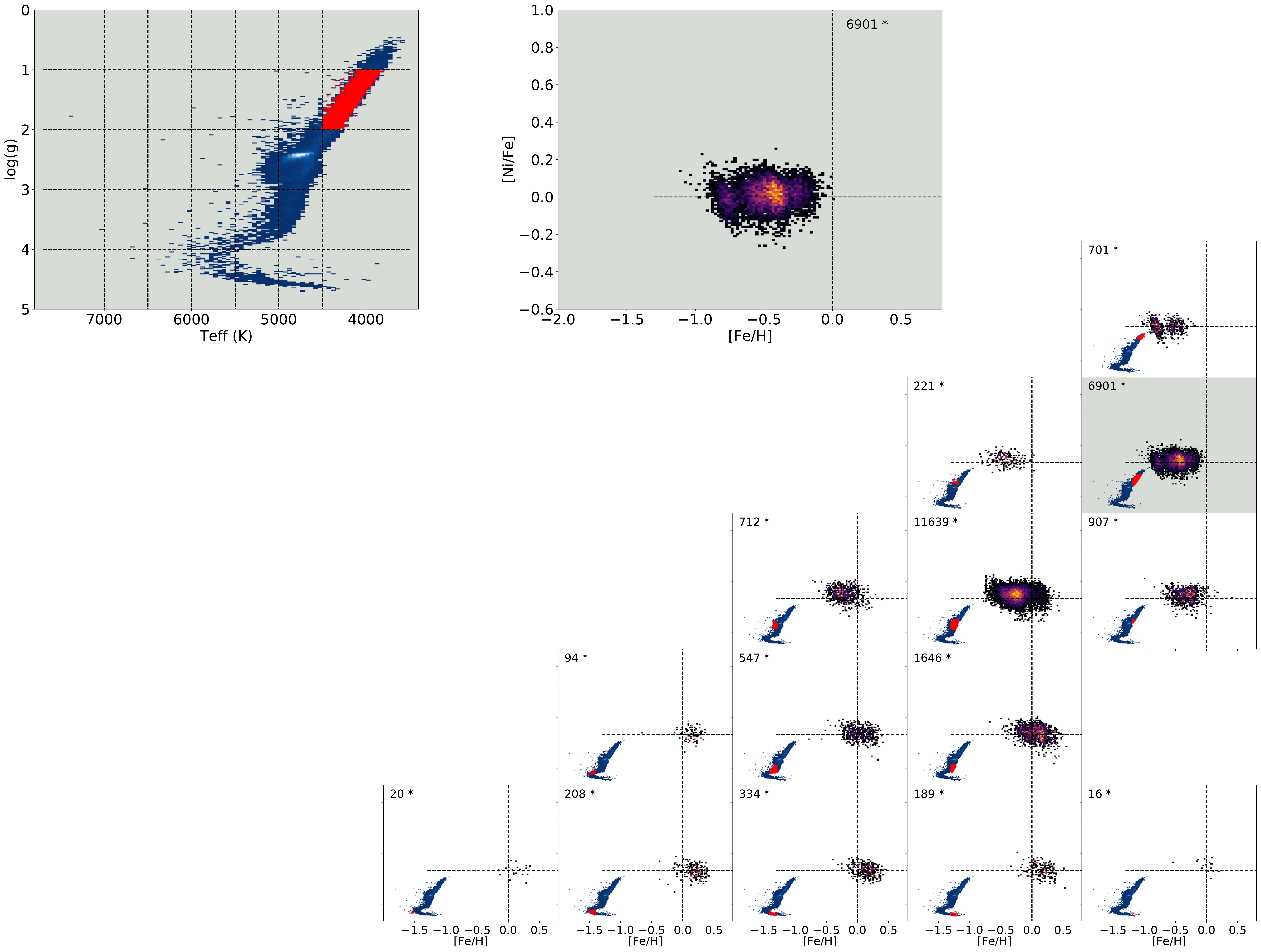}
\caption{\label{fig:Fig21_Gauguin_Ni_vs_FeH}Same as Figure \ref{fig:Fig19_Gauguin_alpha_vs_FeH}, but showing [Al/Fe] as a function 
of [Fe/H] for the \qhigh sample.  In total, [Ni/Fe] and \feh\ abundances for $24\,135$ stars are shown.}
\end{center}
\end{figure*}

Abundance trends for the $\alphafe$, $[\text{Al/Fe}]$ and $[\text{Ni/Fe}]$ can nicely be followed by grouping 
them by \logg\ and \teff\ in the Kiel diagram. This is done in Figure \ref{fig:Fig19_Gauguin_alpha_vs_FeH} 
for $\alphafe$ and for $[\text{Al/Fe}]$ and $[\text{Ni/Fe}]$ in Figures \ref{fig:Fig20_Gauguin_Al_vs_FeH} and 
\ref{fig:Fig21_Gauguin_Ni_vs_FeH}, respectively. In these diagrams we bin stars of the BD40 sample by their 
\texttt{BDASP} \logg\ and \teff\ in bins of size 1 dex and 500 K, respectively. We start with $4<\logg\leq 5$ 
and $4 000 < \teff \leq 4500$ in the lower right panel with \teff increasing towards the left, and \logg\ 
decreasing going upwards. The leftmost plot includes all stars hotter than 7000\,K. Each panel shows the 
$\alphafe$ vs $\feh$ relation, and in addition show an icon of the Kiel diagram in blue, with the respective 
sub sample marked in red. 

The \alphafe\ vs \feh\ relation across the \teff-\logg\ plane is shown in Figure \ref{fig:Fig19_Gauguin_alpha_vs_FeH}. 
The figure nicely demonstrate how the \texttt{GAUGUIN} derived abundances can track the systematically different behavior 
for different Galactic populations. For giants we probe predominantly the low metallicity regime, which shows a 
successively increasing $\alpha$ overabundance with decreasing metallicity and the transition to a plateau at 
$\alphafe \approx 0.4$. Main sequence stars, on the other hand, mainly test the thin disk behavior of the extended solar suburb.

Indeed, while we, as demonstrated in the next section, clearly see several populations of stars in terms of their 
combined chemical and kinematical properties, only in very few areas in the \teff-\logg\ plane do we simultaneously 
see several components, most notably at 
{$\teff<5000$~K} and $\logg <2$, \ie\ for red giants. There is also a trend that 
the slope of the $\alphafe-\feh$-relation is steeper for giants than it is on the main sequence.
When comparing these findings with those of other surveys we have to alert the reader to properly take into account 
the different selection effects. For example APOGEE predominantly focuses on giants and has a large fraction of their 
targets at low Galactic latitudes (thanks to the NIR nature of this survey), while this area is almost completely 
excluded by the survey design of \rave. We also note that bright giants are a much rarer population in the \rave\ sample, 
reflecting the relatively bright magnitude limit of \rave\ compared with other ongoing surveys. \rave\ is basically 
dominated by two populations, main sequence stars and red clump stars, as indicated by the density scale in Figure 
\ref{fig:Fig19_Gauguin_alpha_vs_FeH}.
 
While not an $\alpha$ element, aluminum is also predominantly formed in massive stars {\citep{thielemann1985}} and released to the ISM via type-II 
supernovae. Therefore, similar abundance trends as observed for $\alpha$ elements are expected. Indeed, while the scatter 
is considerably larger - we fit only very few lines in the CaT region, while for \alphafe\ we basically make use of the 
full spectrum - similar trends to those of $\alpha$ elements can be observed (see Figure \ref{fig:Fig20_Gauguin_Al_vs_FeH}), 
in particular a relative-to-solar over-abundance of 
{Al} for metal-poor giant stars, and a systematic trend of decreasing 
aluminum abundance for increasing metallicity.  For the brightest red giants in our sample, as for \alphafe{,} only 
aluminum-enriched very low metallicity stars can be found in the sample, indicative that we trace the halo and metal-weak 
thick disk component. A kinematical analysis of this subset (see Section \ref{sec:science2}) shows that {these stars} are 
indeed on highly eccentric and inclined orbits.

As an iron group element, we would expect (within the accuracy expected by \rave) nickel to basically follow the same 
trends as the iron abundance, \ie, $\nife\approx 0$. This is indeed the case as illustrated by Figure 
\ref{fig:Fig21_Gauguin_Ni_vs_FeH}, which shows systematic changes in the overall abundance as we move from red giants 
via red clump stars to main sequence stars, but the relative abundance between Ni and Fe basically remains constant.

\subsection{Comparison of \rave\  with the APOGEE and GALAH surveys}
\label{subsec:Galah_val}

APOGEE and GALAH are two high-resolution spectroscopic campaigns currently 
underway. APOGEE has  $R=22,500$ resolving power in the NIR, and mainly focuses on giant 
stars in the Galactic disk. Most of the publicly released APOGEE data cover the Northern 
hemisphere, so there is little overlap with \rave\ so far. The joint sample of \rave~and APOGEE DR16  \citep{SDSS-DR16} -- with the \rave\ 
quality constraints defined above and with abundances flagged by the APOGEE consortium as being reliable -- 
amounts to 4859 objects.

GALAH is a high resolution ($R=28,000$) spectroscopic survey at optical wavelengths 
using the HERMES spectrograph and the 2dF fiber positioner facility at the AAO 3.9m 
telescope. The GALAH 2nd data release \citep{GalahDR2}, published shortly before Gaia DR2, provides 
{stellar atmospheric parameters} and abundances for up to 342,682 stars in the Southern Hemisphere. 
The \rave\ and GALAH data sets have 21,534 stars in common. Of these, 13,254 stars have 
a high \SNR\ (\texttt{snr\_med\_sparv}~$\geq 20$) \rave\ spectrum and a satisfactory quality  GALAH spectrum 
(flag\_cannon~$=0$). Only these stars are used in the discussion below, with the further requirement that  
flag\_x\_fe~$=0$ when discussing individual element abundances \Xh.   

\begin{figure*}
\begin{center}
\includegraphics[width=0.75\textwidth]{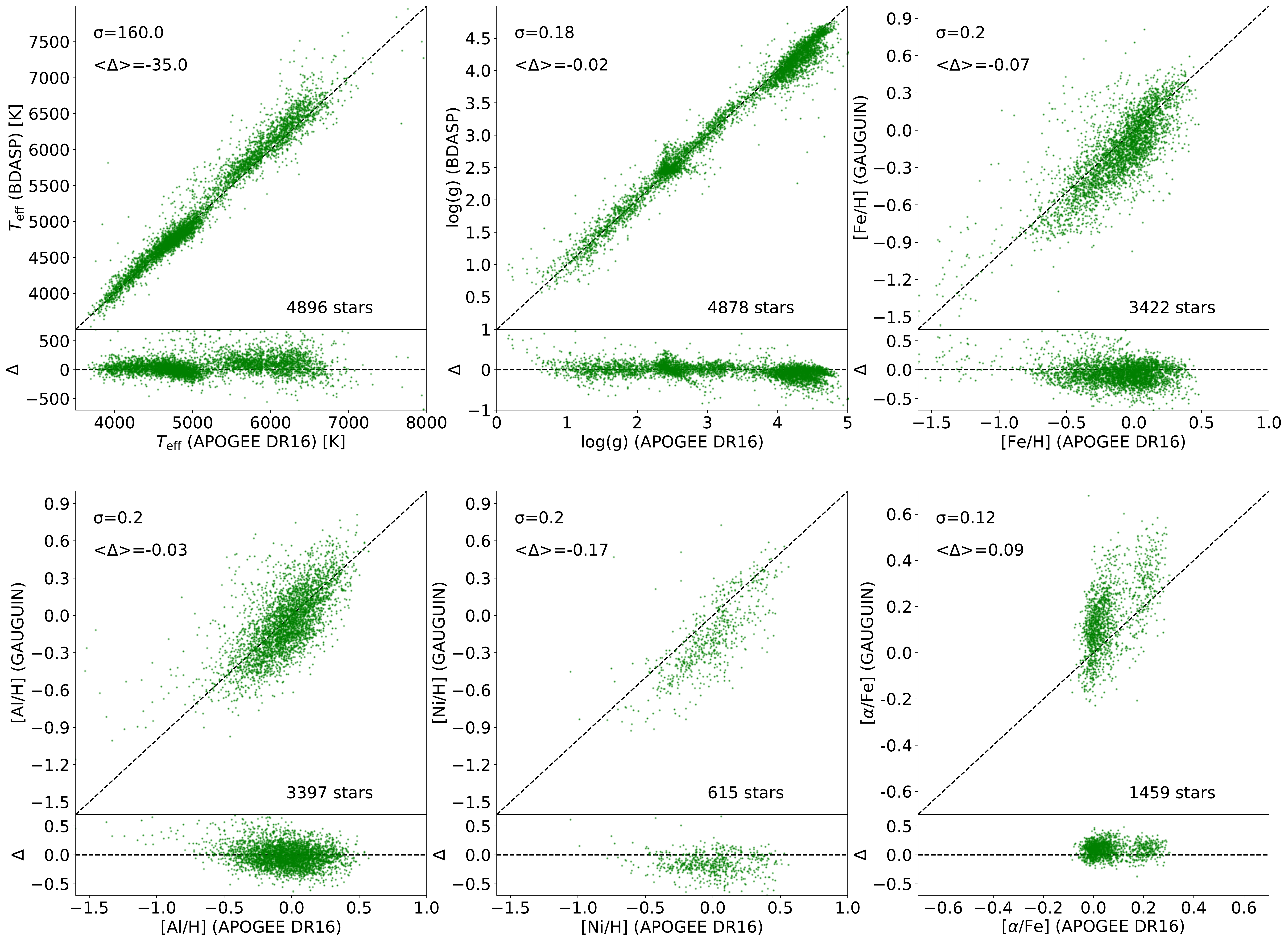}
\includegraphics[width=0.75\textwidth]{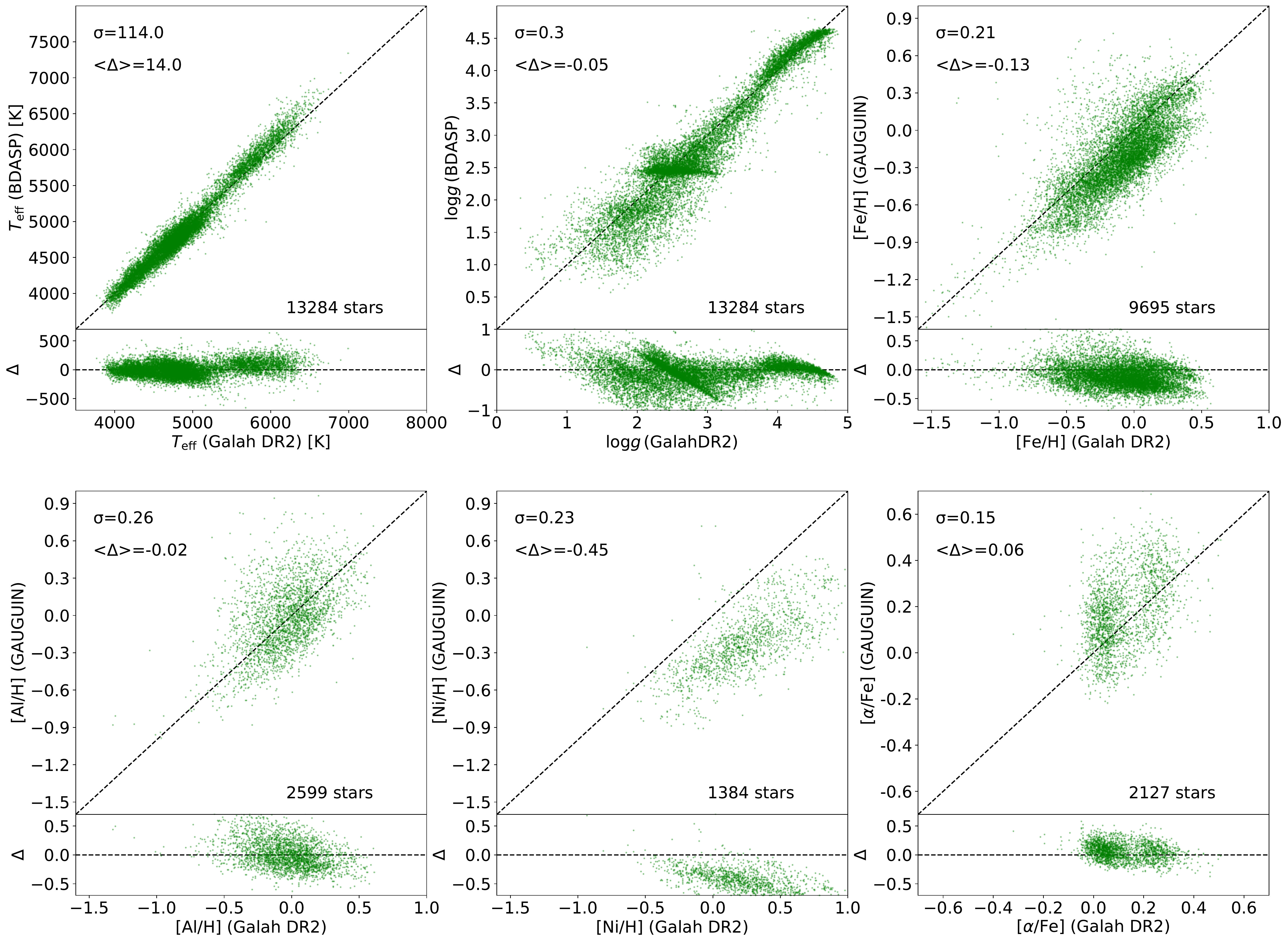}
\caption{\label{fig:DR6_vs_APOGEE}Upper two rows: comparison of the {stellar atmospheric parameters} 
and 
elemental  abundances between \rave\ \texttt{BDASP} and APOGEE DR16, for the stars in common. 
Lower two rows: Same as the upper panels,  but now for the comparisons  
between \rave\ \texttt{BDASP} and GALAH DR2. Element abundances for 
\rave\ were obtained with the \texttt{GAUGUIN} pipeline using \texttt{MADERA} inputs for \teff\ and \logg. 
The number of stars in common is shown in the lower right corner of each panel.}
\end{center}
\end{figure*}

It is illustrative to compare the outcome of the medium resolution \rave\ data 
with the considerably higher resolution APOGEE and GALAH data. When performing these 
comparisons one should, however, be aware that -- unlike \texttt{BDASP} -- neither the APOGEE nor 
the GALAH data pipeline has yet made use of the Gaia DR2 parallaxes. However, both 
surveys employed asteroseismic information from Kepler/K2 for constraining \logg.

Figure \ref{fig:DR6_vs_APOGEE} compares \teff, \logg, and the abundances for Fe, Al, Ni, and $\alpha$ 
between \rave\ (\texttt{BDASP}) and APOGEE DR16 (upper two rows) and GALAH 
DR2 (lower two rows). Note that we plot all stars in common, irrespective of their 
possible classification. This simplification does not affect the general statistics 
of the 
comparison samples appreciably, but could be important when comparing individual 
objects, as discussed in DR6-1, Section 4. 

The comparison shows an excellent agreement in the derived values for \logg\ and \teff. 
Indeed, the advantage of having Gaia DR2 parallax information available for \rave\ 
results in \logg\ estimates that are at least comparable to those derived with higher 
resolution spectroscopy. We compared \rave\ results also with new unpublished GALAH 
values of \logg\ and \feh, which make use of Gaia astrometry. This new information makes the
\logg\ values of both datasets almost identical ($\Delta = 0.03$, $\sigma=0.10$) and 
decreases the iron abundances for GALAH by $\sim 0.1$~dex. For some elements like Fe, 
Al, and 
$\alpha$, the trends between \rave\ and GALAH and between \rave\ and APOGEE appear 
qualitatively to be very similar, exhibiting a slight tilt in the residual in the sense 
that \rave\  tends to underestimate abundances for Solar-type stars. Furthermore, 
unlike the 
\rave/APOGEE sample, the \rave/GALAH sample includes both dwarfs and giants, with each 
component having  slightly different systematics, visible in a slight bimodality 
($\feh>0$ vs $\feh<0$). Note that the expected modest decrease in
\feh\ derived from GALAH when the surface gravity is constrained astrometrically affects 
also the abundances of other elements, as [X$/$H]$=$[X$/$Fe]$+$[Fe$/$H]. As a 
result, the comparison presented in Figure~\ref{fig:DR6_vs_APOGEE} seems rather 
conservative, with a suggestion that \rave\ values could show an even better agreement 
with high resolution surveys when Gaia DR2 results will be used throughout. This is 
encouraging for future applications on medium resolution data, like those that will 
come from WEAVE \citep{WEAVE}, 4MOST \citep{4MOST}, or the Gaia RVS spectrometer. Considering 
the apparent differences between the two high-resolution surveys we feel confident that 
our pipelines have extracted the maximum possible from the \rave\ spectra, given the 
limitation of resolution and \SNR. 

\subsection{Comparison of \rave\  {stellar atmospheric parameters} with asteroseismically calibrated 
parameters}\label{subsec:seismo_val}

\begin{figure}
\begin{center}
\plotone{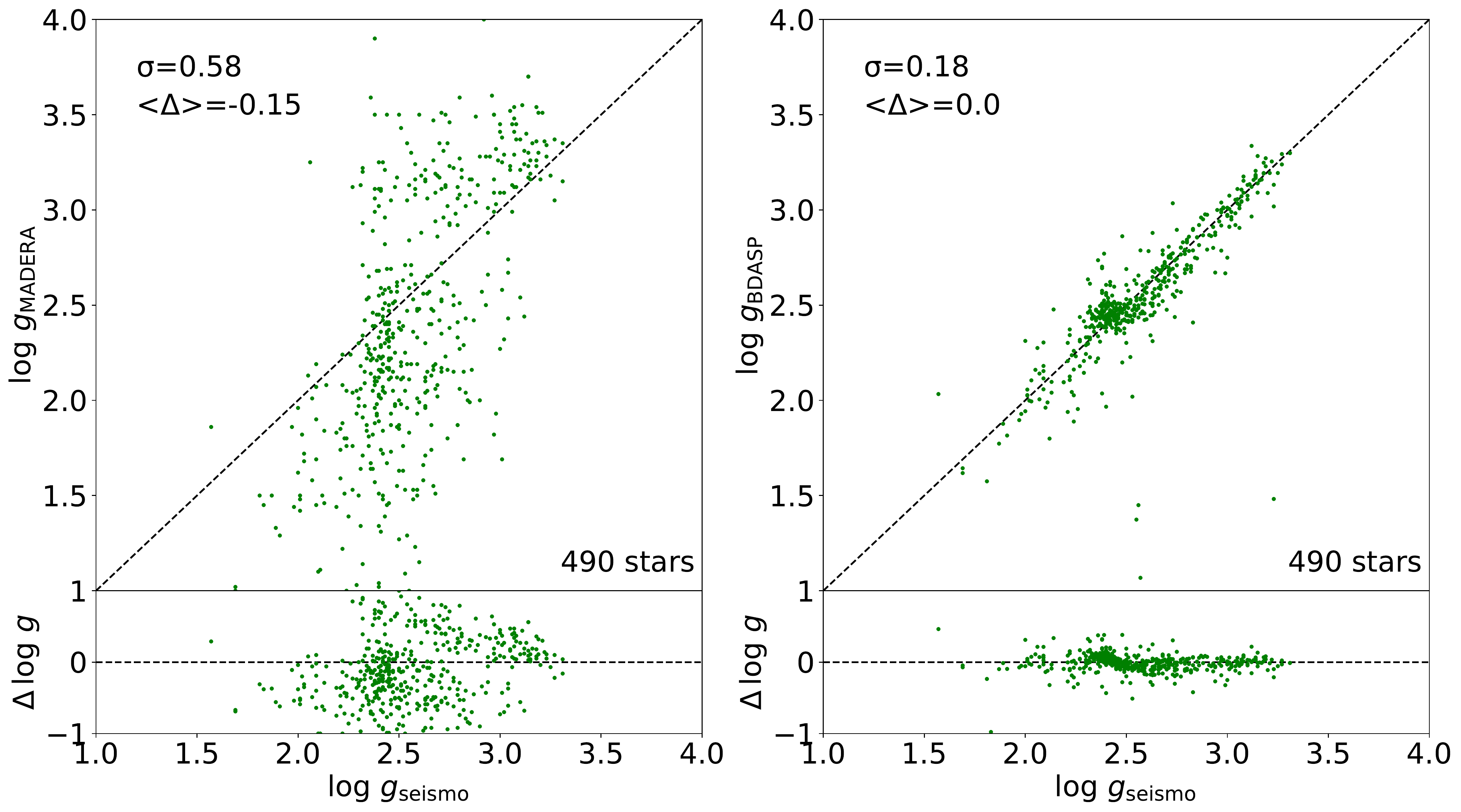}

\caption{\label{fig:seismo_logg}Comparison of the surface gravities \logg\ derived with 
the \texttt{MADERA} (left) and \texttt{BDASP} (right) pipeline vs asteroseismically derived values for 490 stars of the K2 campaign.}
\end{center}
\end{figure}

Figure ~\ref{fig:seismo_logg} compares the asteroseismic \logg\ derived using 
equation~\ref{eq:seismologg} and  \logg\ obtained from the \texttt{MADERA} and \texttt{BDASP} pipelines. 
This comparison shows the impact of Gaia DR2 parallaxes on the derived 
{stellar atmospheric parameters}, which remove the \logg-\teff\ degeneracy of the atmospheric {stellar atmospheric parameters} 
derived using only spectroscopy.

\subsection{Validation with repeat observations}\label{subsec:repeat_val}

A further way to validate the quality of the \rave\ data products is to compare the  
parameters derived for multiple observations of the same object (see Section 2.7 in DR6-1). In the following analysis we calculate for each property 
$W$ under consideration (\teff, \logg, \alphafe, individual element abundances) and for each 
star $k$ that has $N^k_\mathrm{repeat}>1$ observations that fulfill the quality 
threshold,  the difference between the observation $i$ ($1\leq i\leq 
N^k_\mathrm{repeat}$) derived properties $W_i^k$ and the mean $\bar{W}^k$ for the 
respective repeat sequence. We then analyze the distribution function of $\Delta 
W_i^k=W_i^k-\bar{W}^k$ over all stars $k$ and observations $i$. The distribution 
function is then  approximated by a combination of two Gaussians using a least-squares 
fit.

\begin{figure*}
\begin{center}
\plotone{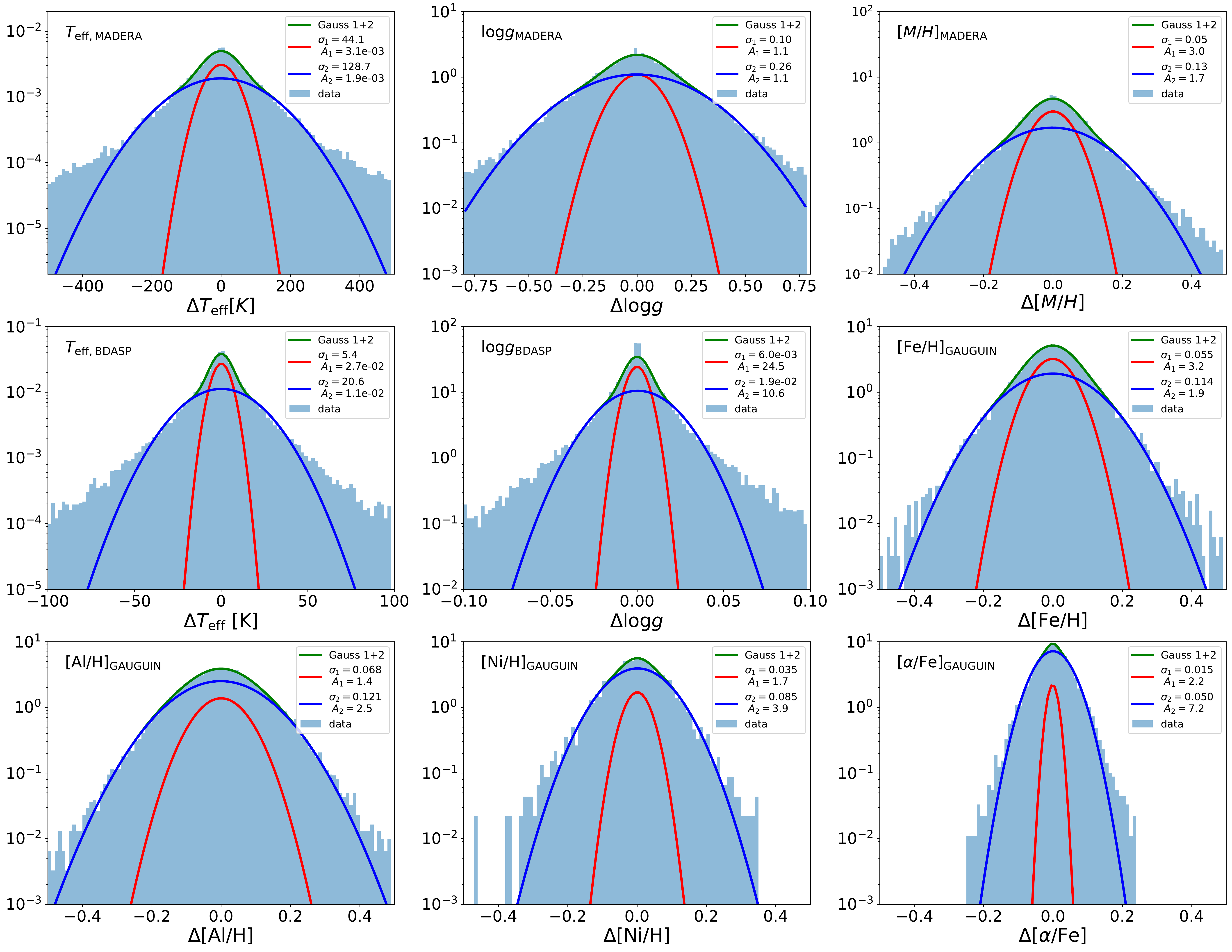}
\caption{\label{fig:repeat_stellar_params}From upper left to lower 
right: Differences in the \texttt{MADERA} estimated \teff, \texttt{MADERA} estimated \logg, \texttt{MADERA} estimated 
\mh, \texttt{BDASP} estimated \teff, \texttt{BDASP} estimated \logg\, \texttt{GAUGUIN} estimated \feh, \alh, \nih, and \alphafe\ for stars in the MD20 and BD20 sample with 
more than one observation, respectively. For the \alphafe\ determination, the \qhigh\ cut has been applied.}
\end{center}
\end{figure*}

Figure \ref{fig:repeat_stellar_params} shows the 
distribution function for  \teff, \logg, and the elemental abundances for the MD20/BD20 sample in addition
to a fit of the distribution with two Gaussians.

Overall the distributions show very similar 
behaviour: a core region that is well fit by two Gaussians, plus a wide exponential 
wing, which includes, however, only a few percent of the overall sample. The prominence 
of the wing and also the width of the wider Gaussian increases if we move from high 
(MD40) to lower quality constraint samples (like MD00). Occasionally, a spike at 
$\Delta W=0$ can be observed, in particular for the \texttt{MADERA} pipeline, reflecting the 
tendency of this pipeline to assign values close to the vertices of the spectral 
template grid (``pixelization''). Those spikes are excluded from the fitting procedure.  

The distribution for the \texttt{MADERA} values 
for \teff, \logg, and \mh,  in the repeat sequence gives a result  
consistent with  the 
differences between the \texttt{MADERA} value and the external data set, though overall the errors 
appear to be systematically smaller, reinforcing that a considerable uncertainty 
has to be assumed also for the {stellar atmospheric parameters} of the high-resolution sample, in 
particular for \logg. The uncertainties for the individual elements are also consistent 
with the errors quoted in Section \ref{sec:GAUGUIN} and \ref{subsubsec:GAUGUIN_val}.

The on-first-sight surprisingly small variance in the \texttt{BDASP} values reflects the fact that 
these properties are predominantly not determined by \rave\ spectroscopic data.  \texttt{BDASP} 
uses as temperature input $T_\mathrm{eff, IRFM}$, which is primarily determined by 
photometric data (which are the same for all members of a repeat sequence) and only 
weakly depends on the input \logg\ from the \texttt{MADERA} pipeline. Variations thus come in only 
via \mh. \logg\ is mainly determined by Gaia DR2 parallax information in the Bayesian 
framework, so again only very weakly dependent on the \texttt{MADERA} input and thus on the 
spectral information.


\begin{deluxetable*}{rlccll}
\tablecaption{\label{tab:DR6_MADERA}\texttt{DR6\_MADERA} catalog description.} 
\tablehead{\colhead{Col}  &  \colhead{Format}   &   \colhead{Units} &   \colhead{NULL} & \colhead{Label}      &               \colhead{Explanations}}
\startdata
1       & char     & -          & N     & \texttt{rave\_obs\_id  }               & \rave\ spectrum designation                                                  \\
2      & float        & K          & Y     & \texttt{teff\_madera }                     & Effective temperature                                       \\
3      & float        & K          & Y     & \texttt{teff\_cal\_madera }                     & Calibrated effective temperature                                         \\
4      & float        & K          & Y     & \texttt{teff\_error\_madera }                    & Error in effective temperature                                   \\
5      & float        & dex        & Y     & \texttt{logg\_madera}                      & log surface gravity                                                   \\
6      & float        & dex        & Y     & \texttt{logg\_cal\_madera }                     & Calibrated log surface gravity                                                    \\
7      & float        & dex        & Y     & \texttt{logg\_error\_madera}                     & Error in log surface gravity                                             \\
8      & float        & dex        & Y     & \texttt{m\_h\_madera }                      & Uncalibrated metallicity $\mh_{\rm u} $                                            \\
9      & float        & dex        & Y     & \texttt{m\_h\_cal\_madera  }                   & Calibrated metallicity $\mh_{\rm DR6}   $                                           \\
10      & float        & dex        & Y     & \texttt{m\_h\_error\_madera}                      & Error in metallicity \mh                                         \\
11      & float        & -          & Y     & \texttt{algo\_conv\_madera}                 & Quality flag for \texttt{MADERA} [0..4] \tablenotemark{a}    \\
12      & float        & -        & Y     & \texttt{chisq\_madera}            & $\chi^2$ of the best fit                \\
13      & float        & -          & Y     & \texttt{snr\_madera}                       & \SNR\  employed in the MADERA pipeline           \\
14      & float        & -        & Y     & \texttt{chisq\_madera}            & $\chi^2$ of the best fit                \\
\enddata
\tablenotetext{a}{
Flag of the MADERA stellar parameter pipeline:     
   $0 =$~pipeline converged. 
   $1 =$~no convergence.
   $2 =$~MATISSE oscillated between two values and the mean was calculated. 
   $3 =$~results of MATISSE at the boundaries or outside the grid and the DEGAS value was adopted.
   $4 =$~the metal-poor giants with \texttt{snr\_madera}$<$20 were re-run by DEGAS with a scale factor (\ie, internal parameter of DEGAS) of 0.40.}
\end{deluxetable*}


\begin{deluxetable*}{rlccll}
\tablecaption{\label{tab:DR6_IRFM}\texttt{DR6\_IRFM} catalog description.} 
\tablehead{\colhead{Col}  &  \colhead{Format}   &   \colhead{Units} &   \colhead{NULL} & \colhead{Label}      &               \colhead{Explanations}}
\startdata
1       & char     & -          & N     & \texttt{rave\_obs\_id}                 & \rave\ spectrum designation                                                    \\
2      & float        & K          & Y     & \texttt{teff\_irfm}                     & Temperature from infrared flux method                                          \\
3      & float        & K          & Y     & \texttt{teff\_error\_irfm}                      & Internal error on \texttt{teff\_irfm}              \\
4      & float        & mas          & Y     & \texttt{rad\_irfm}                      & Angular diameter  from infrared flux method                                          \\
5      & float        & mas         & Y     & \texttt{rad\_error\_irfm}                     & Internal error on \texttt{rad\_irfm\_ir}              \\
6      & char        & -          & N     & \texttt{method\_irfm}                    & \texttt{IRFM} flag\tablenotemark{a}                                   
\enddata
\tablenotetext{a}{Cross-identification flag as follows:
   $IRFM$:~Temperature derived from infrared flux method. 
   $CTRL$:~Temperature computed via color-$T_{\rm eff}$ relations. 
   $NO$:~No temperature derivation possible.}
\end{deluxetable*}


\begin{deluxetable*}{rlccll}
\tablecaption{\label{tab:DR6_BDASP}\texttt{DR6\_BDASP} catalog description. All parameter are determined using the BDASP pipeline.} 
\tablehead{\colhead{Col}  &  \colhead{Format}   &   \colhead{Units} &   \colhead{NULL} & \colhead{Label}      &               \colhead{Explanations}}
\startdata
1  &  char  &  -  &  N  &  \texttt{rave\_obs\_id}  &  \rave\ spectrum designation \\
2  &  float  &  pc  &  N  &  \texttt{distance\_bdasp}  &  Heliocentric distance estimate  \\
3  &  float  &  pc  &  N  &  \texttt{distance\_error\_bdasp}  &  Heliocentric distance uncertainty  \\
4  &  float  &  yr  &  N  &  \texttt{age\_bdasp}  &  Age estimate  \\
5  &  float  &  yr  &  N  &  \texttt{age\_error\_bdasp}  &  Age uncertainty  \\
6  &  float  &  K  &  N  &  \texttt{teff\_bdasp}  &  \teff\ estimate  \\
7  &  float  &  K  &  N  &  \texttt{teff\_error\_bdasp}  &  \teff\ uncertainty  \\
8  &  float  &  dex  &  N  &  \texttt{logg\_bdasp}  &  \logg\ estimate  \\
9  &  float  &  dex  &  N  &  \texttt{logg\_error\_bdasp}  &  \logg\ uncertainty \\
10  &  float  &  $M_\odot$  &  N  &  \texttt{mass\_bdasp}  &  Mass estimate  \\
11  &  float  &  $M_\odot$  &  N  &  \texttt{mass\_error\_bdasp}  &  Mass uncertainty  \\
12  &  float  &  -  &  N  &  l\texttt{og\_a\_v\_bdasp}  &  Log of extinction ($A_V/{\rm mag}$) estimate  \\
13  &  float  &  -  &  N  &  \texttt{log\_a\_v\_error\_bdasp}  &  Log of extinction ($A_V/{\rm mag}$) uncertainty  \\
14  &  float  &  dex  &  N  &  \texttt{m\_h\_bdasp}  &  [M/H] estimate  \\
15  &  float  &  dex  &  N  &  \texttt{m\_h\_error\_bdasp}  &  [M/H] uncertainty  \\
16  &  float  &  mas  &  N  &  \texttt{parallax\_bdasp}  &  Parallax estimate  \\
17  &  float  &  mas  &  N  &  \texttt{parallax\_error\_bdasp}  &  Parallax uncertainty  \\
18  &  float  &  mag  &  N  &  \texttt{dist\_mod\_bdasp}  &  Distance modulus estimate  \\
19  &  float  &  mag  &  N  &  \texttt{dist\_mod\_error\_bdasp}  &  Distance modulus uncertainty  \\
20  &  float  &  mag  &  N  &  \texttt{a\_v\_inf\_prior\_bdasp}  &  Prior on extinction ($A_V$) at infinity used  
\enddata
\end{deluxetable*}


\begin{deluxetable*}{rlccll}
\tablecaption{\label{tab:DR6_Seismo}\texttt{DR6\_Seismo}, the asteroseismically calibrated red giant catalog description.} 
\tablehead{\colhead{Col}  &  \colhead{Format}   &   \colhead{Units} &   \colhead{NULL} & \colhead{Label}      &               \colhead{Explanations}}
\startdata
1       & char     & -          & N     & \texttt{rave\_obs\_id}                 & \rave\ spectrum designation                                                     \\
2       & float     & K          & N     & \texttt{teff\_seismo}                       & Effective temperature     \\
3       & float     & K          & N     & \texttt{teff\_error\_seismo}                      & Error in effective temperature                                                     \\
4       & float     & \dex          & N     & \texttt{logg\_seismo}                       & $\logg_S$                                                    \\
5       & float     & \dex          & N     & \texttt{logg\_error\_seismo}                       & Uncertainty in $\logg_S$                                                    \\
6       & float     & \dex          & N     & \texttt{m\_h\_seismo}                       & $\Mh_S$                                                    \\
7       & float     & \dex          & Y     & \texttt{m\_h\_error\_seismo}                       & Uncertainty in $\Mh_S$                                                    \\
8      & float     & \dex         & Y     & \texttt{fe\_h\_seismo}                 & $\feh_S$                                                    \\
9       & float     & \dex          & Y     & \texttt{fe\_h\_error\_seismo}                 & Uncertainty in $\feh_S$                                                    \\
10       & float     & \dex          & Y     & \texttt{mg\_h\_seismo}                 & $\mgh_S$                                                    \\
11       & float     & \dex          & Y     & \texttt{mg\_h\_error\_seismo}                 & Uncertainty in $\mgh_S$  \\                       12       & float     & -         & Y     & \texttt{chisq\_seismo}                 & $\chi^2$ of the best fit                                                    
\enddata
\end{deluxetable*}


\begin{deluxetable*}{rlccll}
\tablecaption{\label{tab:DR6_GAUGUIN}\texttt{DR6\_GAUGUIN} catalog description.}
\tablehead{\colhead{Col}  &  \colhead{Format}   &   \colhead{Units} &   \colhead{NULL} & \colhead{Label}      &               \colhead{Explanations}}
\startdata
1   & char     & -         & N     & \texttt{rave\_obs\_id}             & \rave\ spectrum designation             \\
2   & float    & dex       & N     & \texttt{alpha\_de\_gauguin}        & $\alphafe$ estimate from \texttt{GAUGUIN} \tablenotemark{a}       \\
3   & float    & dex       & N     & \texttt{alpha\_fe\_error\_gauguin} & $\alphafe$ error from \texttt{GAUGUIN}           \\
4   & float    & -         & N     & \texttt{alpha\_fe\_chisq\_gauguin}  & $\chi^2$ of the best fit                \\
6   & float    & dex       & N     & \texttt{fe\_h\_gauguin}         & [Fe/H] estimate from \texttt{GAUGUIN}                              \\
7   & float    & dex       & N     & \texttt{fe\_h\_error\_gauguin}  & [Fe/H] error from \texttt{GAUGUIN}                                 \\
8   & int      & -         & N     & \texttt{fe\_h\_nl\_gauguin}     & Number of spectral lines used to derive [Fe/H]            \\
9   & float    & -         & N     & \texttt{fe\_h\_chisq\_gauguin}   & $\chi^2$ of the best line fit                             \\
10   & float    & dex       & N     & \texttt{al\_h\_gauguin}         & [Al/H] estimate from \texttt{GAUGUIN}                              \\
11   & float    & dex       & N     & \texttt{al\_h\_error\_gauguin}  & [Al/H] error from \texttt{GAUGUIN}                                 \\
12   & int      & -         & N     & \texttt{al\_h\_nl\_gauguin}     & Number of spectral lines used to derive [Al/H]            \\
13   & float    & -         & N     & \texttt{al\_h\_chisq\_gauguin}   & $\chi^2$ of the best line fit                             \\
14   & float    & dex       & N     & \texttt{ni\_h\_gauguin}         & [Ni/H] estimate from \texttt{GAUGUIN}                              \\
15   & float    & dex       & N     & \texttt{ni\_h\_error\_gauguin}  & [Ni/H] error from \texttt{GAUGUIN}                                 \\
16   & int      & -         & N     & \texttt{ni\_h\_nl\_gauguin}     & Number of spectral lines used to derive [Ni/H]            \\
17   & float    & -         & N     & \texttt{ni\_h\_chisq\_gauguin}   & $\chi^2$ of the best line fit                            
\enddata
\tablenotetext{a}{This table is valid for both catalogues: \texttt{DR6\_GAUGUIN\_MADERA} and \texttt{DR6\_GAUGUIN\_BDASP}.}
\end{deluxetable*}


\section{The sixth \rave\ public data release: catalog presentation II}\label{sec:FDR}

\rave\ DR6 spectra and the derived quantities are made available through a data base 
accessible via \texttt{doi:10.17876/rave/dr.6/} (for details see paper DR6-1). Since key words and unquoted identifiers are case insensitive, in SQL, in general lower case identifiers are used in the data base. The two main identifiers are 
\texttt{rave\_obs\_id} and \texttt{raveid}: the former, \texttt{rave\_obs\_id}, is the 
unique identifier denoting the observation of a particular spectrum -- the name is a composite of the observing date, field name, and fiber number allocated to the star on that occasion. 

\texttt{raveid} is the unique identifier of the target star, the name being a 
composite of the targets Galactic coordinates in the J2000.0 system. Consequently, 
objects that have several observations have the same \texttt{raveid} for all, but differ 
in their \texttt{rave\_obs\_id}. 

For convenience we also provide a set of FITS, CSV, and HDF files of
the overall \rave\ catalog, featuring key variables sufficient for the 
majority of applications of the \rave\ survey. These data are organized in 16 
files according to the pipeline employed; the content for 10 of these files is briefly 
described in the following paragraphs and associated tables, for the remaining 6 we refer to paper DR6-1. We avoid duplication 
of variable entries in the different files, with the exception of 
\texttt{rave\_obs\_id}, which can be used to link the contents of the various catalogs.


\subsection{Which \rave\ DR6 data products to use}

The detailed use of the \rave\ DR6 data products and the quality criteria to be applied 
depend on the particular science case under consideration. A general recommendation can
thus not be made. Considering our experience working with \rave\ data over the past 15 
years, and the various tests we performed in particular in the context of this data release,
we recommend as a starting point in particular in the context of Galactic dynamics and 
Galactic archeology applications:
\begin{itemize}
    \item To use the \texttt{BDASP} values for \logg, \teff, and the distance of the star, and the 
    calibrated \texttt{MADERA} value for the overall metallicity \mh. In particular \texttt{BDASP} properties are recommended for the identification of subpopulations (dwarfs, giants, red-clump stars).
    \item For abundance ratios we recommend the $\alpha$-enhancement and Al, Fe, and Ni values
    derived using \texttt{GAUGUIN} for \logg, \teff, and \mh\ taken from \texttt{MADERA}, as the \texttt{MADERA}/\texttt{GAUGUIN} combination provides an internally consistent pure spectroscopic framework. The use of \texttt{BDASP} derived parameters  as input parameters for a determination of the $\alpha$-enhancement $\alphafe$ should be taken with caution, as it introduces systematic inconsistencies in the templates used for the determination of $\alphafe$ and the determination of the metallicity. 
    \item For applications where a purely spectroscopically derived stellar parameter is 
    sought for, we recommend the calibrated \texttt{MADERA} values for \logg, \teff, and \mh, 
    and $\alpha$-enhancements based on \texttt{MADERA} input parameters.
\end{itemize}


\subsection{The \rave\  DR6 catalog of  {stellar atmospheric parameters}}

Stellar atmospheric parameters are derived using three pipelines and, consequently, 
assembled in three catalogs: \texttt{DR6\_MADERA} (\texttt{doi:10.17876/rave/dr.6/006}, 
Table \ref{tab:DR6_MADERA}), \\
\texttt{DR6\_IRFM} (\texttt{doi:10.17876/rave/dr.6/007}, 
Table \ref{tab:DR6_IRFM}), and \texttt{DR6\_BDASP} 
(\texttt{doi:10.17876/rave/dr.6/008}, Table \ref{tab:DR6_BDASP}). 
\texttt{DR6\_BDASP} also provides an improved distance estimate 
and extinction measure combining Gaia, spectroscopic and photometric data, as well as 
estimates for the mass and the age of the respective star based on a Bayesian 
isochrone comparison.

\subsection{The \rave\ DR6 asteroseismically calibrated Red Giant catalog}\label{subsec:DR6_seismo}

The asteroseismically calibrated red giant catalog is provided in the 
\texttt{DR6\_Seismo} file (\texttt{doi:10.17876/rave/dr.6/013}), Table \ref{tab:DR6_Seismo}).
We recommend to use \texttt{DR6\_Seismo} only for targets that are classfied with $\mathtt{flag1}=$`n' 
and for which the difference between $\logg_S$ and $\logg_u$ is less then 0.5\,\dex. 


\subsection{The \rave\  DR6 catalog of  element abundances and $\alpha$ enhancements}

The abundances of the non-iron group elements (Al, Fe, and Ni) and of the $\alpha$ enhancement $\alphafe$ derived with the 
pipeline \texttt{GAUGUIN} and \texttt{MADERA} input are provided 
(\texttt{doi:10.17876/rave/dr.6/009}, 
Table \ref{tab:DR6_GAUGUIN}). The analogous table for 
\texttt{BDASP} input (see comment above) 
can be found in (\texttt{doi:10.17876/rave/dr.6/010}.

For backward compatibility with the DR4 and DR5 data releases, we also provide a file 
with chemical abundances derived using the CDR pipeline \citep{boeche2011} using \teff, 
\logg, and \mh\ of the \texttt{MADERA} (\texttt{DR6\_CDR\_MADERA}, \texttt{doi:10.17876/rave/dr.6/011}) 
and \texttt{BDASP} (\texttt{DR6\_CDR\_BDASP}, \texttt{doi:10.17876/rave/dr.6/012}) pipeline as input, 
but in general recommend the use of elemental abundances from the \texttt{GAUGUIN} pipeline.


\subsection{The \rave\  DR6 catalog of orbits}\label{subsec:cat_orbits}

\texttt{DR6\_Orbits} (\texttt{doi:10.17876/rave/dr.6/014}, Table \ref{tab:DR6_Orbits}) contains information on the orbits of the 
\rave\ stars, obtained under the assumption of a given Milky Way mass model (Section \ref{sec:Orbits}). \bigskip

The \rave\ DR6 data release is complemented by two files cross-matching \rave\ DR6 with Gaia DR2 
(\texttt{DR6\_GaiaDR2}, \texttt{doi:10.17876/rave/dr.6/015}) and with a suite of other catalogs 
including Tycho-2, 2MASS, WISE, APASS9, and SKYMAPPER (\texttt{DR6\_XMatch}, 
\texttt{doi:10.17876/rave/dr.6/016}).

\startlongtable
\begin{deluxetable*}{rlccll}
\tablecaption{\label{tab:DR6_Orbits}\texttt{DR6\_Orbits} catalog description.} 
\tablehead{\colhead{Col}  &  \colhead{Format}   &   \colhead{Units} &   \colhead{NULL} & \colhead{Label}  &  \colhead{Explanations}}
\startdata
1  &  char  &  -  &  N  &  \texttt{rave\_obs\_id}  &  \rave\ target designation \\
2  &  float  &  kpc  &  N  &  \texttt{helio\_x}  &  Heliocentric X position \\
3  &  float  &  kpc  &  N  &  \texttt{helio\_x\_plus}  &  -- positive uncertainty \\
4  &  float  &  kpc  &  N  &  \texttt{helio\_x\_minus}  &  -- negative uncertainty \\
5  &  float  &  kpc  &  N  &  \texttt{helio\_y}  &  Heliocentric Y position \\
6  &  float  &  kpc  &  N  &  \texttt{helio\_y\_plus}  &  -- positive uncertainty \\
7  &  float  &  kpc  &  N  &  \texttt{helio\_y\_minus}  &  -- negative uncertainty \\
8  &  float  &  kpc  &  N  &  \texttt{helio\_z}  &  Heliocentric Z position \\
9  &  float  &  kpc  &  N  &  \texttt{helio\_z\_plus}  &  -- positive uncertainty \\
10  &  float  &  kpc  &  N  &  \texttt{helio\_z\_minus}  &  -- negative uncertainty \\
11  &  float  &  ${\rm km}\,{\rm s}^{-1}$  &  N  &  \texttt{helio\_vx}  &  Heliocentric velocity in X direction \\
12  &  float  &  ${\rm km}\,{\rm s}^{-1}$  &  N  &  \texttt{helio\_vx\_plus}  &  -- positive uncertainty \\
13  &  float  &  ${\rm km}\,{\rm s}^{-1}$  &  N  &  \texttt{helio\_vx\_minus}  &  -- negative uncertainty \\
14  &  float  &  ${\rm km}\,{\rm s}^{-1}$  &  N  &  \texttt{helio\_vy}  &  Heliocentric velocity in Y direction \\
15  &  float  &  ${\rm km}\,{\rm s}^{-1}$  &  N  &  \texttt{helio\_vy\_plus}  &  -- positive uncertainty \\
16  &  float  &  ${\rm km}\,{\rm s}^{-1}$  &  N  &  \texttt{helio\_vy\_minus}  &  -- negative uncertainty \\
17  &  float  &  ${\rm km}\,{\rm s}^{-1}$  &  N  &  \texttt{helio\_vz}  &  Heliocentric velocity in Z direction \\
18  &  float  &  ${\rm km}\,{\rm s}^{-1}$  &  N  &  \texttt{helio\_vz\_plus}  &  -- positive uncertainty \\
19  &  float  &  ${\rm km}\,{\rm s}^{-1}$  &  N  &  \texttt{helio\_vz\_minus}  &  -- negative uncertainty \\
20  &  float  &  kpc  &  N  &  \texttt{galcyl\_r}  &  Galactocentric cylindrical radius \\
21  &  float  &  kpc  &  N  &  \texttt{galcyl\_r\_plus}  &  -- positive uncertainty \\
22  &  float  &  kpc  &  N  &  \texttt{galcyl\_r\_minus}  &  -- negative uncertainty \\
23  &  float  &  kpc  &  N  &  \texttt{galcyl\_z}  &  Height above the Galactic plane \\
24  &  float  &  kpc  &  N  &  \texttt{galcyl\_z\_plus}  &  -- positive uncertainty \\
25  &  float  &  kpc  &  N  &  \texttt{galcyl\_z\_minus}  &  -- negative uncertainty \\
26  &  float  &  deg  &  N  &  \texttt{galcyl\_phi}  &  Galactocentric azimuth \\
27  &  float  &  deg  &  N  &  \texttt{galcyl\_phi\_plus}  &  -- positive uncertainty \\
28  &  float  &  deg  &  N  &  \texttt{galcyl\_phi\_minus}  &  -- negative uncertainty \\
29  &  float  &  ${\rm km}\,{\rm s}^{-1}$  &  N  &  \texttt{galcyl\_vr}  &  Velocity in Galactocentric cylindrical radial direction \\
30  &  float  &  ${\rm km}\,{\rm s}^{-1}$  &  N  &  \texttt{galcyl\_vr\_plus}  &  -- positive uncertainty \\
31  &  float  &  ${\rm km}\,{\rm s}^{-1}$  &  N  &  \texttt{galcyl\_vr\_minus}  &  -- negative uncertainty \\
32  &  float  &  ${\rm km}\,{\rm s}^{-1}$  &  N  &  \texttt{galcyl\_vz}  &  Velocity perpendicular to the Galactic plane \\
33  &  float  &  ${\rm km}\,{\rm s}^{-1}$  &  N  &  \texttt{galcyl\_vz\_plus}  &  -- positive uncertainty \\
34  &  float  &  ${\rm km}\,{\rm s}^{-1}$  &  N  &  \texttt{galcyl\_vz\_minus}  &  -- negative uncertainty \\
35  &  float  &  ${\rm km}\,{\rm s}^{-1}$  &  N  &  \texttt{galcyl\_vphi}  &  Velocity in Galactocentric azimuth \\
36  &  float  &  ${\rm km}\,{\rm s}^{-1}$  &  N  &  \texttt{galcyl\_vphi\_plus}  &  -- positive uncertainty \\
37  &  float  &  ${\rm km}\,{\rm s}^{-1}$  &  N  &  \texttt{galcyl\_vphi\_minus}  &  -- negative uncertainty \\
38  &  float  &  kpc  &  N  &  \texttt{min\_galcyl\_r}  &  Minimum Galactocentric cylindrical radius on orbit \\
39  &  float  &  kpc  &  N  &  \texttt{min\_galcyl\_r\_plus}  &  -- positive uncertainty \\
40  &  float  &  kpc  &  N  &  \texttt{min\_galcyl\_r\_minus}  &  -- negative uncertainty \\
41  &  float  &  kpc  &  N  &  \texttt{max\_galcyl\_r}  &  Maximum Galactocentric cylindrical radius on orbit \\
42  &  float  &  kpc  &  N  &  \texttt{max\_galcyl\_r\_plus}  &  -- positive uncertainty \\
43  &  float  &  kpc  &  N  &  \texttt{max\_galcyl\_r\_minus}  &  -- negative uncertainty \\
44  &  float  &  kpc  &  N  &  \texttt{max\_galcyl\_z}  &  Maximum height above the Galactic plane on orbit \\
45  &  float  &  kpc  &  N  &  \texttt{max\_galcyl\_z\_plus}  &  -- positive uncertainty \\
46  &  float  &  kpc  &  N  &  \texttt{max\_galcyl\_z\_minus}  &  -- negative uncertainty \\
47  &  float  &  kpc  &  N  &  \texttt{min\_galsph\_r}  &  Minimum Galactocentric spherical radius on orbit \\
48  &  float  &  kpc  &  N  &  \texttt{min\_galsph\_r\_plus}  &  -- positive uncertainty \\
49  &  float  &  kpc  &  N  &  \texttt{min\_galsph\_r\_minus}  &  -- negative uncertainty \\
50  &  float  &  kpc  &  N  &  \texttt{max\_galsph\_r}  &  Maximum Galactocentric spherical radius on orbit \\
51  &  float  &  kpc  &  N  &  \texttt{max\_galsph\_r\_plus}  &  -- positive uncertainty \\
52  &  float  &  kpc  &  N  &  \texttt{max\_galsph\_r\_minus}  &  -- negative uncertainty \\
53  &  float  &  kpc  &  N  &  \texttt{mean\_galcyl\_r}  &  Orbit averaged Galactocentric cylindrical radius \\
54  &  float  &  kpc  &  N  &  \texttt{mean\_galcyl\_r\_plus}  &  -- positive uncertainty \\
55  &  float  &  kpc  &  N  &  \texttt{mean\_galcyl\_r\_minus}  &  -- negative uncertainty \\
56  &  float  &  ${\rm km}^2\,{\rm s}^{-2}$  &  N  &  \texttt{energy}  &  Orbital energy \\
57  &  float  &  ${\rm km}^2\,{\rm s}^{-2}$  &  N  &  \texttt{energy\_plus}  &  -- positive uncertainty \\
58  &  float  &  ${\rm km}^2\,{\rm s}^{-2}$  &  N  &  \texttt{energy\_minus}  &  -- negative uncertainty \\
59  &  float  &  ${\rm kpc}\,{\rm km}\,{\rm s}^{-1}$  &  N  &  \texttt{angmom}  &  Angular momentum about Galactic $z$ axis \\
60  &  float  &  ${\rm kpc}\,{\rm km}\,{\rm s}^{-1}$  &  N  &  \texttt{angmom\_plus}  &  -- positive uncertainty \\
61  &  float  &  ${\rm kpc}\,{\rm km}\,{\rm s}^{-1}$  &  N  &  \texttt{angmom\_minus}  &  -- negative uncertainty \\
62  &  float  &  -  &  N  &  \texttt{eccentricity}  &  Eccentricity \\
63  &  float  &  -  &  N  &  \texttt{eccentricity\_plus}  &  -- positive uncertainty \\
64  &  float  &  -  &  N  &  \texttt{eccentricity\_minus}  &  -- negative uncertainty \\
65  &  float  &  ${\rm kpc}\,{\rm km}\,{\rm s}^{-1}$  &  N  &  \texttt{jr}  &  Radial action \\
66  &  float  &  ${\rm kpc}\,{\rm km}\,{\rm s}^{-1}$  &  N  &  \texttt{jr\_plus}  &  -- positive uncertainty \\
67  &  float  &  ${\rm kpc}\,{\rm km}\,{\rm s}^{-1}$  &  N  &  \texttt{jr\_minus}  &  -- negative uncertainty \\
68  &  float  &  ${\rm kpc}\,{\rm km}\,{\rm s}^{-1}$  &  N  &  \texttt{jz}  &  Vertical action \\
69  &  float  &  ${\rm kpc}\,{\rm km}\,{\rm s}^{-1}$  &  N  &  \texttt{jz\_plus}  &  -- positive uncertainty \\
70  &  float  &  ${\rm kpc}\,{\rm km}\,{\rm s}^{-1}$  &  N  &  \texttt{jz\_minus}  &  -- negative uncertainty \\
\enddata
\end{deluxetable*}


\section{Science Applications}\label{sec:science}
The following section presents some first science applications for \rave\ DR6. The main aim here is less to demonstrate 
particularly new results but rather to demonstrate, using well-established features, the capabilities and limits within the \rave\ data set for Galactic archeology applications.

\subsection{Tomography of the volume probed by \rave}\label{subsec:MW_Tomography}

In a first science application we show the changes in the iron abundance, $\alpha$ 
enhancement, and kinematics throughout the volume probed by the \rave\ survey (for work 
with previous data releases, often employing a considerably smaller sample of stars, 
see, \eg, \cite{williams2013}, \cite{boeche2013}, \cite{kordopatis2013b}, 
\cite{wojno2016}, \cite{wojno2018}, or \cite{carillo2018}. We employ the full \qhigh\ sample for giants ($\logg<3.5$). 
\texttt{BDASP}  {stellar atmospheric parameters} and abundances using \texttt{GAUGUIN} with \texttt{MADERA} input are used through this section. 

\begin{figure*}
\begin{center}

\includegraphics[width=0.8\textwidth]{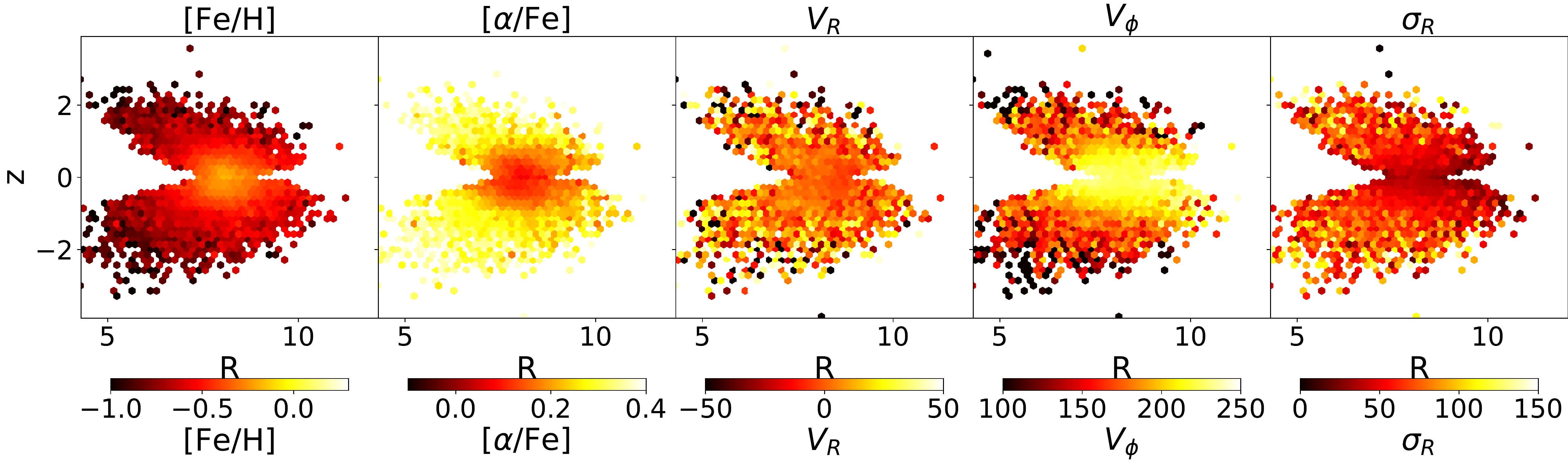}
\includegraphics[width=0.8\textwidth]{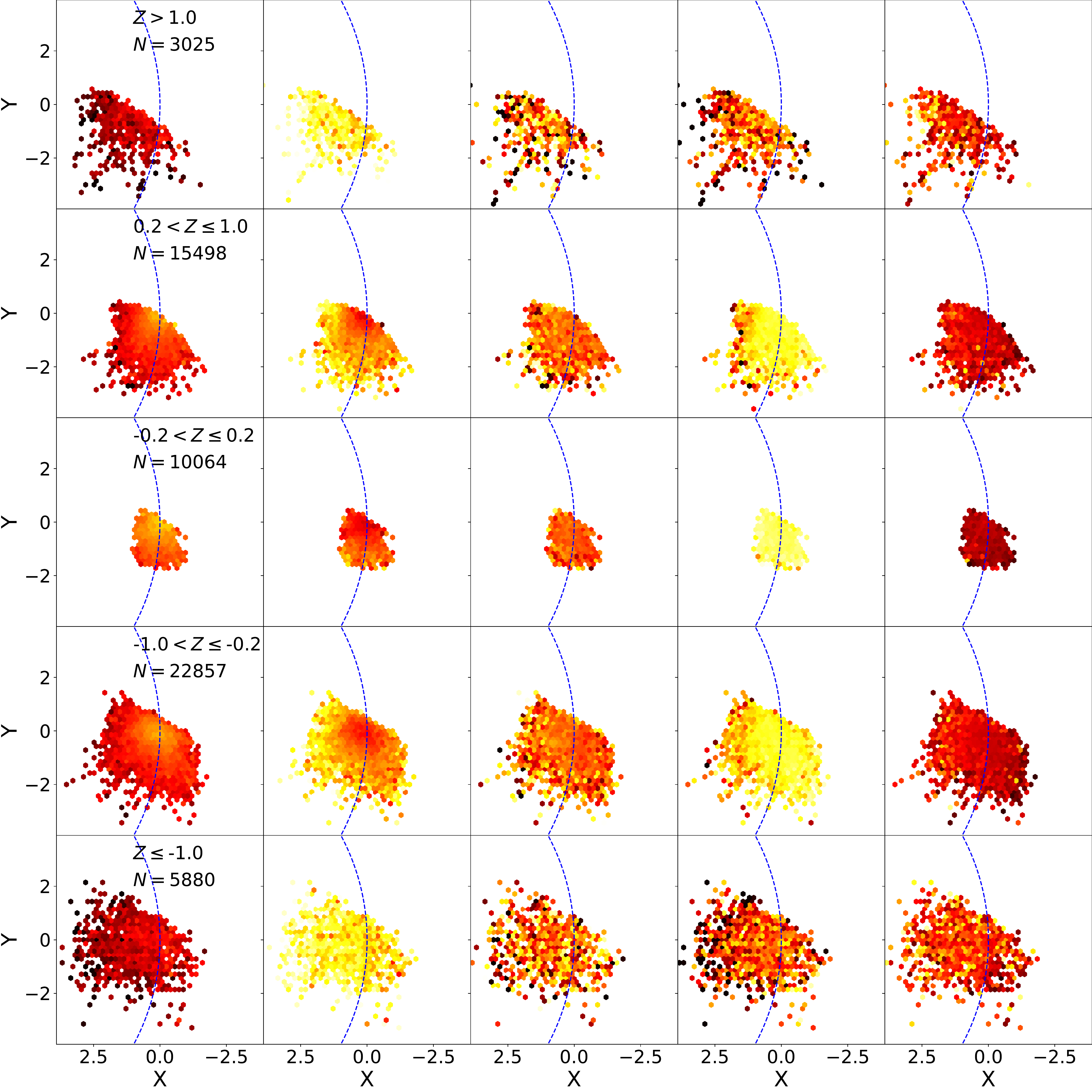}
\caption{\label{fig:MW_Tomography}Tomography of the volume probed by \rave. Top row: 
projection into the Galactocentric R-z plane; next rows from top to bottom: slice in the 
X-Y plane focused on the {local standard of rest (LSR)} for stars with $Z> 1$\,kpc, $0.2<Z \leq 1.0$\,kpc, 
$-0.2 < Z \leq 0.2$\,kpc, $-1<Z \leq 0.2$\,kpc, and $ Z \leq -1$\,kpc, respectively. The columns show (from left to right) the median of hexbins of size 
$0.16$ kpc for the iron abundance \feh, the $\alpha$-enhancement \alphafe, the 
Galactocentric radial velocity $V_R$, the tangential velocity $|V_\phi|$, and the radial 
velocity dispersion $\sigma_R$, respectively. The dotted curve in the X-Y plots indicates the solar circle. }
\end{center}
\end{figure*}

The top row of Figure \ref{fig:MW_Tomography} shows a projection of \rave\ stars onto 
the $R$-$Z$ plane of the Galactocentric cylindrical coordinate system. Stars are binned 
using hexagonal bins of size $0.16$\,kpc. The upper left plot shows the clear decrease of 
the average iron abundance as we move from the Galactic plane to larger heights above the 
Galactic plane. Simultaneously, the composition becomes more $\alpha$ enriched (second 
plot from the left). The change in abundances coincides with a change in kinematics: 
The average tangential velocity decreases (second plot from the right), while the 
disk is hotter, resulting in a higher radial velocity dispersion $\sigma_R$ 
(rightmost plot). The center plot in the top row shows the Galactocentric radial velocity, exhibiting a mild negative radial velocity gradient, consistent with the finding by 
\cite{siebert2011a} using \rave\ DR3.

The next 5 rows dissect the Milky Way in slices and focus on the distribution in the 
X-Y plane for each of these slices. The slices are for $Z>1\,$kpc (1st row) 
and $Z\leq -1\,$kpc (bottom row). In between, slices for $-1 < Z\leq -0.2\,$kpc 
(second row), $-0.2 < Z \leq 0.2\,$kpc (third row), and $0.2 < Z\leq 1\,$kpc (forth row) are shown, respectively. The shrinking size of the slice in 
the X-Y plane reflects the double-cone structure of the \rave\ survey volume, created 
by the exclusion of low Galactic latitude fields, in particular towards the Galactic 
center.

Each individual slice for \feh, \alphafe, $v_\phi$, and $\sigma_R$ shows a relatively 
homogeneous structure, with the changes with Z as presented in the previous paragraph, 
reflecting the disk like structure of the Milky Way. Only the central three slices 
exhibit some apparent radial gradient in the iron abundance with the distance from the 
Sun, an immediate result again of the 
\rave\ survey geometry: owing to the double conical layout of the survey volume, stars 
more distant in the central slices are predominantly from larger heights above the 
Galactic plane and thus have on average lower abundances than stars in the immediate 
solar neighborhood. The middle column again shows a mild outward directed radial 
velocity gradient, indicative of a non-axisymmetric gravitational potential.

\subsection{Relation between chemical abundances and kinematics}\label{sec:science2}

In this section we repeat the analysis of \cite{boeche2013} with the \rave\ DR6 data 
set. Unlike \cite{boeche2013}, we, however, only apply a very weak quality cut, namely 
the full \qhigh\ giant sample with determinations of \feh\ and 
$\alphafe=($[Mg/Fe]$+$[Si/Fe]$)/2$. Consequently, the underlying sample could be 
increased 
by more than a factor of five from 9,131 in \cite{boeche2013} to $~40,000$ stars, respectively.

As in \citet{gratton2003} or in \citet{boeche2013} we begin with a kinematical decomposition into a so-called 
\emph{thin-disk component}, \emph{dissipative-collapse component} and \emph{accretion component}, based on the eccentricity $e$, 
the tangential velocity in cylindrical coordinates $V_{\phi}$ and the maximum 
altitude from the Galactic plane $\text{Z}_{\text{max}}$. The thin-disc component consists of stars with a low eccentricity ($e<0.25$)
and low maximum altitude ($Z_{\text{max}}<0.8\,$kpc). The dissipative-collapse component, is composed of 
stars with higher eccentricity $e>0.25$, 
traveling higher above the Galactic plane $(\text{Z}_{\text{max}}>0.8\,$kpc), 
with $V_{\phi}>40\,$\kms. This component is mainly composed of thick disc and halo stars. The accretion~component is 
mainly composed of halo stars and accreted stars. We adopted the criteria $V_{\phi}<40\,$\kms, meaning that such 
stars are slowly-rotating or even counter-rotating with respect to the Galactic disc. Based on the \qhigh\ sample, 
we have at hand $39\,130$ stars of the thin disc component, $12\,354$ stars 
in the dissipative-collapse component, and $1\,931$ stars in the accretion component. 

\begin{figure*}
\begin{center}
\plotone{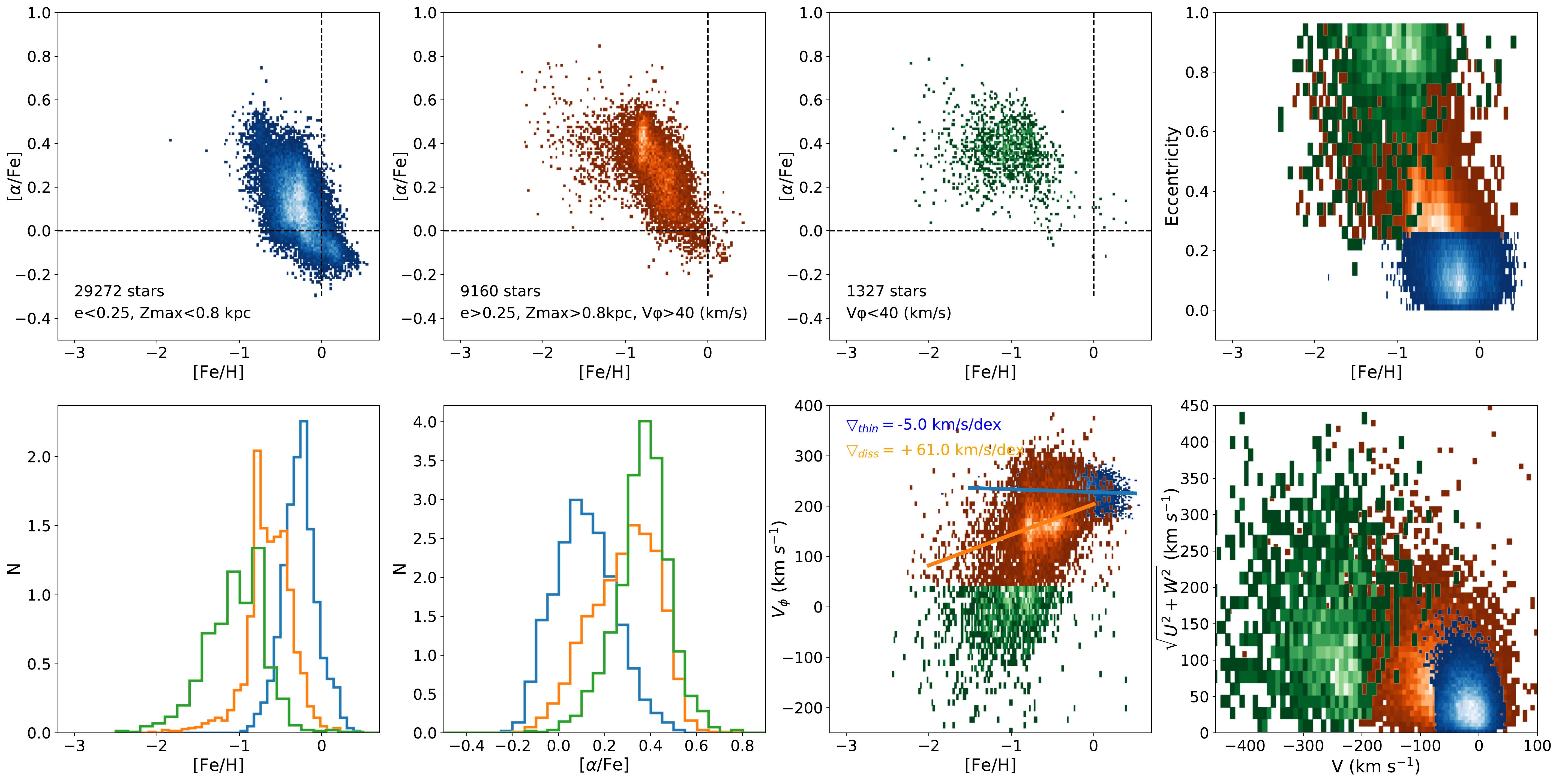}
\caption{\label{fig:thin_thick_halo}Top row: 
\alphafe\,vs.\,{\feh} density maps of the thin-disc component (blue), 
the dissipative-collapse component (orange), and the accretion component 
(green). Furthermore the orbital eccentricity is shown against the {iron abundance} for the three Galactic components. 
Bottom row: $\feh$ and $\alphafe$ distributions are shown, complemented by the gradient in the tangential velocity $V_\phi$ and the Toomre diagram.}
\end{center}
\end{figure*}

Figure \ref{fig:thin_thick_halo} shows the $\alphafe$ pattern for these 
three components. The thin-disc component is mainly confined at {$\feh>-1\,$dex}, 
and shows an increase of its $\alphafe$ with decreasing {$\feh$}. The 
dissipative-collapse component has a large range {in iron abundance}, 
with very few metal-rich stars and extends to very low metallicities down to $\feh\approx-3\,$dex. 
Its $\alphafe$ sequence is narrower, with an increased scatter towards the metal-poor end (owing to halo stars). 
The accretion~component is mostly composed of 
metal-poor stars, in the range {$-2.5<\feh<-0.5$}, and is $\alphafe$-rich, but no 
trend is observed with ${\feh}$. The thin-disc component is the 
most metal-rich component ($\langle{\feh}\rangle=-0.18\,$dex), while 
the dissipative-collapse component and the accretion~component have 
an average metallicity of $\langle{\feh}\rangle=-0.64\,$dex and 
$\langle{\feh}\rangle=-1.16\,$dex, respectively. The mean 
$\alphafe$ increases with decreasing metallicity, 
\ie, $\langle\alphafe\rangle=+0.11\,$dex for the thin-disc component, 
$\langle\alphafe\rangle=+0.26\,$dex for the dissipative-collapse component 
and $\langle\alphafe\rangle=+0.37\,$dex for the accretion component. 

In a Toomre diagram, we clearly see that the dissipative-collapse component 
shows typical kinematics of thick disc stars, but also halo-like 
kinematics (overlapping with the accretion component). We notice the 
presence as well of thin-disc like stars belonging to the 
dissipative-collapse component. The eccentricity 
increases with decreasing metallicity, and the 
dissipative-collapse component and the accretion component 
show considerable overlap (likely due to halo stars). 
We characterized the gradients of $V_{\phi}$ as a function 
of $\mh$ in both thin-disc and dissipative-collapse components. 
In the thin-disc component, we measure a weak anti-correlation 
($\nabla=-6\kms/\text{dex}$), while a strong correlation is visible in the 
dissipative~collapse~components ($\nabla=+55\kms/\text{dex}$). Such gradients are 
consistent with previous works, like for example \citet{lee2011} 
with SEGUE data or \citet{wojno2018} for RAVE.

\begin{figure}
\begin{center}
\plotone{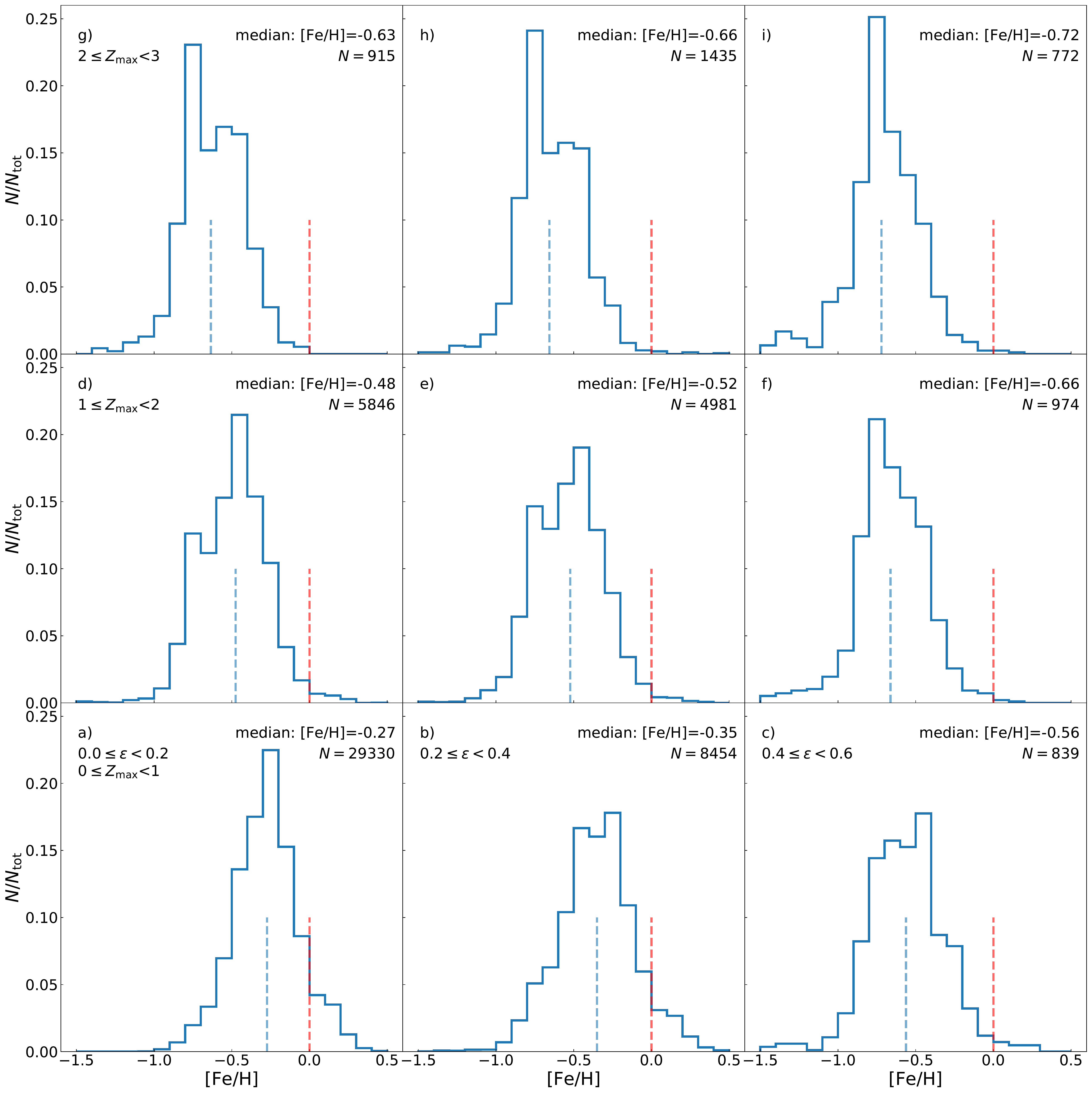}
\caption{\label{fig:Fe_vs_Orbit}
Distribution of iron abundance \feh\ as a function of orbital parameters: stars 
are grouped according to their eccentricity $\varepsilon$ (from left to right) and 
according to their maximum height above the Galactic plane $Z_\mathrm{max}$ (from 
bottom to top). The blue dashed vertical line indicates the median of the distribution, 
the red dotted line corresponds to the solar value.}
\end{center}
\end{figure}

\begin{figure}
\begin{center}
\plotone{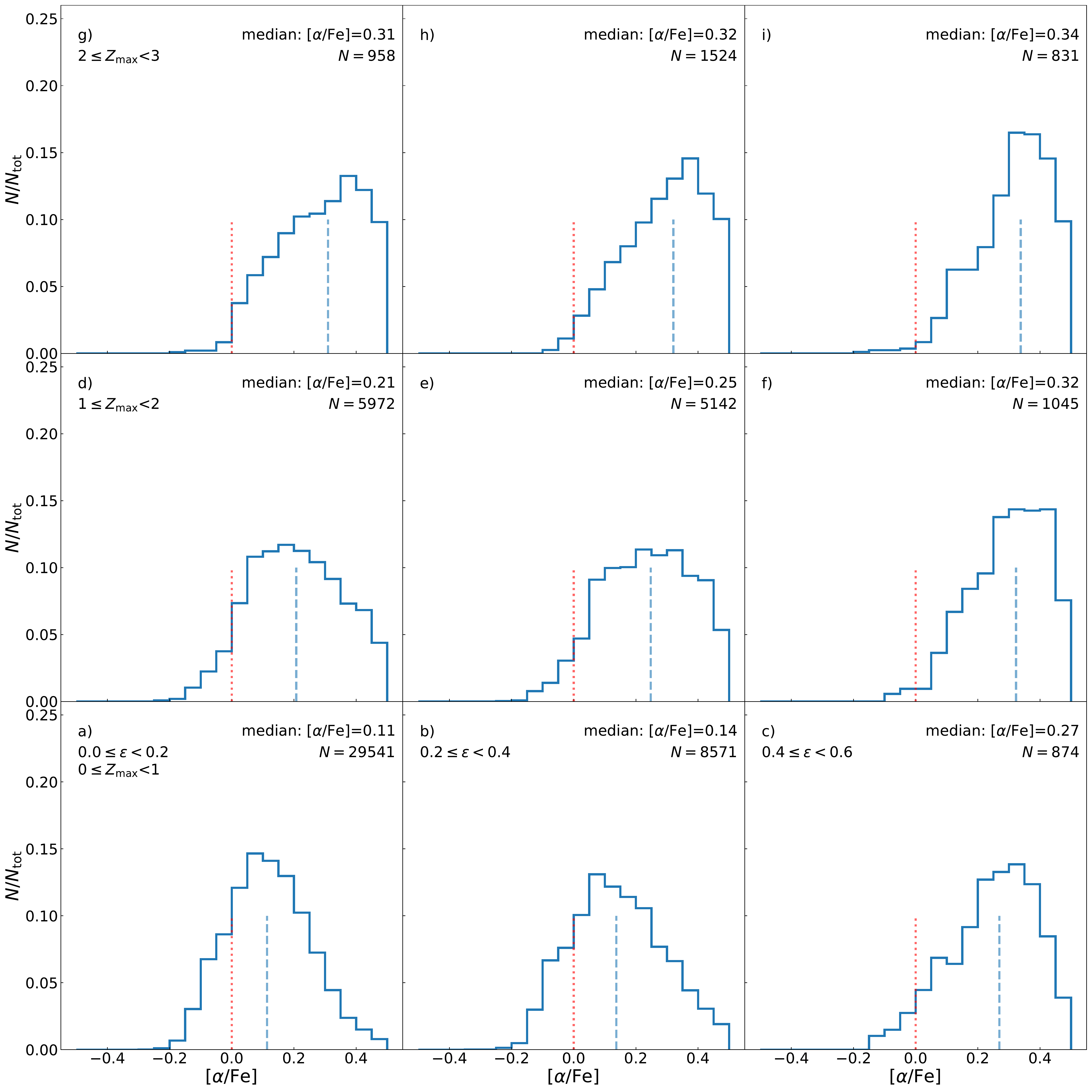}
\caption{\label{fig:alpha_vs_Orbit}
$\alpha$ enhancement \alphafe\ for the stellar samples shown in Figure 
\ref{fig:Fe_vs_Orbit}. The blue dashed vertical line indicates the median of the 
distribution, the red dotted line corresponds to the solar value.}
\end{center}
\end{figure}

\begin{figure}
\begin{center}
\plotone{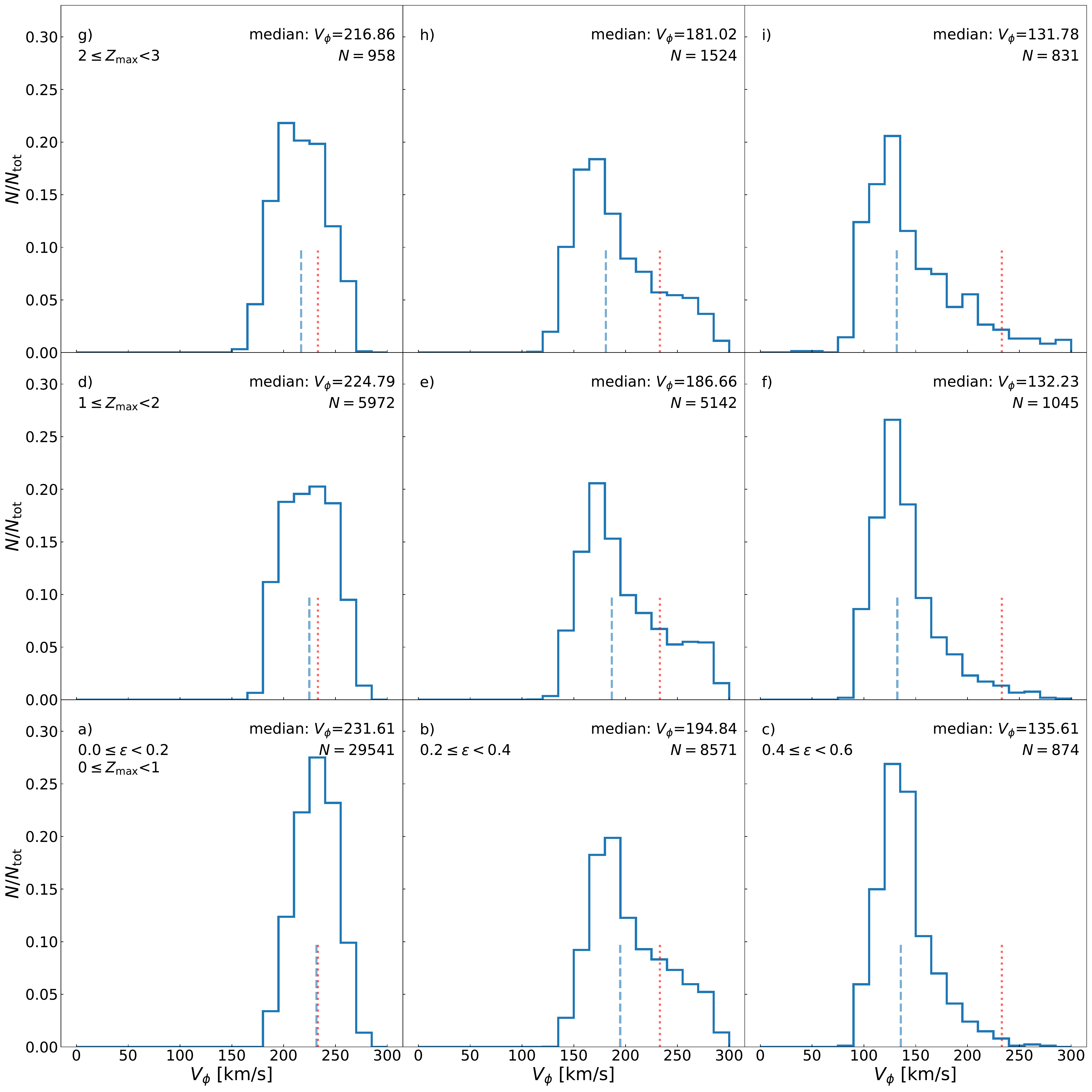}
\caption{\label{fig:vphi_vs_Orbit}
Distribution of tangential velocities $v_\phi$ for the stellar samples shown in Figure 
\ref{fig:Fe_vs_Orbit}. The blue dashed vertical line indicates the median of the 
distribution, the red dotted line corresponds to the circular velocity of the LSR.}
\end{center}
\end{figure}

\begin{figure}
\begin{center}
\plotone{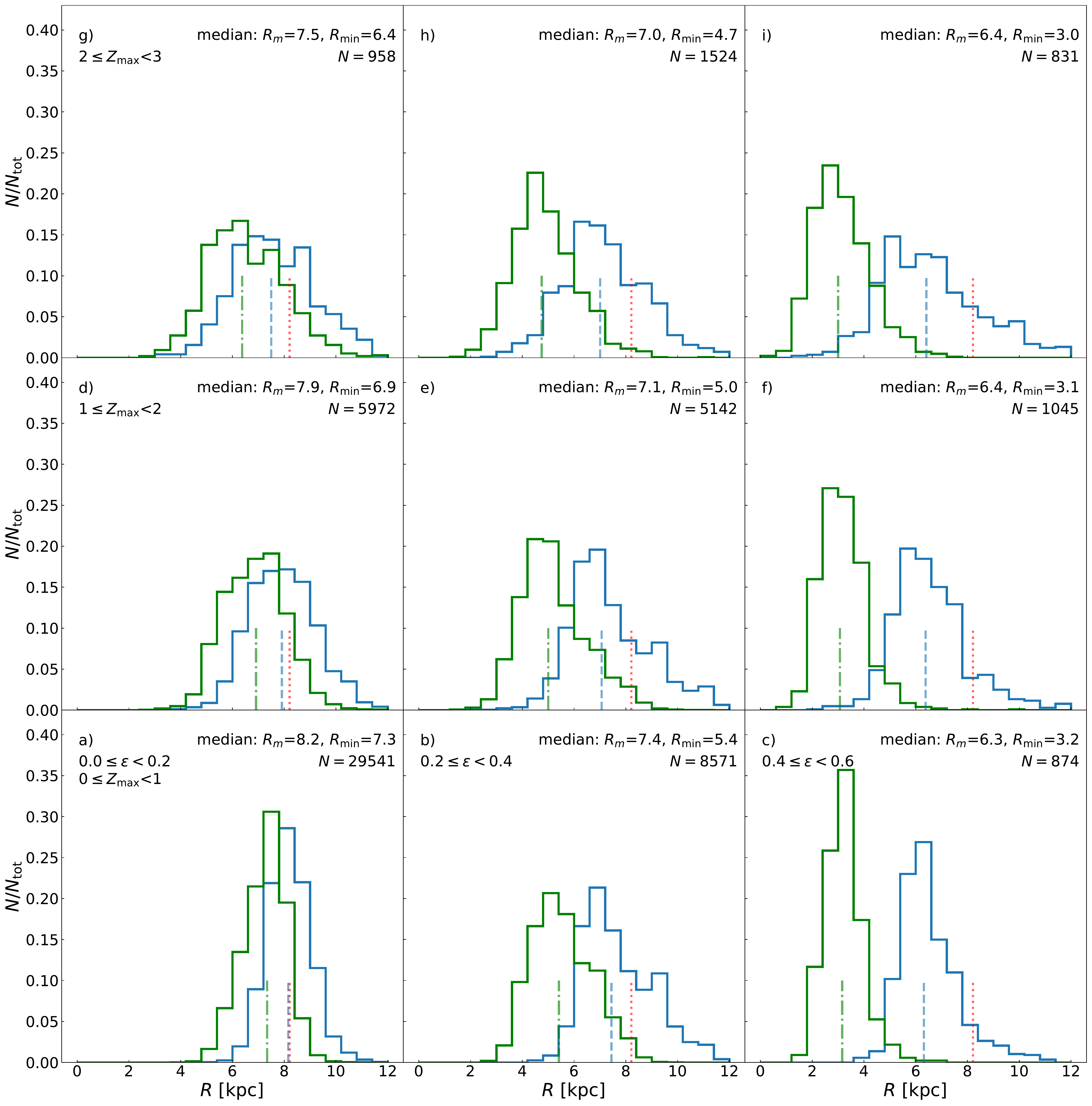}
\caption{\label{fig:Rm_vs_Orbit}
Distribution of the mean Galactocentric cylindrical radius $R_m$ (blue) and 
the minimum Galactocentric cylindrical radius $R_\mathrm{min}$ (green) for the stellar 
samples shown in Figure \ref{fig:Fe_vs_Orbit}. The blue dashed vertical line indicates 
the median of the $R_m$ distribution, the green dashed-dotted vertical line the median 
of the $R_\mathrm{min}$ distribution, and the red dotted line the radius of the solar 
circle, respectively.}
\end{center}
\end{figure}

We continue the strategy of \cite{boeche2013} and analyze the orbits of disk giants in 
the $\epsilon-Z_{\max}$ plane. We first separate the giant population into 9 different bins 
with 3 ranges in orbital eccentricity ($0\leq \epsilon<0.2$, $0.2\leq \epsilon<0.4$, and $0.4\leq 
\epsilon<0.6$) and 3 ranges in maximum altitude, $Z_{\max}$, reached by a star in its orbit 
($0\,\mathrm{kpc}\leq Z_{\max} < 1\,\mathrm{kpc}$, $1\,\mathrm{kpc}\leq Z_{\max} < 
2\,\mathrm{kpc}$, and $2\,\mathrm{kpc}\leq Z_{\max} < 3 \,\mathrm{kpc}$).
We then investigate the distributions of the iron abundance \feh\ (Figure 
\ref{fig:Fe_vs_Orbit}), of the $\alpha$ enhancement \alphafe\ (Figure 
\ref{fig:alpha_vs_Orbit}), and of the tangential velocity $|V_\phi|$ (Figure 
\ref{fig:vphi_vs_Orbit}), as well as of the mean and minimum 
Galactocentric cylindrical radius $R_m$ and $R_{\min}$ for these 9 bins 
(Figure \ref{fig:Rm_vs_Orbit}). Stars on orbits with $\epsilon>0.6$ or 
$Z_{\max} >3$\,kpc are not considered in these plots, as we mainly focus on the thin and thick disk. 

We first concentrate on the stars in panel (a), \ie, those on orbits closest to local 
disk kinematics. Indeed, stars in this bin have a tangential velocity distribution that peaks near 
227\,\kms, at a mean Galactocentric distance of about 8\,kpc. The abundances peak at 
$\feh\approx -0.280$\,\dex\ (Figure~\ref{fig:Fe_vs_Orbit}), i.e. somewhat below the 
solar value and are slightly $\alpha$-enhanced ($\alphafe = 0.11$) compared to the immediate solar 
neighborhood (Figure~\ref{fig:alpha_vs_Orbit}). This reflects the fact that our sample consists of giant stars, which 
are therefore more distant than a few hundred pc.
Indeed, a sample consisting of dwarf stars 
with the same orbital constraints exhibits a median abundance of $\feh=-0.06$\,\dex\ and an 
$\alpha$-enhancement of $\alphafe=0.1$\,\dex.

As we go from less to more eccentric orbits (panels a $\to$ c) and from lower to higher 
height above the plane (panels a$\to$ g) or both (panels a$\to$ i), the distribution in 
the mean orbital radius $R_m$ and in particular in the minimum distance $R_\mathrm{min}$ 
becomes broader and more skewed (Figure \ref{fig:Rm_vs_Orbit}), 
peaking at considerably smaller Galactocentric cylindrical radii. The median tangential 
velocity decreases from about 230\,\kms\ to less than 135\,\kms\ (panel i) as expected 
from the increasing asymmetric drift with increasing in velocity dispersion (see previous section). The \feh\ distribution 
moves to successively lower values and reaches only $\feh\approx -0.75$ in panel i, 
but the abundance mixture is now considerably more $\alpha$-enriched with $\alphafe
\approx 0.34$ (Figure \ref{fig:alpha_vs_Orbit}), indicating that the sample is 
increasingly dominated by thick disk and halo stars.

\begin{figure*}
\begin{center}
\plotone{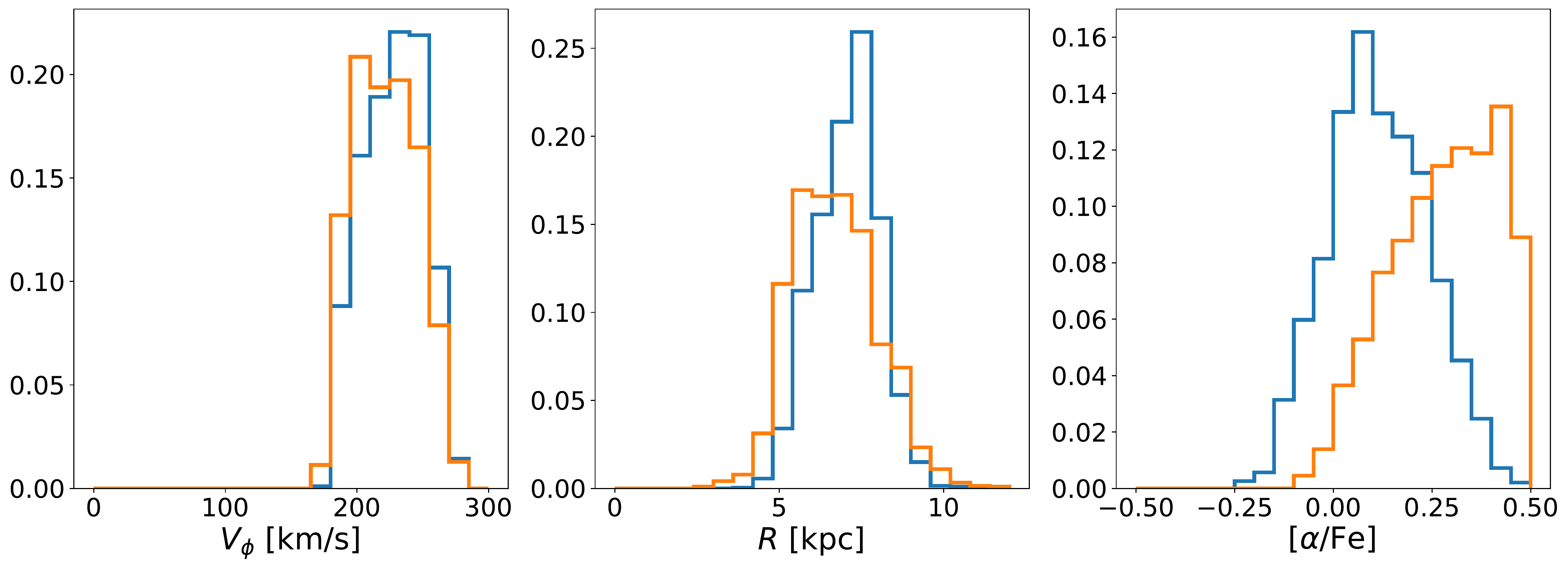}
\caption{\label{fig:low_vs_high_Fe}
Distribution of $v_\phi$, $R_m$, and \alphafe\ for stars in Figure \ref{fig:Fe_vs_Orbit} 
panel c, separated by those with $\feh<-0.5$\,\dex\ (blue) and $\feh>-0.4$\,\dex\ (orange). }
\end{center}
\end{figure*}

If we focus on the more eccentric orbital bins (c, f, and i), the distribution in 
velocity and radius is considerably more skewed, as expected, and the metallicity 
distribution exhibits a tail towards higher abundances (Figure \ref{fig:Fe_vs_Orbit}). 
Owing to the high eccentricity, a significant fraction of the stars come from the inner 
disk, also resulting in a 
broader distribution in \feh, indicating that this is a superposition of at least two 
populations, a low \feh\ one with all the chemical characteristics of the thick disk, 
and a higher \feh\ one more similar to the inner thin disk. Indeed, Figure 
\ref{fig:low_vs_high_Fe} shows that while samples with $\feh < -0.5$\,\dex\ and with 
$\feh>-0.4$\,\dex\ have little difference in their respective distribution of $R_m$ and 
$V_\phi$, they exhibit quite different patterns in \alphafe, consistent with the 
findings in \cite{boeche2013}, but now reproduced for a much larger fraction of the \rave\ DR6 giant sample, as represented by the \qhigh\ sample.

\begin{figure*}
\begin{center}
\plotone{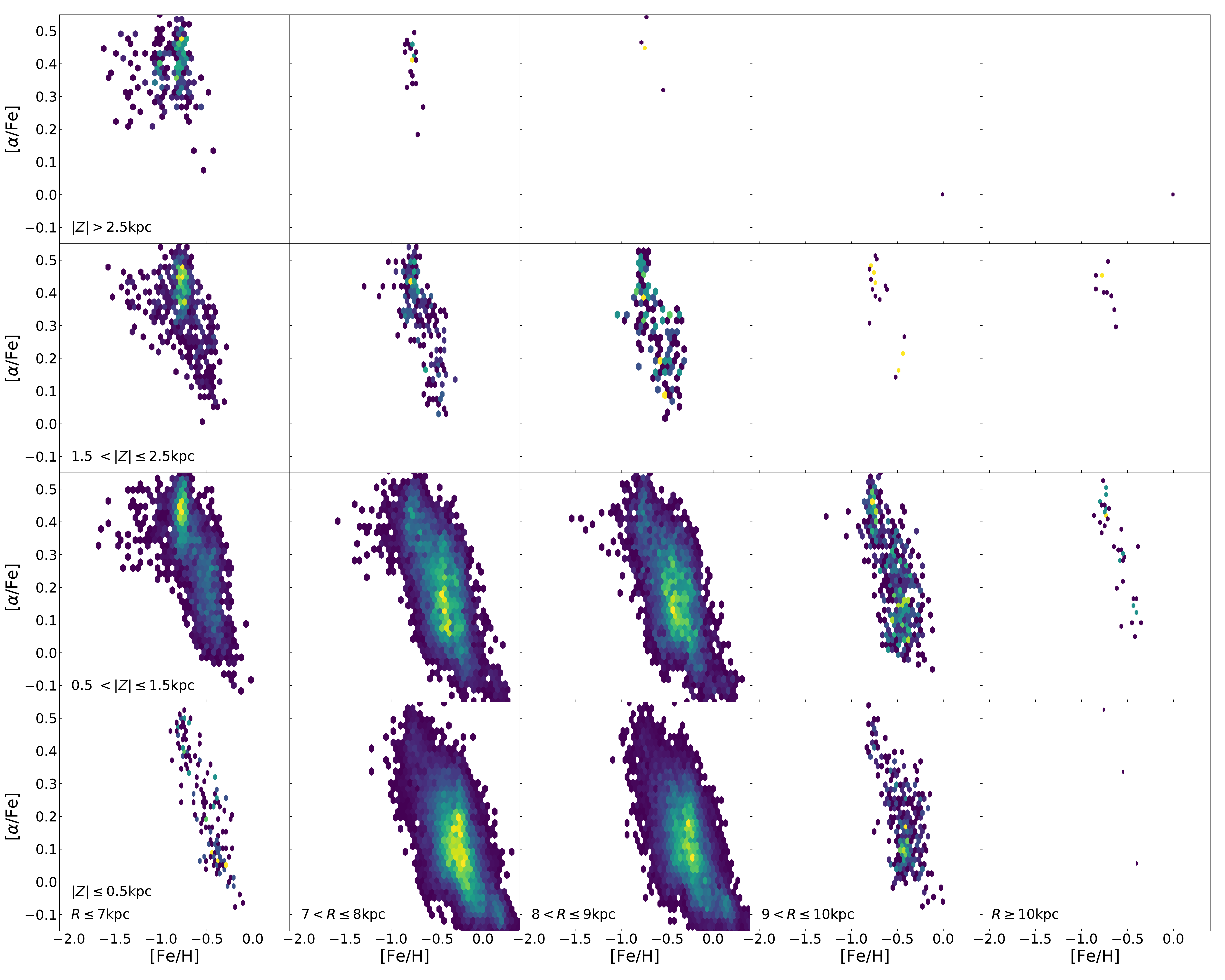}
\caption{\label{fig:tomo_actual} Tomographic plot of the $\alpha$ enhancement \alphafe\ \emph{vs.} \feh\ of the \qhigh\ 
sample of $59,012$ giant stars, separated by Galactocentric coordinates $R$ and $Z$.}
\end{center}
\end{figure*}

\begin{figure*}
\begin{center}
\plotone{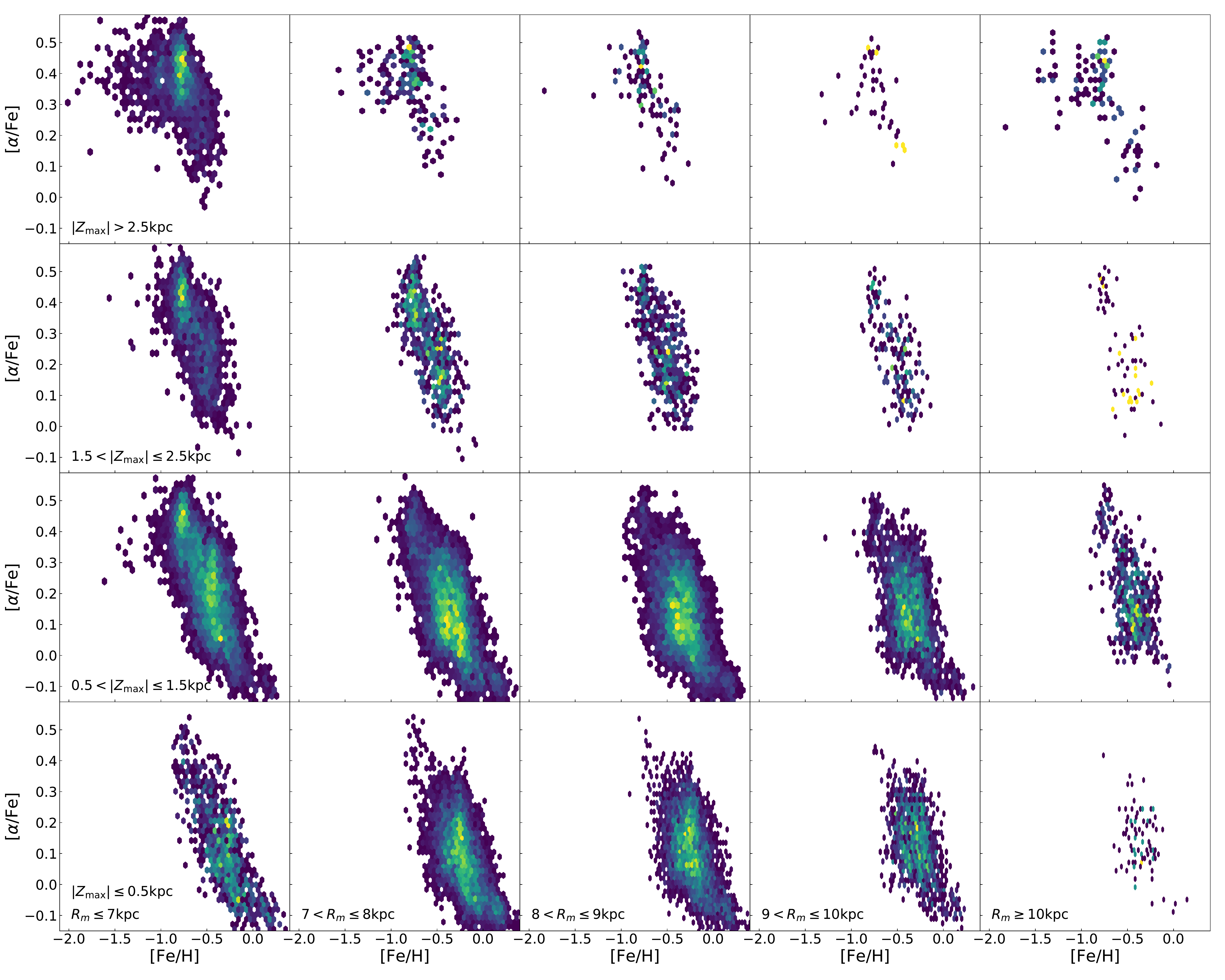}
\caption{\label{fig:tomo_orbits} Same as Figure \ref{fig:tomo_actual}, but shown for the mean radius $R_m$ and maximum 
height $Z_{\rm max}$ of the stars orbit,
for $58\,985$ giants. }
\end{center}
\end{figure*}

We conclude our discussion by taking a tomographic look at the \alphafe\ vs.~Fe relation and its 
systematic changes as a function of location in the Galactic disk following the analysis of \citet{hayden2015}, 
which is based on higher resolution APOGEE data (see, in particular, their Figure 4). Figure \ref{fig:tomo_actual} 
shows a similar representation for the RAVE DR6 \qhigh\ giant sample. As we move away from the Galactic plane, 
we systematically depopulate the high metallicity branch of the disk and start populating the very metal poor and 
very $\alpha$ enriched region of the \alphafe\ vs Fe relation. The respective distributions are, of course, a 
superposition of the chemical properties of the local environment convolved with that of other regions brought in 
by stars on highly eccentric and/or highly inclined orbits. Thanks to the astrometry provided by Gaia DR2, we 
are in the position to disentangle these effects, e.g. rather than plotting the \alphafe\ vs \feh\ relation at the 
actual position space, by using the average orbital radius (or guiding radius) $R_m$ and the maximum height 
above the Galactic plane ($Z_{\rm max}$) as the respective spatial coordinates (see Figure \ref{fig:tomo_orbits}). 
A comparison with Figure \ref{fig:tomo_actual} reveals that the stars currently located in the \rave\ volume are 
drawn from a considerably larger volume, showing a clear transition from the local disk to the more metal-poor disk 
at larger heights above the plane to the more halo-type population seen at the largest distances. Also the systematic 
dependence of metallicity and $\alpha$ enrichment on the eccentricity of the orbit (and/or vice versa) results in clear 
shifts towards metal poorer and more $\alpha$ enriched populations.

\section{Summary and conclusions}\label{sec:conclude}
The \rave\ final data release concludes a more-than-15 year effort to provide a 
homogeneous data set for Galactic archeology studies. \rave\ DR6 presents spectra and 
radial velocities for individual stars in the magnitude range $9<I<12$\,mag. The spectra 
cover a wavelength range of $8410-8795\,$\AA\ at an average resolution of $R\approx 7500$. The 
\rave\ catalog can be accessed via \texttt{doi.org/10.17876/rave/dr.6/001}. The typical 
\SNR\ of a \rave\ star is 40, and the typical uncertainty in radial velocity is 
$< 2$\,\kms. For the majority of the 518,387 \rave\ spectra, reliable atmospheric 
parameter can be derived with two different pipelines. The \texttt{MADERA} pipeline, based on the 
algorithms of MATISSE and DEGAS, derives  {stellar atmospheric parameters} purely spectroscopically with 
uncertainties in \teff, \logg, and \mh\ of 300\,K, 0.7\,\dex, and 0.2\,\dex, 
respectively. The Bayesian pipeline \texttt{BDASP} makes use of Gaia DR2 parallaxes, resulting 
in less-biased \teff\ estimates (compared to photometrically derived temperatures with the 
infrared flux method) for temperatures between $5200 - 6000$\,K  and  substantially 
improved estimates of \logg\ with an uncertainty of about 0.2\,\dex. \texttt{BDASP} also provides 
an improved distance estimate and extinction measure combining Gaia, spectroscopic, and 
photometric data, as well as estimates for the mass and the age of the respective star 
based on isochrone fitting. The  new pipeline, \texttt{GAUGUIN}, provides reliable 
$\alpha$ enhancements $\alphafe$ down to low metallicity stars with $\mh \la -1$ with 
an uncertainty of less than $0.2$\,\dex.  Two science applications regarding the chemo-dynamical 
structure of the volume probed by \rave\ demonstrate the potential of \rave\ for 
applications in the area of Galactic archeology.


\acknowledgments
Major scientific projects like the \rave\ survey are made possible by the contributions of many, in particular those of graduate 
students and postdocs. This final data release is published in memory of one of the first and most active student participants 
in \rave\, Gregory R. Ruchti (1980 - 2019), whose life was taken far too early.  His enthusiasm and  dedication were key elements 
of the success  of the \rave\ collaboration and his  contributions live on in the discoveries that are enabled by the \rave\ data.

Funding for \rave\  has been provided by: the Leibniz-Institut f\"{u}r Astrophysik 
Potsdam (AIP); the Australian Astronomical Observatory;  the Australian National 
University; the Australian Research Council; the French National Research Agency (Programme National Cosmology et Galaxies 
(PNCG) of CNRS/INSU with INP and IN2P3, co-funded by CEA and CNES); the 
German Research Foundation (SPP 1177 and SFB 881); the European Research Council 
(ERC-StG 240271 Galactica); the Istituto Nazionale di Astrofisica at Padova; The Johns 
Hopkins University; the National Science Foundation of the USA (AST-0908326); the W. M. 
Keck foundation; the Macquarie University; the Netherlands Research School for Astronomy; 
the Natural Sciences and Engineering Research Council of Canada; the Slovenian Research 
Agency (research core funding no. P1-0188); the Swiss National Science Foundation; 
the Science \& Technology Facilities 
Council of the UK; Opticon; Strasbourg Observatory; and the Universities of Basel, 
Groningen, Heidelberg, and Sydney. PJM is supported by grant 2017-03721 from the Swedish Research Council. LC is the recipient 
of the ARC Future Fellowship FT160100402. RAG acknowledges the support from the PLATO CNES grant. SM would like to acknowledge 
support from the Spanish Ministry with the Ramon y Cajal fellowship number RYC-2015-17697. MS thanks the Research School of Astronomy \& Astrophysics in Canberra for 
support through a Distinguished Visitor Fellowship. RFGW thanks the Kavli Institute for 
Theoretical Physics and the Simons Foundation for support as a Simons Distinguished 
Visiting Scholar. This research was supported in part by the National Science Foundation 
under Grant No. NSF PHY-1748958 to KITP.

This work has made use of data from the European Space Agency (ESA) mission
{\it Gaia} (\url{https://www.cosmos.esa.int/gaia}), processed by the {\it Gaia}
Data Processing and Analysis Consortium (DPAC,
\url{https://www.cosmos.esa.int/web/gaia/dpac/consortium}). Funding for the DPAC
has been provided by national institutions, in particular the institutions
participating in the {\it Gaia} Multilateral Agreement.

Based on data products from observations made with ESO Telescopes at the La Silla Paranal 
Observatory under programme ID 188.B-3002.

This work has also made use of observations obtained with the Apache Point Observatory 
3.5-meter telescope, which is owned and operated by the Astrophysical Research Consortium.

The reference grids and learning phases used to run MATISSE and 
\texttt{GAUGUIN} were computed with the high-performance computing facility
SIGAMM, hosted by OCA.

\software{HEALPix \citep{healpix}, IRAF \citep{iraf}, Matplotlib \citep{matplotlib}, numpy \citep{numpy}, pandas \citep{pandas}, RVSAO \citep{rvsao}}

\appendix

\section{Calibration and validation samples}\label{sec:external}

{As outlined in Section \ref{subsubsec:calibration}, the output of the \texttt{MADERA} pipeline is calibrated against a sample of reference stars in order to minimize possible systematic effects. The calibration was performed against the stellar atmospheric parameters \teff, \logg, and \mh, where for the observational data \feh\ was used as a calibrator.
The sample of calibration stars is detailed in Section 4 of \cite{kordopatis2013}  and Section 6.2 of \cite{kunder2017}. This sample consists of follow-up spectroscopy of \citep{ruchti2011} and Fulbright (in preparation), complemented by data in the PASTEL catalog \footnote{\texttt{http://pastel.obs.u-bordeaux1.fr}}, CFLIB \footnote{\texttt{http://www.noao.edu/cflib}} \citep{valdes2004} and Gaia benchmark stars of \citep{jofre2014}. 
Since not a suitable number of \rave targets that had literature data available could be identified for the full range of surface gravities, metallicities and effective temperatures, additional high-resolution spectra with coverage of the \rave\ wavelength range were binned down to Rave-resolution, analyzed with the MADERA pipeline, and added as additional calibration points.  
This happened as part of the work performed in pervious data releases, in particular DR4 \citep{kordopatis2013}. The calibration library consists of in total 1384 spectra  covering a broad range of stellar atmospheric parameters in the Kiel diagram (see Figure \ref{fig:kiel_compare}).}

{For the validation of the chemical abundance pipeline \texttt{GAUGUIN} (and the pipeline by \cite{boeche2011} for previous data releases), we compiled a list of individual elemental abundances available from high resolution data for stars in the \rave\ data base. These include data from \cite{ruchti2011} (383 spectra), Fulbright (in preparation, 178 spectra),\cite{adibekyan2012} (153 spectra), \cite{reddy2003} (5 spectra), \cite{reddy2006} (18 spectra), \cite{soubiran2005} (97 spectra), \cite{valenti2005} (72 spectra), Gaia-ESO \cite{gilmore2012} (71 spectra), and \cite{bensby2014} (113 spectra), as well as other field (204) and cluster (75) stellar spectra as detailed in \cite{kunder2017}. Since there are a number of common target in the reference data set, the complete validation data set contains 1369 spectra for 948 unique stars. When \alphafe\ was given in the respective catalog, this value was adopted, otherwise \alphafe\ was estimated from the average of \sife\ and \mgfe. If only one of the two abundances \sife\ and \mgfe\ as available, this value was used as a proxy for \alphafe.}

\bibliographystyle{aasjournal}
\bibliography{sample}

\end{document}